\documentclass[sigconf,nonacm]{acmart}

\renewcommand\footnotetextcopyrightpermission[1]{}
\usepackage[table]{xcolor}
\usepackage[flushleft]{threeparttable}
\usepackage{float}

\newcommand{\toolname}{\textit{UncertaintyVis}}

\begin{document}

\title{\toolname{}: Preserving Linguistic Uncertainty in Automated Text-to-Chart Generation}

\author{Songheng Zhang}
\email{shzhang.2021@phdcs.smu.edu.sg}
\affiliation{%
  \institution{Singapore Management University}
  \city{Singapore}
  \country{Singapore}}

\author{Emily Aurelia}
\email{eaurelia.2025@phdcs.smu.edu.sg}
\affiliation{%
  \institution{Singapore Management University}
  \city{Singapore}
  \country{Singapore}}

\author{Anthony Tang}
\email{tonyt@smu.edu.sg}
\affiliation{%
  \institution{Singapore Management University}
  \city{Singapore}
  \country{Singapore}}

\renewcommand{\shortauthors}{Zhang et al.}

\begin{abstract}
Data-rich documents pair narrative text with quantitative claims, and authors
routinely qualify those claims with linguistic uncertainty markers such as
``nearly,'' ``approximately,'' or ``at least.'' Automated text-to-chart systems
discard these markers, producing visualizations that appear definitive even
when the source text expresses hedged or incomplete knowledge. Readers may then
over-interpret precision and misjudge author intent. We present \toolname{}, a
system that preserves linguistic uncertainty during automated chart generation.
A formative corpus analysis of 211 uncertainty expressions across 12 documents
and 8 domains yielded a four-category taxonomy: Surface Form Normalization,
Precision Boundaries, Inferential Derivation, and Non-Inferable Gaps. We mapped
each category to chart-specific visual encodings that signal uncertainty without
disturbing the spatial integrity readers rely on, and implemented an end-to-end
pipeline pairing large language model text analysis with uncertainty-aware
rendering. In a two-part study with 12 participants, readers matched charts to
source text with 85\% accuracy and text to charts with 76\%. Uncertainty-aware
visualizations trended toward lower cognitive demand (effect sizes 0.460 and
0.769 for mental demand and effort), and 75\% of participants preferred them to
plain text, describing explicit uncertainty encodings as a basis for verifying
data claims. Encoding effectiveness varied by chart type: bar and pie encodings
performed consistently, while line chart encodings require redesign.
\end{abstract}


\begin{teaserfigure}
  \centering
  \includegraphics[width=1\linewidth]{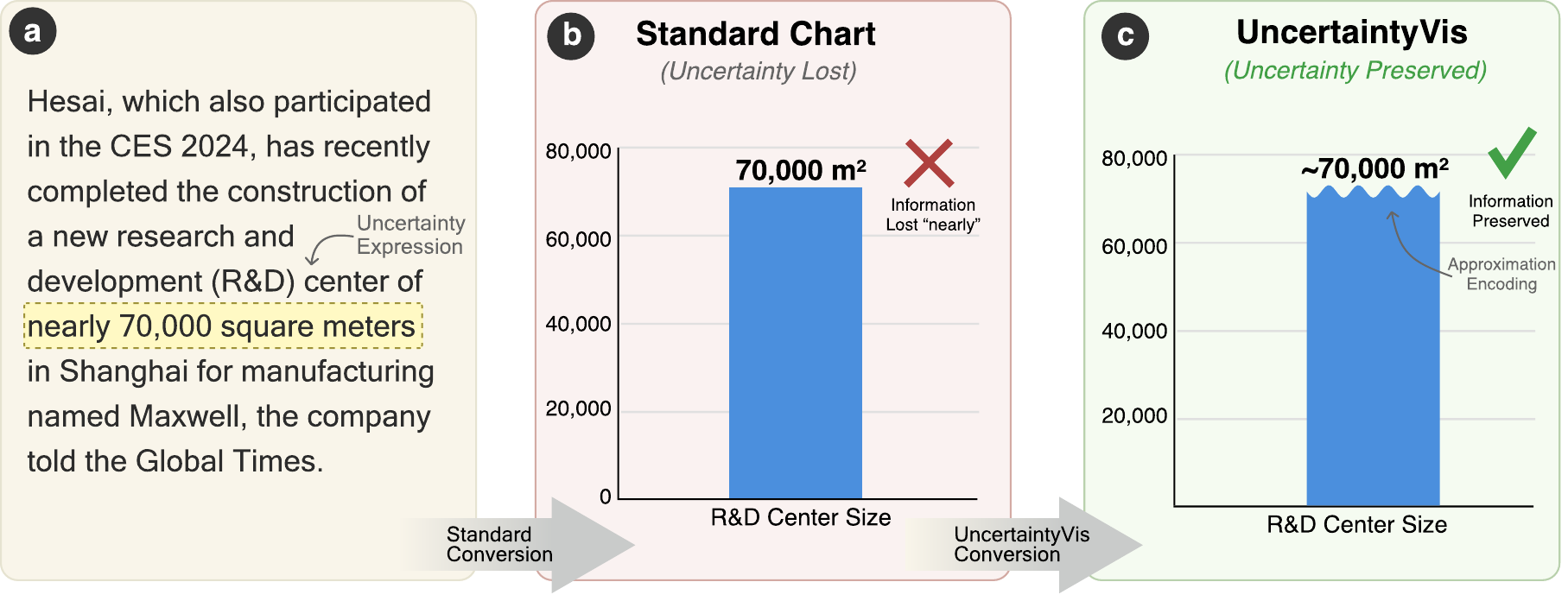}
  \caption{\textbf{Current text-to-chart systems discard textual uncertainty
  from source text.} a, Source text describing Hesai's R\&D center uses the
  approximation marker ``nearly'' before ``70,000 square meters,'' expressing
  deliberate imprecision about the facility size. b, Current text-to-chart
  conversion extracts only the numerical value (70,000
  m\textsuperscript{2}), losing the uncertainty marker ``nearly'' and its
  semantic meaning. c, \toolname{} preserves the uncertainty through dual
  encoding: the tilde symbol ($\sim$70,000 m\textsuperscript{2}) maintains
  the uncertainty marker textually, while the wiggly top on the bar provides
  a visual representation of the imprecise boundary. This approach ensures
  readers interpret the data as approximate rather than exact.}
  \Description{Three-panel figure comparing source text containing an
  approximation marker, a conventional chart that drops the marker, and the
  UncertaintyVis chart that preserves it through a tilde and a wiggly bar top.}
  \label{fig:teaser}
\end{teaserfigure}

\maketitle

\section{Introduction}
Data-rich documents, integrating textual narratives with quantitative data, are essential communication tools across scientific research, journalism, policy-making, and business analytics. 
Medical reports combine patient histories with laboratory values, policy briefs blend demographic statistics with legislative proposals, and data journalism articles interweave investigative narratives with empirical evidence. 
These documents serve diverse audiences, from researchers synthesizing findings to policymakers making evidence-based decisions to public readers interpreting current events. 
Effective interpretation requires understanding both the narrative context and the quantitative relationships within the data.

To enhance reader comprehension of data embedded in documents, several automated visualization systems extract numerical data and generate charts to reveal patterns and relationships~\cite{Zou2025GistVisAG, Zhang2024ChartifyTextAC, Masson2023CharagraphIG}. 
For example, GistVis~\cite{Zou2025GistVisAG} generates word-scale visualizations that embed charts directly within text, while ChartifyText~\cite{Zhang2024ChartifyTextAC} converts textual descriptions into standalone chart. 
These systems extract structured data from prose, identify appropriate chart types, and render visual representations that make implicit data patterns explicit. 
By transforming textual data into visual forms, these tools promise to make data-rich documents more accessible and interpretable.

However, these automated approaches treat all textual data as precise and certain, systematically discarding the linguistic uncertainty markers that authors deliberately include to qualify their claims (Figure~\ref{fig:teaser}). Consider phrases like "nearly 70,000 square meters"(Figure~\ref{fig:teaser}a)~\cite{Zou2025GistVisAG, Masson2023CharagraphIG}. 
Such expression conveys important semantic information about data precision, inferential confidence, and completeness~\cite{Hullman2019WhyAD, BenitoSantos2021EvaluatingAT}. 
Current systems strip away these qualifications, generating charts that appear definitive even when the underlying text expresses uncertainty~\cite{Hullman2019WhyAD}. 
This creates a fundamental mismatch between how humans communicate data with nuanced qualifications and how systems process it~\cite{BenitoSantos2021EvaluatingAT, Thawani2021RepresentingNI} . 
The consequence is semantic information loss: \textbf{readers viewing these charts may over-interpret precision, misunderstand author intent, and make decisions based on false certainty}~\cite{Joslyn2021VisualizingUF, Greis2018UncertaintyVI, Kale2020VisualRS, Hullman2019WhyAD}.

In this work, we address this gap by developing \toolname{}, a system that preserves linguistic uncertainty when automatically generating visualizations from data-rich documents. 
We began with a formative corpus analysis of 12 documents across 8 domains such as medical, political, sports, and policy domains to identify how authors express uncertainty through language (Figure~\ref{fig:corpus-analysis}a). 
From this analysis, we developed a four-category taxonomy of linguistic uncertainty:  Inferential Derivation (logically derivable but not explicitly stated values),  Precision Boundaries (approximations and ranges), Surface Form Normalization (varied numerical formats like "5.2 million" versus "5,200,000"), and Non-Inferable Gaps (fundamentally missing data) (Figure~\ref{fig:corpus-analysis}d)~\cite{BenitoSantos2021EvaluatingAT}.
We then designed visual encoding strategies that map each uncertainty type to appropriate chart-specific visual treatments, such as squiggly top precision boundaries and interactive elements for inferential derivation, while maintaining the spatial integrity essential for accurate chart reading (Figure~\ref{fig:taxonomy-overivew})~\cite{Boukhelifa2012EvaluatingSA, Maceachren2012VisualS, Wood2012SketchyRF, Correll2018ValueSuppressingUP}. 
Finally, we implemented an end-to-end system combining large language model-based text analysis with uncertainty-aware visualization generation (Figure~\ref{fig:reading-interface}).

We evaluate \toolname{} through a two-part user study with 12 participants, examining three interdependent components simultaneously (Figure~\ref{fig:study-design}). 
First, we assess the validity of our taxonomic approach by testing whether the four uncertainty categories produce visual encodings that readers can reliably identify and interpret. 
Second, we examine the effectiveness of our visual encoding strategies by measuring how accurately participants match linguistic uncertainty expressions to their corresponding visual treatments and vice versa. 
Third, we evaluate the practical utility of the \toolname{} system by measuring whether uncertainty-aware visualizations improve comprehension, reduce cognitive workload, and increase reader confidence when engaging with data-rich documents.
The first evaluation uses bidirectional matching tasks to test encoding and decoding accuracy: participants match charts to source text descriptions and vice versa. 
The second evaluation employs document reading tasks, comparing comprehension outcomes when readers encounter uncertainty-enhanced visualizations versus plain text. 
We measure matching accuracy and perceived semantic suitability in Part 1, and comprehension accuracy, reading efficiency, cognitive workload, and user preference in Part 2. 
Our results reveal how uncertainty-aware visualizations influence readers' ability to accurately interpret linguistic uncertainty while maintaining efficient comprehension.

Our evaluation yields three key findings. 
First, our visual encoding strategies successfully communicate uncertainty semantics to readers: chart-to-text matching achieves 85\% accuracy and text-to-chart matching achieves 76\%, demonstrating that participants can reliably interpret uncertainty encodings once presented (Figure~\ref{fig:eval-part1-results}). 
Encoding effectiveness varies substantially by chart type, with bar and pie chart encodings succeeding consistently while line chart encodings require fundamental redesign, providing concrete targets for future iteration (Table~\ref{tab:text-to-chart-performance}). 
Second, uncertainty-aware visualizations produce a meaningful trend toward reduced cognitive demand during document reading, with medium to large effect sizes on perceived mental demand and effort (effect sizes 0.460 and 0.769 respectively), suggesting practical cognitive benefits that larger-scale studies should investigate further (Figure~\ref{fig:eval-NASA}). 
Third, 75\% of participants express strong preference for uncertainty-aware visualizations over plain text, reporting that explicit uncertainty encodings function as trust infrastructure that enables them to verify data claims and calibrate their interpretations of document content (Figure~\ref{fig:preference-distribution}). 
Together, these findings demonstrate that preserving linguistic uncertainty in automated chart generation is both technically feasible and practically beneficial, while identifying specific design refinements needed for robust deployment.

This work makes three primary contributions to uncertainty visualization and automated chart generation. 
\begin{enumerate}
    \item We present a taxonomy characterizing how uncertainty is expressed linguistically in data-rich documents, extending beyond traditional statistical uncertainty to linguistic uncertainty communicated through word choice. 
    \item We provide a framework of visual encoding strategies that map linguistic uncertainty categories to chart-specific visual encodings, grounded in uncertainty visualization principles while respecting fundamental chart interpretation practices. 
    \item We implemented a proof-of-concept system and conducted an empirical evaluation to examine how preserving author intent in automated visualization pipelines influences both the feasibility and the benefits for reader comprehension of data-rich documents.
\end{enumerate}

\section{Related Work}

\subsection{Automated Chart Generation from Text}

Automated visualization systems can readily generate charts from well-structured data sources such as tables and JSON files, but extracting quantitative insights from narrative text presents substantial challenges~\cite{zhu2020survey, Thawani2021RepresentingNI, Zhang2023AdaVisAA, Wang2023LLM4VisEV, Zhou2020Table2ChartsRC}. 
The extraction challenge operates across three distinct dimensions. 
First, format heterogeneity means identical values appear in varied textual representations: a single sentence may state 'enrollment reached 2.5M in 2020 versus 2,500,000 in 2019,' requiring systems to recognize these as equivalent. 
Second, linguistic entanglement means numerical values are embedded within textual structures that carry semantic meaning beyond the number itself, demanding contextual interpretation rather than simple pattern matching. 
Third, interpretive ambiguity means the same expression can carry different meanings depending on domain conventions and authorial intent, making reliable extraction dependent on broader narrative understanding. 
Together, these challenges mean that traditional data parsing methods fail on unstructured text~\cite{Gpfert2022MeasurementEW, Almasian2023CQEAC}, motivating the development of more sophisticated extraction approaches that can serve readers across scientific~\cite{Masson2023CharagraphIG, Zou2025GistVisAG}, journalistic, and policy domains who would otherwise need to manually identify and interpret numerical data across dense documents.

Recent systems have made significant progress in addressing these extraction challenges. 
Charagraph~\cite{Maceachren2012VisualS} enables users to generate interactive charts from data-rich paragraphs, allowing readers to explore statistical information embedded in prose. 
For example, from text stating "the university has 10,000 students, with 6,000 studying engineering," Charagraph can generate a bar chart comparing total and engineering enrollment. 
GistVis~\cite{Zou2025GistVisAG} extends this concept by embedding word-scale visualizations directly within textual narratives, creating in-situ visualizations that enhance the reading experience without requiring readers to navigate away from the text. 
Rather than relying on rigid rule-based approaches, GistVis utilizes Large Language Models (LLMs) to obtain quantitative information more flexibly, representing a significant advance in extracting quantitative facts from unstructured text~\cite{Dagdelen2024StructuredIE, Jiao2023InstructAE}.

Despite these advances, current text-to-chart systems fail to address the critical challenge of representing linguistic uncertainty expressed in the source text~\cite{Zou2025GistVisAG, Masson2023CharagraphIG, Zhang2024ChartifyTextAC}. 
When converting textual narratives to charts, these systems extract specific data points but systematically discard linguistic qualifications that authors include to convey data limitations. 
Figure~\ref{fig:teaser} illustrates this contrast directly: when source text states that \textit{Hesai constructed an R\&D center of 'nearly 70,000 square meters'} a conventional text-to-chart system extracts only the numerical value 70,000 and renders a standard bar chart, discarding the approximation marker 'nearly' and the imprecision it communicates (Figure~\ref{fig:teaser}b). 
However, the desired chart should instead preserve this linguistic uncertainty through a dedicated visual encoding, representing the approximate boundary (Figure~\ref{fig:teaser}c). 
This contrast reveals a fundamental limitation of current systems: by treating all textual data as certain and complete, they ignore the qualified statements, potentially misleading readers about the reliability of visualized information~\cite{Hullman2019WhyAD, Joslyn2021VisualizingUF}. 
This creates a pressing need for visualization methods that explicitly preserve and represent linguistic uncertainty expressed in data-rich documents.

\subsection{Uncertainty Visualizations}
Uncertainty visualization is a well-established research area with numerous techniques for representing the statistical confidence and measurement uncertainty associated with quantitative data~\cite{Hullman2019InPO, Kamal2021RecentAA}. 
These methods employ two primary strategies. 
The first strategy directly represents statistical uncertainty using visual cues such as transparency/color~\cite{Correll2018ValueSuppressingUP,Maceachren2012VisualS}, fuzziness/sketchiness~\cite{Boukhelifa2012EvaluatingSA, Wood2012SketchyRF}, or size variations~\cite{padilla2020uncertainty} to depict confidence levels of data points. 
The second strategy uses statistical methods to impute missing values, and also visualizes the resulting uncertainty through techniques like error bars, probability density plots, and confidence intervals~\cite{Song2019WheresMD, Alsufyani2024VisualizationOM, Kale2019HypotheticalOP, Kay2016WhenI}. 
These approaches effectively convey statistical variance, measurement error, and confidence bounds for quantitative data.

However, these established techniques cannot address uncertainty expressed through ambiguous language in textual narratives~\cite{Hullman2019InPO}. 
Existing uncertainty visualization methods are designed for numerical inputs\cite{Castro2021ExaminingEI, padilla2020uncertainty}. 
They require quantitative measures of uncertainty such as standard deviations or confidence levels. 
They cannot process linguistic markers like "approximately 15\%" or interpret phrases like "significant improvement" that lack explicit numerical bounds. 
Linguistic uncertainty represents a fundamentally different challenge: rather than quantifying statistical confidence, authors use word choice to qualify their statements.
Current visualization techniques provide no systematic methods for categorizing these linguistic uncertainty types or translating them into visual encodings, creating a significant research gap at the intersection of text analysis and uncertainty representation.

\subsection{Linguistic Uncertainty in Natural Language Processing}

In natural language processing, researchers have examined how quantitative information is expressed in text and developed methods to extract it despite linguistic ambiguity~\cite{Gpfert2022MeasurementEW}. 
This research focuses on data-rich documents such as scientific papers, news articles, and business reports, where textual narratives mix with quantitative information~\cite{Dagdelen2024StructuredIE, Chen2021FinQAAD, Zhu2021TATQAAQ, Alonso2018QuantitativeIE, Liberatore2024QuantitativeIE}. 
Researchers have identified several key dimensions of linguistic uncertainty. 
Surface form normalization addresses challenges where numbers appear in varied formats—for instance, "2.5M in 2020 versus 2,500,000 in 2019" requires recognizing that different representations encode the same value~\cite{Gpfert2022MeasurementEW}. 
Semantic ambiguity arises when interpretation depends on broader context, such as understanding that "significant improvement" implies an increase without stating specific magnitude. 
Domain-specific challenges include specialized terminology, informal abbreviations, and jargon variations across fields. 
To address these challenges and accurately extract quantitative information, researchers have developed rule-based systems~\cite{Madaan2016NumericalRE}, machine learning approaches\cite{Foppiano2019AutomaticIA}, and more recently, large language models that can reason about numerical content in context\cite{Dagdelen2024StructuredIE}.

These NLP approaches, however, are fundamentally misaligned with visualization needs because they aim to eliminate uncertainty rather than preserve it for human interpretation~\cite{Gpfert2022MeasurementEW, Almasian2023CQEAC}. 
The goal of NLP extraction systems is to minimize ambiguity and maximize extraction accuracy for downstream computational tasks, converting "approximately 15\%" into a precise "15\%" value. 
In contrast, visualization for human readers should communicate uncertainty as valuable information, preserving the author's deliberate choice to write "approximately" rather than presenting a false sense of precision. 
This fundamental mismatch in objectives has consequences: no systematic taxonomy exists for categorizing uncertainty types relevant to visualization, and no established methods exist for translating linguistic uncertainty markers into visual encodings that readers can interpret. 
Our work addresses this gap by developing \toolname{}, a framework that provides both a taxonomy of linguistic uncertainty in data-rich documents and corresponding visualization techniques that preserve rather than eliminate the semantic information authors encode through their word choices.

\section{Formative Study: Understanding Uncertainty in Text}

\begin{figure*}
    \centering
    \includegraphics[width=\linewidth]{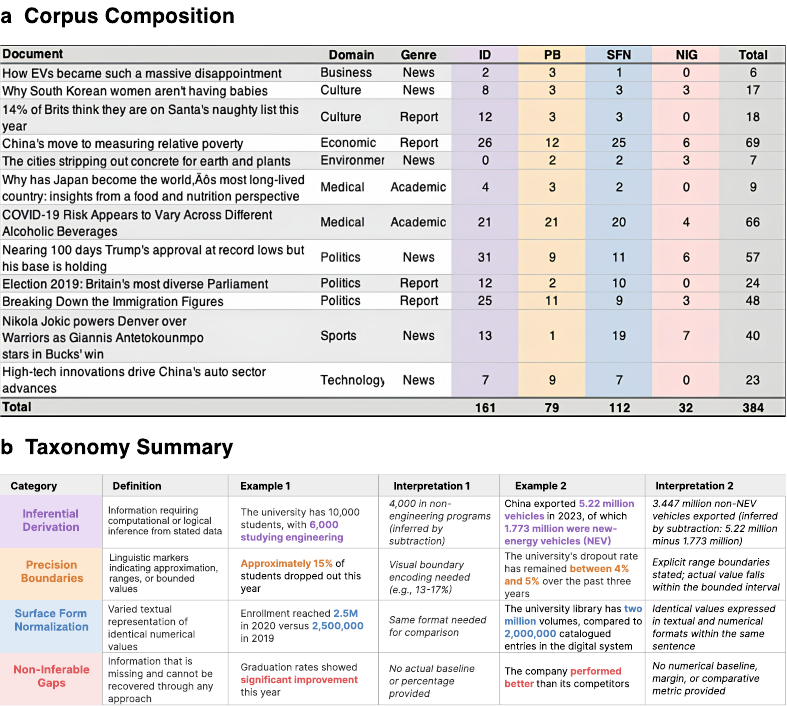}
    \caption{\textbf{We analyzed 211 uncertainty expressions across 12 documents spanning 8 domains and 3 genres to establish empirical foundations for the taxonomy.} a, Document composition table showing uncertainty category counts for each document, ordered by domain and genre. b, Taxonomy summary table defining each uncertainty category with representative examples.}
    \label{fig:corpus-analysis}
\end{figure*}

Current automated chart generation systems treat all textual data as definitive and complete, thereby overlooking linguistic expressions authors use to convey uncertainty, approximation, and qualified claims~\cite{Zou2025GistVisAG, Zhang2024ChartifyTextAC, Masson2023CharagraphIG}. 
This approach systematically eliminates critical semantic information embedded in natural language. 
Expressions like "approximately," "more than," and "between X-Y" carry essential uncertainty semantics that should directly inform visualization design decisions\cite{Hullman2019WhyAD}. 
Systematic understanding of how uncertainty is expressed in data-rich documents is fundamentally essential for developing effective uncertainty-preserving visualization approaches.

To empirically establish this systematic understanding, the formative study addresses four interconnected  questions. 
First, what distinct types of uncertainty expressions exist in data-rich documents across different domains and genres? 
Second, how frequently do these uncertainty types occur, and which distributional patterns emerge? 
Third, in what ways do uncertainty expressions vary between different document types—such as news articles, academic papers, and reports—and topical domains? 
Finally, how frequently do multiple uncertainty types co-occur within individual expressions, and what implications does this have for designing composable visual encoding strategies? 
These questions provide the empirical foundation for developing a comprehensive taxonomy that can guide uncertainty-aware visualization design.

\subsection{Corpus Development}
The corpus development strategy prioritized purposive sampling with theoretical grounding to ensure comprehensive coverage of uncertainty expressions without domain-specific bias and limited generalization. 
We established three core design principles. 
First, ensuring systematic domain representation across multiple domains to avoid genre-specific limitations. 
Second, selecting materials with high data density and rich quantitative content suitable for uncertainty analysis. 
Third, maintaining equilibrium between analytical depth and domain coverage to enable both detailed qualitative coding and generalizable findings across diverse document types.

The final corpus comprised 12 carefully curated documents strategically selected across 8 distinct and complementary topical domains: politics, medical, environment, sports, business, culture, economic, and technology. 
The collection included news articles, academic papers, and reports to ensure genre diversity, with each document meeting strict selection criteria of rich numerical information and diverse uncertainty expression patterns. 
Document sources were drawn from authoritative publications, research institutions, and established media outlets, ensuring both credibility and representativeness of professional data-rich writing across multiple authoring contexts and institutional perspectives (Figure~\ref{fig:corpus-analysis}a).

\subsection{Methodology}
The methodology began with established taxonomic categories from quantitative information extraction (QIE) research, which has systematically categorized challenges in accurate identification and extraction of definitive numerical values from unstructured text~\cite{Thawani2021RepresentingNI, Gpfert2022MeasurementEW, Almasian2023CQEAC, Foppiano2019AutomaticIA}. 
However, QIE frameworks fundamentally focus on precise extraction of quantitative facts, treating uncertainty expressions as extraction obstacles to overcome rather than meaningful semantic information to preserve. 
This created a critical paradigmatic gap for uncertainty-preserving visualization goals: QIE aims to eliminate ambiguity for computational processing, while \toolname{} requires preserving and interpreting uncertainty semantics as valuable information to inform appropriate visual encoding strategies for human interpretation.

We systematically adapted QIE categories through a reframing process that reconceptualized extraction challenges as uncertainty preservation opportunities. 
For example, where QIE research identifies linguistic modifiers and approximation markers such as "around" and "nearly" as obstacles to accurate numerical extraction, we reconceptualized these as valuable precision boundary indicators that carry essential semantic meaning for visualization design~\cite{Davidov2010ExtractionAA}. 
This adaptation process involved three systematic steps. 
First, extending existing categories to address uncertainty-specific phenomena, such as adding inferential derivation for logically derived quantitative information. 
Second, removing extraction-oriented categories that focused purely on computational extraction rather than semantic interpretation. 
Third, developing new definitional frameworks that prioritize uncertainty semantics over extraction accuracy.

To ensure category stability and applicability, we implemented multi-round iterative coding validation across all 12 documents using systematic refinement protocols. 
Each coding round involved systematic category application followed by refinement analysis: initial application of adapted categories to identify boundary issues and classification ambiguities, followed by systematic category refinement and definitional clarification based on encountered coding challenges and definitional gaps. 
This iterative process continued through successive coding iterations until consistent categorization was achieved without additional category modifications. 
Quality assurance was maintained through explicit decision rules and clear categorization criteria for each uncertainty type, ensuring that uncertainty expressions could be reliably classified according to explicit criteria, thereby ensuring consistent categorization.

\subsection{Taxonomy Development}
The systematic adaptation and validation process yielded four distinct but complementary uncertainty categories that comprehensively capture different dimensions of uncertainty expressions in data-rich documents (Figure~\ref{fig:corpus-analysis}b)~\cite{BenitoSantos2021EvaluatingAT}. 
These categories operate as a complementary rather than mutually exclusive framework, enabling simultaneous classification when uncertainty expressions exhibit multiple characteristics within a single textual instance. 
\textbf{Inferential Derivation} captures quantitative information that is not explicitly stated but can be calculated or logically deduced from other values present in the document. 
This category encompasses two distinct inferential mechanisms: mathematical calculations and logical entity inferences. 
In mathematical calculations, a value is derived through arithmetic operations on stated quantities. 
For example, 'the university has 10,000 students, with 6,000 studying engineering' allows the reader to infer that 4,000 students study non-engineering fields through subtraction (Figure~\ref{fig:corpus-analysis}b). 
In logical entity inferences, a category or group is contextually derived from stated information. 
For example, 'China exported 5.22 million vehicles in 2023, of which 1.773 million were new-energy vehicles' allows the reader to infer that 3.447 million non-new-energy vehicles were exported as the complementary remainder.

\textbf{Precision Boundaries} captures numerical statements that explicitly qualify their own precision limits through linguistic markers indicating approximation, ranges, or directional comparisons. 
Authors use this category to communicate deliberate imprecision rather than precise values. 
Approximation markers include expressions such as 'approximately 15\% of students dropped out this year,' 'around 1.4 million residents,' and 'nearly 70,000 square meters' (Figure~\ref{fig:teaser}a), where the author signals that the stated value is an estimate rather than an exact measurement. 
Range specifications include expressions such as 'the university's dropout rate has remained between 4\% and 5\% over the past three years' and 'the number of infected students ranges from 30 to 50 cases per day,' where the author provides explicit lower and upper bounds (Figure~\ref{fig:corpus-analysis}b). 
Directional comparisons include expressions such as 'more than 8\%,' 'fewer than 200 participants,' 'slightly above the threshold,' and 'well below the national average,' where the author communicates a value relative to a reference point without specifying the exact quantity. 
Across all three forms, the author deliberately chooses imprecise language to reflect genuine uncertainty about the exact value.

\textbf{Surface Form Normalization} captures the variation in textual representations of identical numerical values across documents and authors. 
The same underlying quantity frequently appears in multiple formats within a single document, creating representational inconsistency that requires resolution before visual comparison is possible. 
Numerical and abbreviated forms represent the most common variation: 'enrollment reached 2.5M in 2020 versus 2,500,000 in 2019' presents identical values in abbreviated and fully expanded formats within a single sentence (Figure~\ref{fig:corpus-analysis}b). 
Textual and numerical forms create a second common variation: 'the university library has two million volumes, compared to 2,000,000 catalogued entries in the digital system' expresses the same quantity through written and numerical formats within the same sentence. 
Unit and scale variations produce a third form: 'GDP grew by 1.2 trillion dollars' and 'GDP grew by 1,200 billion dollars' express the same magnitude through different scale choices. 
In all cases, the author makes a formatting choice that carries contextual meaning while representing a value that must be standardized for accurate visual comparison.

\textbf{Non-Inferable Gaps} captures quantitative information that is fundamentally absent from the document and cannot be recovered through any analytical approach, including the other three uncertainty categories. 
This category requires strict separation from information that appears absent but could be recovered through inferential derivation, precision boundaries, or surface form normalization. 
Statements expressing qualitative change without quantitative grounding represent the most common form: 'graduation rates showed significant improvement this year' communicates that change occurred but provides no baseline value, no current value, and no magnitude that could support numerical visualization (Figure~\ref{fig:corpus-analysis}b). 
Relative comparisons without reference values constitute a second common form: 'the company performed better than its competitors' implies a ranking but provides no numerical basis for comparison. 
Incomplete temporal data represent a third common form: 'sales have grown steadily over the past decade' describes a trend without providing the data points needed to construct it. 
In all cases, no value can be derived, estimated, or recovered from the document, making the gap fundamentally irreducible.

\subsection{Corpus Analysis Results}
Analysis of uncertainty expressions across the 12-document corpus reveals distinct frequency patterns among the four taxonomy categories, with inferential derivation and surface form normalization accounting for the majority of uncertainty expressions in data-rich documents  (Figure~\ref{fig:corpus-analysis}a). 
We identified 211 uncertainty expressions within the corpus. 
Inferential derivation emerged as the most frequent category, followed by surface form normalization, precision boundaries, and non-inferable gaps. 
This distribution indicates that certain types of uncertainty expressions occur more commonly in data-rich writing, with the top two frequent categories representing the particularly critical challenges for automated visualization systems to address.

Beyond individual category frequencies, analysis identified substantial co-occurrence of multiple uncertainty types within single uncertainty expressions, with the majority of uncertainty expressions containing two or more uncertainty categories simultaneously. 
The most common combination involves inferential derivation together with precision boundaries and surface form normalization appearing within a single uncertainty expression. 
This pattern typically manifests in statements such as 'the university enrolled around 2.5M students last year, with approximately 900,000 in undergraduate programs,' which simultaneously exhibits an approximation marker ('around'), a format variation ('2.5M' requiring normalization), and a derivable value (the remaining students in non-undergraduate programs inferred through subtraction). 
The second most common pattern pairs inferential derivation with surface form normalization, such as 'the university library holds 1M volumes across its main campus, while the science faculty alone accounts for 800,000 books,' where the abbreviated format '1M' requires normalization while the number of non-science faculty volumes remains derivable through subtraction but is never explicitly stated. 
The third pattern pairs non-inferable gaps with surface form normalization, as seen in statements such as 'the university awarded 3M dollars in scholarships in 2023 compared to 2,000,000 dollars the previous year, though student satisfaction with financial aid showed notable improvement,' where identical values appear in different formats requiring normalization while the magnitude of satisfaction improvement cannot be recovered through any analytical approach.
These co-occurrence patterns demonstrate that uncertainty expressions in data-rich documents are semantically layered rather than categorically isolated.

Analysis across domains and genres reveals preliminary patterns in how different document types use linguistic uncertainty, though these observations should be interpreted with caution given the small number of documents per domain and the absence of length normalization across the corpus. 
Within our sample, news and journalism articles tend to use precision boundaries more frequently, employing approximation markers such as 'approximately,' 'around,' and 'nearly' to qualify reported values. 
Academic papers show a higher tendency toward surface form normalization and inferential derivation.
Business and organizational documents show more mixed patterns, with political documents in our sample using precision boundaries more frequently while economic documents tend toward inferential derivation.

\subsection{Toward Design Requirements for Uncertainty-Aware Visualizations}\label{sec:design-requirement}
The corpus analysis yields three findings that directly motivate design requirements for uncertainty-aware visualization. 
First, uncertainty in data-rich documents operates through four distinct uncertainty categories, each requiring different visual treatment. 
Second, inferential derivation and surface form normalization dominate the corpus, appearing in the majority of analyzed expressions, indicating that these categories represent the most critical challenges for automated visualization systems to address. 
Third, the majority of uncertainty expressions contain two or more co-occurring categories simultaneously, meaning that visualization design must handle layered semantic complexity rather than isolated uncertainty categories. 
Together, these findings motivate six requirements that guide both the visual encoding strategies and the technical implementation of \toolname{}.

\begin{itemize}

\item \textbf{R1}-Visual encodings must be applied selectively, only to data elements that contain identified linguistic uncertainty expressions. 
The corpus analysis shows that uncertainty expressions are distributed unevenly across documents, appearing alongside precise and definitive statements that carry no uncertainty. 
Applying visual uncertainty indicators universally would reduce their communicative value by suggesting that all data carries equal uncertainty, which misrepresents the source text and risks misleading readers about data quality.

\item \textbf{R2}-Visual encodings must preserve the exact numerical values and mathematical relationships present in the source text. 
The corpus analysis revealed that precision boundary expressions such as 'between 4\% and 5\%' and 'approximately 15\%' consistently contain precise numerical anchor points that serve as reference values even while expressing imprecision. 
Visualizations should preserve these anchor values. This will strengthen the quantitative integrity of the chart and enable accurate data comparison.

\item \textbf{R3}-Readers must be able to trace any visual encoding directly back to its original textual expression in the source document. 
The corpus analysis revealed that the majority of uncertainty expressions contain multiple co-occurring categories simultaneously, producing layered visual encodings that risk becoming uninterpretable without clear connections to their textual origins. 
Traceability enables readers to verify how UncertaintyVis interpreted each expression and to assess whether the resulting visual encoding accurately reflects the author's intended meaning.

\item \textbf{R4}-Visualization systems must be capable of representing multiple uncertainty categories simultaneously within a single data element. 
The corpus analysis shows that the majority of uncertainty expressions contain two or more co-occurring categories, as in 'the university enrolled around 2.5M students last year, with approximately 900,000 in undergraduate programs,' which simultaneously exhibits an approximation marker, a format variation requiring normalization, and a derivable complementary value.
Systems that encode only a single uncertainty type per data element would fail to represent the full semantic complexity of such expressions, producing visualizations that partially misrepresent the source text.

\item \textbf{R5}-Uncertainty encodings must function consistently and legibly across all supported chart types. 
Different data characteristics require different visualization forms: categorical comparisons favor bar charts while temporal trends require line charts and compositional relationships suit pie charts. 
An uncertainty encoding strategy that functions only for one chart type would be insufficient for the diversity of quantitative content found in real data-rich documents, limiting the practical applicability of the system.

\item \textbf{R6}-Uncertainty encodings must use intuitive visual conventions that minimize the cognitive burden placed on readers. 
The corpus analysis shows that the majority of uncertainty expressions contain multiple co-occurring uncertainty categories, meaning readers frequently encounter several simultaneous visual encodings within a single chart. 
Encodings that require readers to learn entirely novel visual languages would impose prohibitive learning overhead, particularly when multiple uncertainty types appear together. 
Building on established visual conventions wherever possible reduces this burden and supports accurate interpretation without extensive prior training.

\end{itemize}

The first three requirements ensure semantic fidelity: visual encodings must accurately preserve the uncertainty meaning authors express through language. 
The final three requirements ensure system integration: visualizations must handle the complexity, diversity, and cognitive demands of data-rich documents. 
Together, these six requirements guide the visual encoding strategies (Section~\ref{sec:visual-encoding}) and technical implementation (Section~\ref{sec:system}).

\section{Visual Encoding Strategy}\label{sec:visual-encoding}
\begin{figure}
    \centering
    \includegraphics[width=\linewidth]{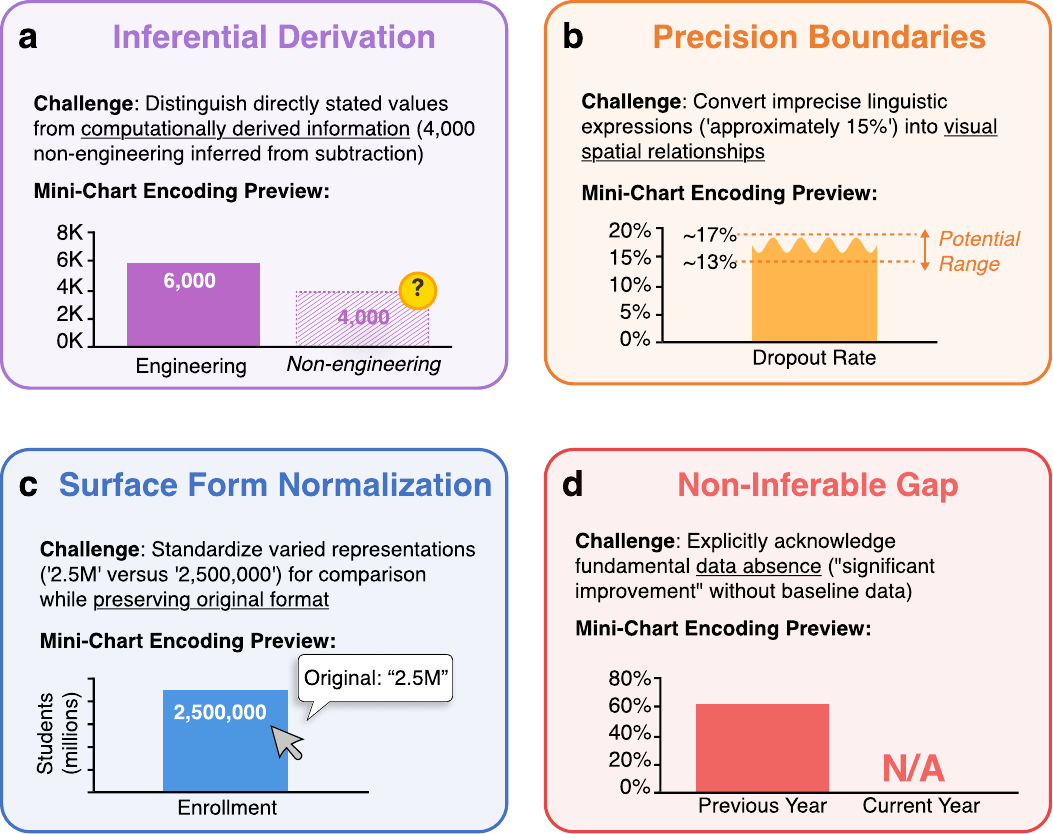}
    \caption{The taxonomy identifies four complementary uncertainty types, each presenting unique visualization challenges. a, Inferential Derivation marks computed information (10,000 total students with 6,000 in engineering → 4,000 non-engineering inferred). Challenge: distinguish stated from derived values. Mini-chart shows sketched bar with yellow question mark for inferred values.  b, Precision Boundaries handles linguistic approximation ("approximately 15\% dropout"). Challenge: convert imprecise language into visual spatial boundaries. Mini-chart shows ~15\% with gradient indicating potential range of 13-17\%. c, Surface Form Normalization addresses varied representations of identical values ("2.5M" versus "2,500,000" enrollment). Challenge: standardize for comparison while preserving original format. Mini-chart shows tooltip revealing "Original: 2.5M".d, Non-Inferable Gaps acknowledges missing information ("significant improvement" without baseline). Challenge: explicitly indicate absence. Mini-chart shows "N/A" label where percentage would appear.}
    \label{fig:taxonomy-overivew}
\end{figure}

\subsection{Category-Specific Encoding Strategies}
Each uncertainty category addresses distinct semantic aspects requiring specialized visual encoding approaches (Figure~\ref{fig:taxonomy-overivew}). 
Inferential Derivation demands provenance marking that differentiates directly extracted from logically derived information. 
Precision Boundaries require spatial boundary encoding that communicates potential value ranges while maintaining mathematical precision (R2). 
Surface Form Normalization necessitates interactive mechanisms that enable value comparison while preserving source format authenticity (R3). 
Non-Inferable Gaps require explicit void indication that acknowledges information limitations honestly. 
This category-specific approach ensures appropriate semantic fidelity while supporting distinct interpretation needs.

\begin{figure}
    \centering
    \includegraphics[width=\linewidth]{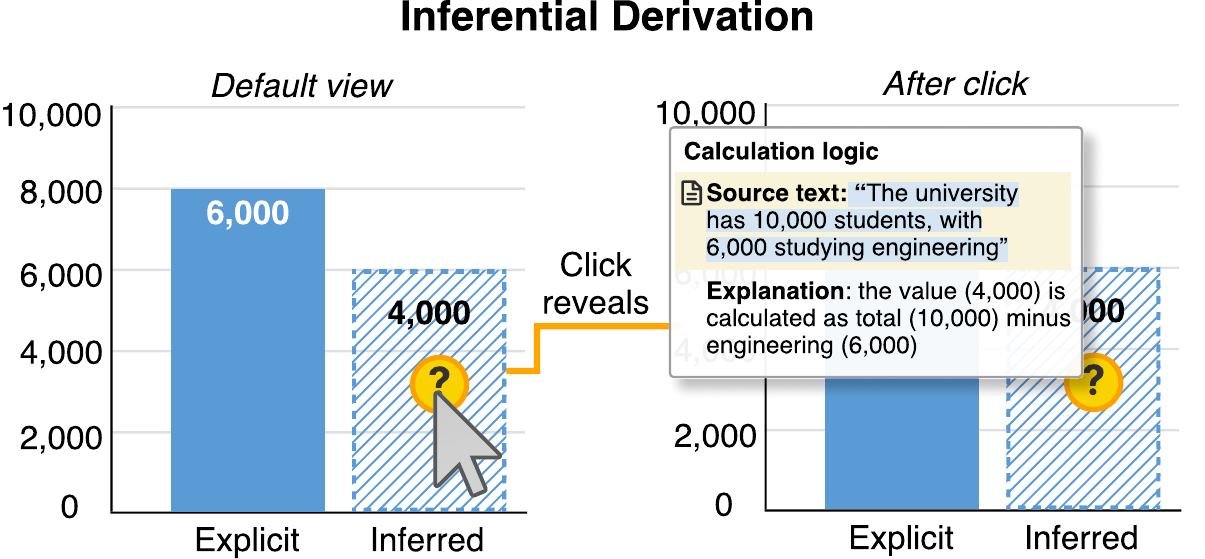}
    \caption{Interactive explainability for inferentially derived values. Left panel shows default visualization state where explicitly stated values appear as solid bars (6,000 engineering students) while inferred values use diagonal hatching pattern with question mark badages (4,000 non-engineering students). Right panel demonstrates the interaction: clicking the question mark reveals a "Calculation logic" tooltip displaying both the source text excerpt ("The university has 10,000 students, with 6,000 studying engineering") and step-by-step explanation of how the inferred value was derived (4,000 = 10,000 total minus 6,000 engineering). This design makes LLM reasoning transparent, allowing readers to verify computational logic and assess inference validity without cluttering the default visualization.}
    \label{fig:inferential-derivation}
\end{figure}

\textbf{Inferential Derivation.} This encoding addresses the need to distinguish explicitly stated information from logically inferred data (Figure~\ref{fig:inferential-derivation}). 
Consider the statement "the university has 10,000 students, with 6,000 studying engineering"—readers can infer 4,000 non-engineering students, but this value was never explicitly stated. 
Our visual solution employs a diagonal hatching pattern combined with interactive yellow question mark badges to create clear provenance indicators~\cite{Boukhelifa2012EvaluatingSA, Wood2012SketchyRF, Edelsbrunner2025VisualizationBC}. 
The hatching pattern overlays inferred visual marks while explicitly stated values maintain standard chart appearance. 
Interactive question mark badges appear adjacent to inferred elements, revealing derivation logic and supporting evidence when clicked (R3). 
This dual-layer approach ensures transparency while maintaining readability, enabling users to access detailed information without overwhelming the initial chart reading experience (R6).

We further distinguish between Entity Inferred and Value Inferred to provide precise transparency about the inference type. 
Entity Inferred occurs when chart categories are contextually derived, such as deriving "non-engineering students" as a complementary category. 
In these cases, axis labels and category indicators receive gray and italic styling. Value Inferred involves mathematical calculations, such as calculating 4,000 non-engineering students from total and engineering values. 
Here, visual marks display the diagonal hatching pattern while category labels retain standard styling (R3). 
When both inference types co-occur, the \toolname{} applies layered encoding with intelligent tooltip filtering to avoid information overload (R4).

\begin{figure*}
    \centering
    \includegraphics[width=\textwidth, height=0.85\textheight, keepaspectratio]{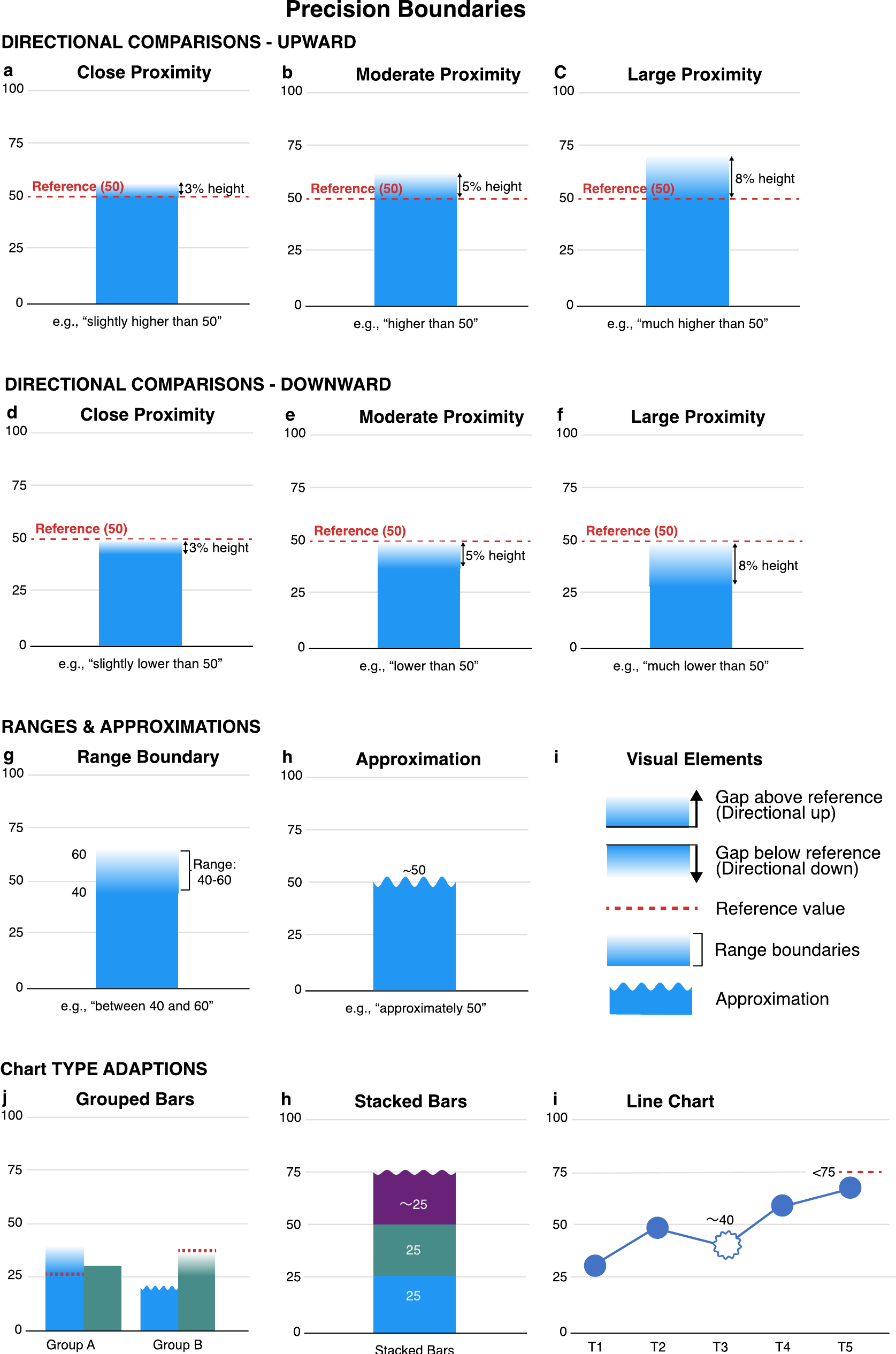}
    \caption{Visual encoding strategies for precision boundaries across comparison types and chart formats. a-c, Upward directional comparisons encode proximity through gradient extension heights: (a)"slightly higher" uses 3\% extension, (b)unmodified comparatives ("higher") use 5\% , (c) and "much higher" uses 8\% . Red dashed lines indicate reference values. d-f, Downward directional comparisons apply the same proximity encoding below the reference line. g, Range boundaries display explicit value spans (e.g., "between 40 and 60") using bracketed bars covering the stated interval. h, Approximation markers (e.g., "approximately 50") appear as squiggly edges on bar tops. i, Visual elements legend summarizing all encoding strategies. j-l, Encoding adaptations across chart types: (j)grouped bars, (k) stacked bars with approximation encoding on the top segment, (l) and line charts with dashed circles indicating approximate data points.}
    \label{fig:precision-boundaries}
\end{figure*}

\textbf{Precision Boundaries.} This encoding addresses linguistic expressions that convey imprecision and directionality (Figure~\ref{fig:precision-boundaries}a-f). 
Different directional expressions carry distinct semantic intensities requiring visual encoding that preserves these distinctions. 
We establish three proximity levels based on semantic degree: close proximity expressions like "slightly greater than" map to 3\% spatial difference from reference value, moderate proximity expressions such as "greater than" correspond to 5\% difference, while large proximity expressions like "substantially greater" translate to 8\% spatial extension. 
This semantic-to-spatial mapping ensures visual distance accurately reflects linguistic intensity (R2, R3), enabling users to interpret comparative magnitude directly from spatial relationships.

Precision boundaries also encompass bounded range expressions and approximation indicators, each requiring distinct visual treatments (Figure~\ref{fig:precision-boundaries}g-h). 
Bounded range expressions such as "between 4-5\%" utilize split bar design where distinct markers represent minimum and maximum boundaries, enabling users to visualize the complete potential range (R2). 
Approximation expressions like "approximately 15\%" employ a squiggly line on the bar top that evokes hand-drawn aesthetics to suggest inherent imprecision~\cite{Boukhelifa2012EvaluatingSA, Wood2012SketchyRF}. 
Since bar height represents value, the squiggly line applies only to the top edge. 
This design draws upon established conventions where irregular lines indicate uncertainty, creating an intuitive visual metaphor (R6)~\cite{Boukhelifa2012EvaluatingSL, Maceachren2012VisualS}.

\begin{figure}
    \centering
    \includegraphics[width=\linewidth]{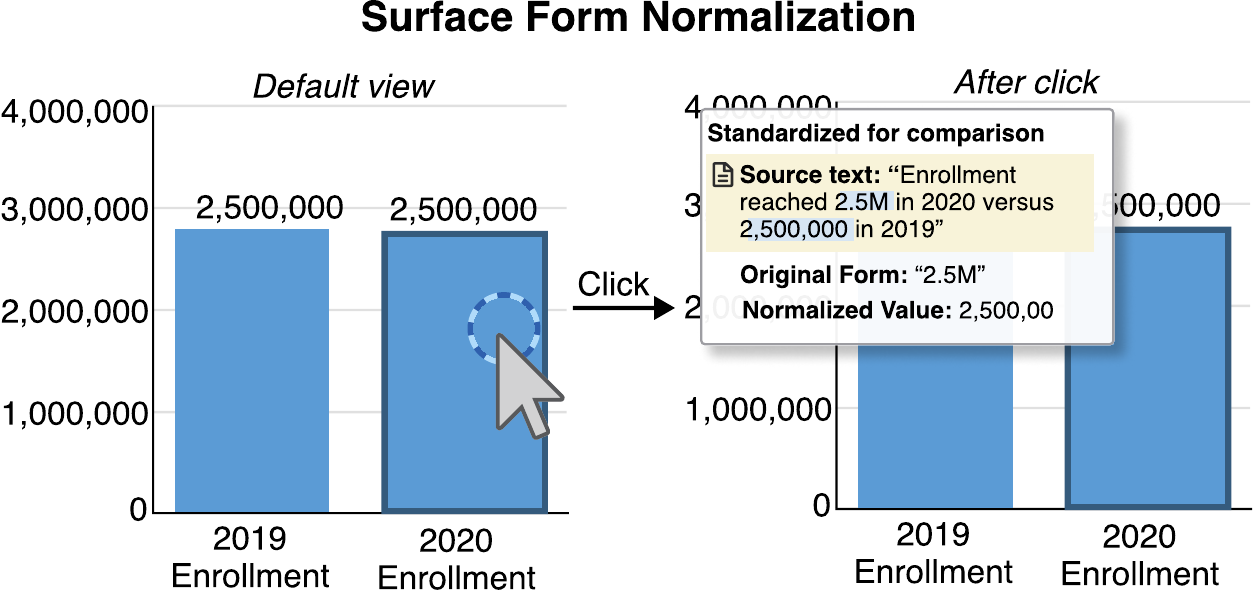}
    \caption{Interactive transparency for surface form normalization in comparative visualizations. Left panel shows default state where enrollment values appear identical (2,500,000 for both years), but the 2020 bar displays a darker and thicker border and dashed circle icon indicating that its value was normalized from a textual representation. Right panel demonstrates the interaction: clicking the bar reveals a "Standardized for comparison" tooltip showing the source text excerpt ("Enrollment reached 2.5M in 2020 versus 2,500,000 in 2019"), the original abbreviated form ("2.5M"), and the normalized numerical value (2,500,000). This design alerts readers to textual variations that may carry semantic meaning while maintaining spatial accuracy in the chart, allowing users to verify the normalization.}
    \label{fig:surface-form-normalization}
\end{figure}

\textbf{Surface Form Normalization.} This encoding addresses textual values requiring conversion to numeric form for visualization, such as "half" to 50, "nine" to 9, or "2.5M" to 2,500,000 (Figure~\ref{fig:surface-form-normalization}). 
Consider "enrollment reached 2.5M in 2020 versus 2,500,000 in 2019", these represent identical values in different formats, requiring standardization for visual comparison. 
This presents a transparency challenge: readers need both standardized numeric values for comparison and understanding of the conversion logic~\cite{Wang2025VizTAEC}. 
We employ darker and thicker borders to mark converted values while providing interactive access to conversion explanations. 
Chart elements derived from textual forms receive a darker and thicker border that signals conversion without disrupting visual hierarchy~\cite{Kavasidis2018ASC}. 
Users can click bordered elements to reveal the original textual expression and conversion logic (R3). This shows the complete transformation: source text, conversion rule, and resulting numeric value. 
This dual-layered approach balances readability with transparency (R6): standardized format enables direct comparison (R2), while the interactive layer preserves connection to original expressions and clarifies the automated process.

\begin{figure}
    \centering
    \includegraphics[width=\linewidth]{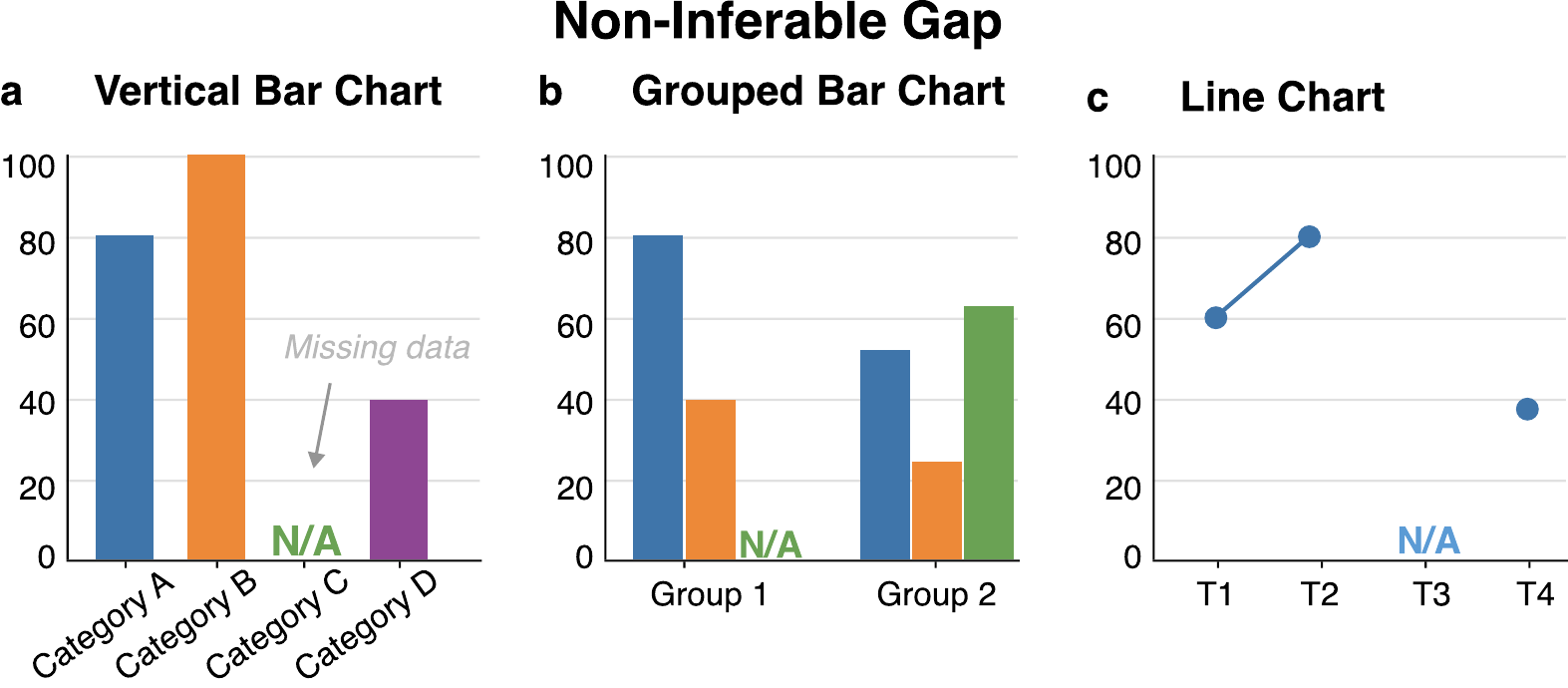}
    \caption{Non-Inferable gap indicated through explicit N/A labeling across chart types. When data is fundamentally missing and cannot be inferred, \toolname{} displays "N/A" labels at the expected data position rather than omitting the category entirely. a, Vertical bar chart shows "N/A" label positioned where the missing bar (Category C) would appear, with an annotation clarifying "Missing data." b, Grouped bar chart applies N/A labeling to specific missing series within groups (Group 1 lacks the green series). c, Line chart displays "N/A" at the x-axis position where a data point is absent (T3), maintaining temporal continuity. }
    \label{fig:non-inferable-gaps}
\end{figure}

\textbf{Non-Inferable Gaps.} This encoding represents the absence of quantifiable data where textual information is fundamentally absent and cannot be logically derived, requiring transparent acknowledgment of limitations (Figure~\ref{fig:non-inferable-gaps})~\cite{Song2019WheresMD, Alsufyani2024VisualizationOM}. 
Consider "graduation rates showed significant improvement this year", without baseline numbers or specific percentages, the improvement magnitude cannot be visualized numerically. 
Our encoding employs "N/A" text labels positioned strategically where expected numerical content would naturally appear, using the same color scheme as data categories to maintain visual consistency (R1). 
These labels communicate clear availability limits, distinguishing truly absent information from inferable gaps (R3). 
This approach prioritizes clear absence indication over visual aesthetic perfection, ensuring users understand exactly what data is unavailable.

\subsection{Chart-Type Adaptations}

\toolname{} supports five chart types: vertical bar, grouped bar, stacked bar, line, and pie charts. 
These five formats cover the primary data structures found in data-rich documents, encompassing categorical comparisons, compositional relationships, hierarchical groupings, and temporal trends. 
As mentioned in R5, uncertainty encodings must function consistently and legibly across all five formats, because different data structures in data-rich documents naturally favor different visualization forms. 
The core challenge is that each chart type possesses distinct visual grammar and spatial constraints that must accommodate uncertainty encodings without compromising the conventions readers rely on to interpret that chart type accurately.

Vertical bar charts serve as the primary foundation for uncertainty encoding design because their discrete rectangular structure accommodates all four uncertainty categories within a single format. 
Surface form normalization applies darker and thicker borders to converted values, allowing readers to identify normalized elements at a glance while clicking reveals the original textual form (Figure~\ref{fig:surface-form-normalization}). 
Precision boundaries appear as squiggly edges on bar tops for approximation expressions (Figure~\ref{fig:precision-boundaries}h) and as bracketed bars spanning explicit range boundaries (Figure~\ref{fig:precision-boundaries}a-g). 
Inferential derivation uses diagonal hatching patterns on bar fills combined with interactive question mark badges that reveal derivation logic on demand (Figure~\ref{fig:inferential-derivation}). 
Non-inferable gaps appear as explicit N/A labels positioned where expected bars would otherwise appear (Figure~\ref{fig:non-inferable-gaps}a).

Grouped and stacked bar charts extend the vertical bar encoding strategy to handle multi-series and compositional data. 
In grouped bar charts, each individual bar within a group receives the appropriate uncertainty encoding independently, allowing different series to carry different uncertainty types simultaneously. 
In stacked bar charts, uncertainty encodings apply at the segment level rather than the whole bar level, ensuring that approximation or inferential uncertainty in one segment does not visually contaminate adjacent segments.

Line charts require the most substantial adaptation because their continuous temporal structure is fundamentally incompatible with the discrete spatial encodings designed for bar charts. 
For approximation expressions, hollow dots with squiggly circular borders mark individual uncertain data points, distinguishing them from standard filled markers while preserving line continuity (Figure~\ref{fig:precision-boundaries}i). 
For precision boundary expressions, directional reference markers appear above and below uncertain data points to indicate bounded ranges without interrupting the temporal progression of the line.

Pie charts require specialized adaptations because their circular geometry and angular segments impose spatial constraints that bar chart encodings cannot directly address.
Squiggly border styling applies along segment boundaries to indicate approximation uncertainty, while question mark badges appear within segment space to signal inferential derivation. 
Non-inferable gaps do not apply to pie charts because pie charts represent compositional data where all segments must sum to a complete whole, while a missing segment would violate the fundamental premise of the format and produce a misleading visual representation.

When multiple uncertainty categories co-occur within a single chart element, a defined precedence order governs how encodings layer without producing visual conflict (R4). Diagonal hatching patterns for inferential derivation take precedence over standard fill colors. 
Interactive question mark badges and other interactive elements supersede default styling. 
Explicit absence indicators such as N/A labels maintain priority over purely visual design elements. 
This precedence structure ensures that the most semantically significant uncertainty type remains visually dominant while secondary encodings remain accessible through interaction rather than cluttering the primary chart reading experience.

\section{\toolname{} System}\label{sec:system}

\begin{figure*}
    \centering
    \includegraphics[width=\linewidth]{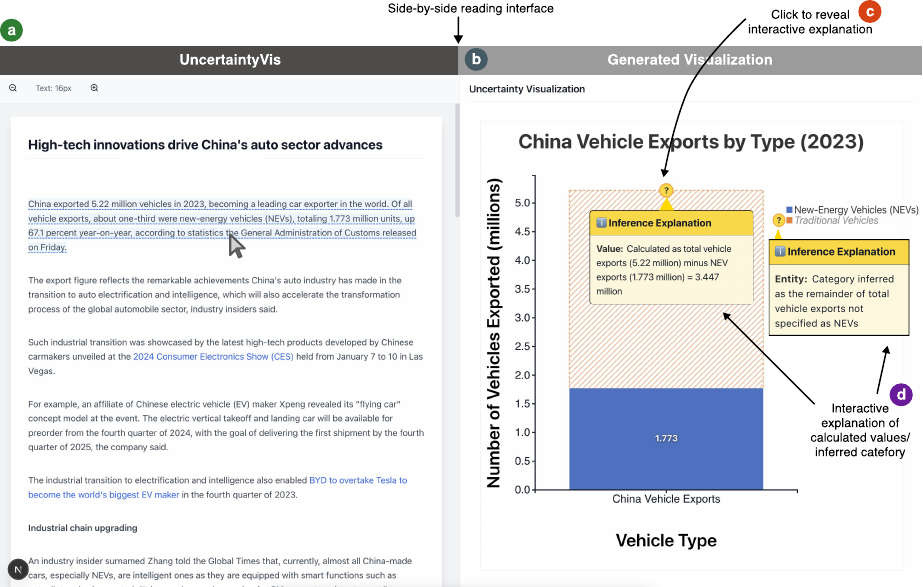}
    \caption{Split-screen reading interface for uncertainty-aware document comprehension. a, Left panel displays source document with underlined text excerpts containing uncertainty expressions (e.g., "about one-third were new-energy vehicles" and "1.773 million units"). b, Right panel shows automatically generated uncertainty-aware visualization (stacked bar chart of China vehicle exports by type) corresponding to the selected excerpt. c, Yellow question mark badges indicate interactive explanation availability for inferred values. d, Clicking badges reveals detailed "Inference Explanation" tooltips showing: (top) calculation logic for derived values (Traditional Vehicles = 5.22 million total - 1.773 million NEVs = 3.447 million), and (bottom) entity classification reasoning for inferred categories. }
    \label{fig:reading-interface}
\end{figure*}

\begin{figure*}
    \centering
    \includegraphics[width=\linewidth]{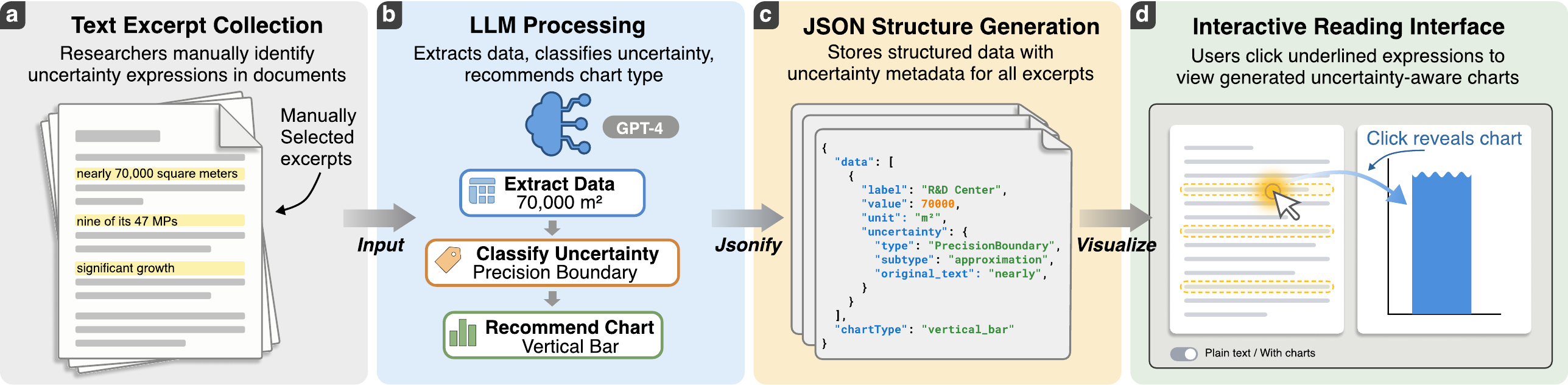}
    \caption{\toolname{} workflow for uncertainty-aware visualization generation. a, Researchers manually identify and highlight text excerpts containing quantitative data with uncertainty expressions across all four taxonomy categories from source documents. b, GPT-4 processes each excerpt to extract numerical data (70,000 m²), classify uncertainty type (Precision Boundary), and recommend appropriate chart format (vertical bar). c, \toolname{} generates structured JSON metadata storing extracted values, uncertainty attributes (type, subtype, original text form), and chart specifications for all excerpts. d, Interactive reading interface displays documents with dashed underlines marking pre-processed excerpts; clicking any underlined expression reveals its corresponding uncertainty-aware visualization in the right panel. Toggle allows switching between plain text and chart-embedded views for comparative evaluation.}
    \label{fig:system-workflow}
\end{figure*}

\subsection{System Overview}

\toolname{} implements a three-stage pipeline that transforms data-rich documents into uncertainty-aware reading experiences displayed in a side-by-side interface (Figure~\ref{fig:system-workflow}). 
In the first stage, researchers manually identify linguistic uncertainty expressions across all four uncertainty categories within source documents (Figure~\ref{fig:system-workflow}a). 
In the second stage, a large language model processes each identified expression to extract structured numerical data, classify the corresponding uncertainty category according to the taxonomy, and recommend an appropriate chart type based on the data structure (Figure~\ref{fig:system-workflow}b-c). 
In the third stage, \toolname{} automatically generates uncertainty-aware visualizations from the structured output, applying category-specific visual encodings that preserve the semantic meaning of each linguistic uncertainty expression (Figure~\ref{fig:system-workflow}d). 
The resulting visualizations appear in the right panel alongside the source document, where readers can interact with visual encodings to explore both the quantitative data and its associated uncertainties (Figure~\ref{fig:reading-interface}).

The rendering process achieves two objectives simultaneously, directly executing the requirements established in Section~\ref{sec:design-requirement}. 
First, quantitative accuracy is preserved through precise spatial encodings: bar heights, line positions, and segment angles directly represent numerical values without distortion, satisfying R2. 
Second, linguistic uncertainty is communicated through secondary visual layers that overlay the primary spatial representations without distorting them, satisfying R1 and R4. 
Inferred values appear with diagonal hatching patterns, approximation expressions receive squiggly borders on bar tops, and surface form normalization cases display darker and thicker borders. 
Together these visual encodings enable readers to interpret both quantitative magnitudes and their associated uncertainty simultaneously at a glance. 
Interactive layers further allow readers to verify how \toolname{} interpreted each uncertainty expression by revealing the original textual form and derivation logic on demand, satisfying R3.

The following subsections detail the large language model based uncertainty-aware pipeline (Section~\ref{sec:pipeline}), the automated visualization generation with uncertainty encoding (Section~\ref{sec:chart-generation}), and the interactive reading interface (Section~\ref{sec:interface}) that together realize these objectives.

\subsection{LLM-Based Uncertainty-Aware Pipeline}\ref{sec:pipeline}
\subsubsection{LLM Processing}

\toolname{} leverages large language models' semantic understanding capabilities~\cite{Dagdelen2024StructuredIE,Jiao2023InstructAE} to simultaneously extract quantitative information from text to structured data form (JSON), classify uncertainty categories according to the taxonomy, and recommend appropriate chart types based on data structure. 
These tasks require interpreting linguistic nuance that rule-based approaches~\cite{Thawani2021RepresentingNI,Gpfert2022MeasurementEW} cannot capture. 
For instance, distinguishing whether "approximately 15\% of students dropped out this year" represents true approximation versus informal expression of a precise value, or determining if "enrollment reached 2.5M in 2020" represents a meaningful stylistic choice versus identical meaning to "2,500,000 in 2019," demands contextual interpretation beyond pattern matching.
When given a text excerpt, the model identifies data points, detects words or phrases indicating uncertainty (such as "approximately," "more than," or varied numerical formats), and organizes this information into structured data. 
This output preserves both the numerical content and the uncertainty characteristics expressed in the original text, ensuring that uncertainty classifications reflect the author's intended meaning rather than applying rigid syntactic rules that may misinterpret context-dependent expressions.

\subsubsection{Uncertainty Metadata Structure}
The LLM's structured output encodes each data point with metadata that specifies which uncertainty categories apply and provides category-specific attributes necessary for faithful visual translation. 
This metadata structure captures semantic nuances that distinguish qualitatively different uncertainty expressions. 
For example, "approximately 15\%" requires different attributes for approximation boundaries than "graduation rates showed significant improvement this year" requires for representing missing numerical values. 
Each data point therefore carries metadata tailored to its specific uncertainty characteristics, enabling the visualization system to select appropriate visual encodings that preserve the author's intended meaning. 
The following paragraphs detail the attributes for each of the four supported uncertainty categories.

\textbf{Inferential Derivation} metadata employs two independent attributes that identify which elements required computational inference. 
The first attribute indicates whether the data point's name was inferred from context rather than explicitly stated. 
The second attribute specifies whether the numerical value required inference. 
For example, when text states "the university has 10,000 students, with 6,000 studying engineering," both the "non-engineering students" category name and its value of 4,000 must be inferred through subtraction and logical reasoning. 
Each inferred element receives a natural language explanation documenting the LLM's reasoning process. 
For the name, the explanation states "Category inferred as remainder of university students not studying engineering," while for the value it provides "Calculated as total students (10,000) minus engineering students (6,000) = 4,000." 
This documented explanation makes the computational interpretation transparent and verifiable rather than opaque, enabling readers to assess the validity of automated inferences.

\textbf{Precision Boundaries} metadata specifies the boundary type that characterizes how text expresses numerical imprecision. 
\toolname{} distinguishes three primary types: directional comparisons such as "less than" or "greater than," bounded ranges like "between 4 and 5 percent," and approximations indicated by terms such as "around" or "approximately." 
For example, "approximately 15\% of students dropped out this year" is classified as an approximation type. 
For directional comparisons, the metadata further classifies linguistic intensity into three proximity levels. 
Close proximity captures mild expressions like "slightly more than 15\%," moderate proximity represents neutral comparisons like "more than 15\%," and large proximity indicates strong intensifiers like "well above 15\%." 
These proximity levels enable visual encodings to reflect how far the actual value likely deviates from the stated reference point based on the author's linguistic choices.

\textbf{Surface Form Normalization} metadata preserves the original textual form in which each value appeared. 
In the enrollment example "enrollment reached 2.5M in 2020 versus 2,500,000 in 2019," \toolname{} records "2.5M" as the original expression alongside the standardized numerical value of 2,500,000. 
This dual representation enables charts to display consistent numerical formats for accurate comparison while maintaining traceability to the author's original choice of expression. 
The ability to compare 2020 and 2019 enrollment using standardized values while preserving the author's deliberate choice to abbreviate one year's data provides essential transparency. 
Readers can access these original forms through interactive mechanisms, revealing how the \toolname{} interpreted and normalized diverse textual formats into comparable numerical values.

\textbf{Non-Inferable Gaps} represent situations where data point names appear in text but no numerical value can be derived. 
\toolname{} encodes these cases with null values, distinguishing true data absence from situations where values could be inferred through other uncertainty categories. 
This distinction maintains semantic fidelity to cases where authors express incompleteness. 
For instance, the statement "graduation rates showed significant improvement this year" mentions that improvement exists but provides no baseline numbers or percentages that would enable quantification of the improvement magnitude. 
By preserving these gaps rather than fabricating values, \toolname{} ensures visualizations accurately reflect the limits of available information as expressed in the source text.

Data points that express no uncertainty have no metadata attributes. 
When text states precise, definitive values without uncertainty markers—such as "15\% of students dropped out this year" without the qualifier "approximately"—the structured output contains only the category name and numerical value, with metadata set to none. 
This clean distinction between certain and uncertain data maintains structural clarity while avoiding unnecessary metadata overhead for straightforward quantitative statements.

\subsubsection{Chart Type Recommendation}
The LLM recommends chart types~\cite{Hu2018VizMLAM, Li2021KG4VisAK} from five supported options: vertical bar, stacked bar, grouped bar, pie, and line charts. 
The model analyzes the structured data's characteristics, matching data patterns to visualization conventions that best convey the underlying relationships. 
The LLM evaluates whether the data represents comparisons across categories, compositions of a whole, temporal trends, or hierarchical groupings. 
For instance, university enrollment by major (engineering versus non-engineering students) suggests vertical or grouped bar charts for category comparison, while dropout rates tracked over multiple years would recommend line charts for temporal trends~\cite{Munzner2014VisualizationAA}.
When data includes percentages that sum to 100\%, the model may recommend pie charts or stacked bars to emphasize compositional relationships. 
\toolname{} constrains recommendations to these five chart types to maintain consistency in uncertainty encoding strategies across different visualization forms, ensuring that readers encounter familiar uncertainty treatments regardless of the underlying chart structure.

\subsubsection{JSON Schema Structure}
The complete structured output format integrates all metadata elements into a unified JSON representation that preserves both quantitative data and uncertainty semantics (Figure~\ref{fig:system-workflow}c). 
We illustrate this structure through the university enrollment example: "The university has 10,000 students, with 6,000 studying engineering." 
The resulting schema contains explicitly stated values including 10,000 total students and 6,000 engineering students, with the latter carrying no metadata since it was directly stated in the text. 
The schema also includes the derived value of 4,000 non-engineering students, calculated through subtraction, which carries Inferential Derivation attributes indicating that both the category name and numerical value required inference. 
The metadata includes explanatory text: "Category inferred as remainder of university students not studying engineering" for the name inference, and "Calculated as total students (10,000) minus engineering students (6,000) = 4,000" for the value inference. 
Additionally, the schema specifies the recommended chart type (vertical bar), axis labels ("Student Type"), and chart title ("University Enrollment by Major"). 
This integrated structure enables the visualization system to access all information needed to generate uncertainty-aware charts: the quantitative data for spatial encoding, the uncertainty classifications for visual styling, and the explanatory text for interactive exploration.

\subsection{Visualization Generation with Uncertainty Encoding} \label{sec:chart-generation}
\subsubsection{Automated Encoding Selection}
\toolname{} applies deterministic mappings between uncertainty metadata and visual encoding techniques defined aforementioned, automatically selecting appropriate visual representations to ensure consistent uncertainty communication across all generated visualizations. 
Each uncertainty category entails specific visual treatments that apply uniformly across contexts. 
For example, "approximately 15\%" always receives squiggly border styling combined with gray coloring regardless of whether it appears in a bar chart, pie chart, or line chart. 
This automated approach applies to all chart types used in the \toolname{} system, eliminating variability in how uncertainty appears across different visualization formats. 
The consistency reduces users' cognitive load in learning and perceiving uncertainty-aware charts, as they encounter the same visual language for uncertainty across diverse document contexts. 
Furthermore, this consistency enables systematic evaluation of how specific visual encodings affect comprehension and supports reliable comparison across different documents and uncertainty types. 
\toolname{} eliminates user control over these encoding choices, ensuring that identical uncertainty classifications produce identical visual treatments regardless of when or by whom the text is processed, maintaining reproducibility and standardization throughout the pipeline.

\subsubsection{Interactive Transparency} 
Beyond these visual encodings, \toolname{} makes computational reasoning transparent through category-specific interactions that reveal how the LLM interpreted uncertain expressions. 
For Inferential Derivation, yellow question mark badges~\cite{Edelsbrunner2025VisualizationBC} appear adjacent to visual marks representing inferred values and around category labels for inferred names. 
When readers click the badge on the bar representing 4,000 non-engineering students, a tooltip displays the derivation logic: "Calculated as total students (10,000) minus engineering students (6,000) = 4,000" for the derived value, and "Category inferred as remainder of university students not studying engineering" for the inferred category name. 
For Surface Form Normalization, hovering over chart elements triggers windows that display the original textual form alongside its normalized numerical representation. 
For example, hovering over the 2020 enrollment bar reveals the transformation from "2.5M" to "2,500,000," making the standardization process explicit. 
These interactions enable readers to inspect the \toolname{}'s interpretations and evaluate whether computational inferences align with their own understanding of the source text, building trust through transparency rather than obscuring automated decision-making~\cite{Sunny2025TrustIT}.

\subsection{Interactive Reading Interface}\label{sec:interface}

\subsubsection{Interface Design}
The interface employs a two-panel layout with original document text on the left and uncertainty-aware visualizations on the right (Figure~\ref{fig:reading-interface})~\cite{August2022PaperPM}. 
This side-by-side arrangement enables readers to access visual representations while maintaining continuous engagement with the textual narrative.
Dashed underlines mark uncertainty expressions in the left panel, signaling which text passages contain quantitative data and uncertainty markers that the \toolname{} can visualize. 
For instance, underlined phrases include "approximately 15\% of students dropped out this year," "enrollment reached 2.5M in 2020," and "the university has 10,000 students, with 6,000 studying engineering." 
Readers click these underlined expressions to immediately generate and display the corresponding chart in the right panel. 
The persistent right panel creates a stable spatial location where readers expect charts to appear, eliminating the need for navigation between different views or windows. 
A toggle control switches between visualization-enhanced and text-only reading modes, enabling readers to compare their comprehension with and without visual support—for instance, comparing interpretation of "approximately 15\%" with versus without the squiggly border visualization that signals approximation. 
Text zoom controls allow readers to adjust font size for comfortable reading across different display sizes. 
These interactions maintain simplicity to minimize cognitive overhead: the click-to-visualize mechanism requires only a single action, and readers can freely explore multiple uncertainty expressions within a document without managing complex interface states. 
This streamlined design prioritizes reading flow over feature complexity, supporting the proof-of-concept goal of demonstrating how uncertainty-aware visualization integrates naturally into document reading workflows.

\subsubsection{Implementation}
The implementation supports real-time uncertainty-aware visualization generation through a distributed architecture. T
he frontend employs React to manage interface components and user interactions, while D3.js handles chart rendering from structured data. 
This separation allows the interface to respond immediately to user actions while delegating visualization construction to a specialized graphics library optimized for data-driven document manipulation. 
The backend uses Python to orchestrate the processing pipeline, interfacing with OpenAI's API to access large language model capabilities for text analysis and structured data extraction. 
When readers click an uncertainty expression—such as "approximately 15\% of students dropped out this year", \toolname{} sends the text excerpt to the backend, which processes it through the LLM and returns structured JSON containing metadata indicating approximation type and the 15\% numerical value. 
The frontend then renders a vertical bar chart with the squiggly border encoding that visually represents the approximation uncertainty. 
This real-time architecture supports dynamic exploration where readers can examine multiple text excerpts within a single reading session without requiring \toolname{} to pre-compute all possible visualizations, maintaining responsiveness while enabling flexible investigation of uncertainty expressions throughout data-rich documents.

\section{User Study}

\begin{figure*}
    \centering
    \includegraphics[width=\linewidth]{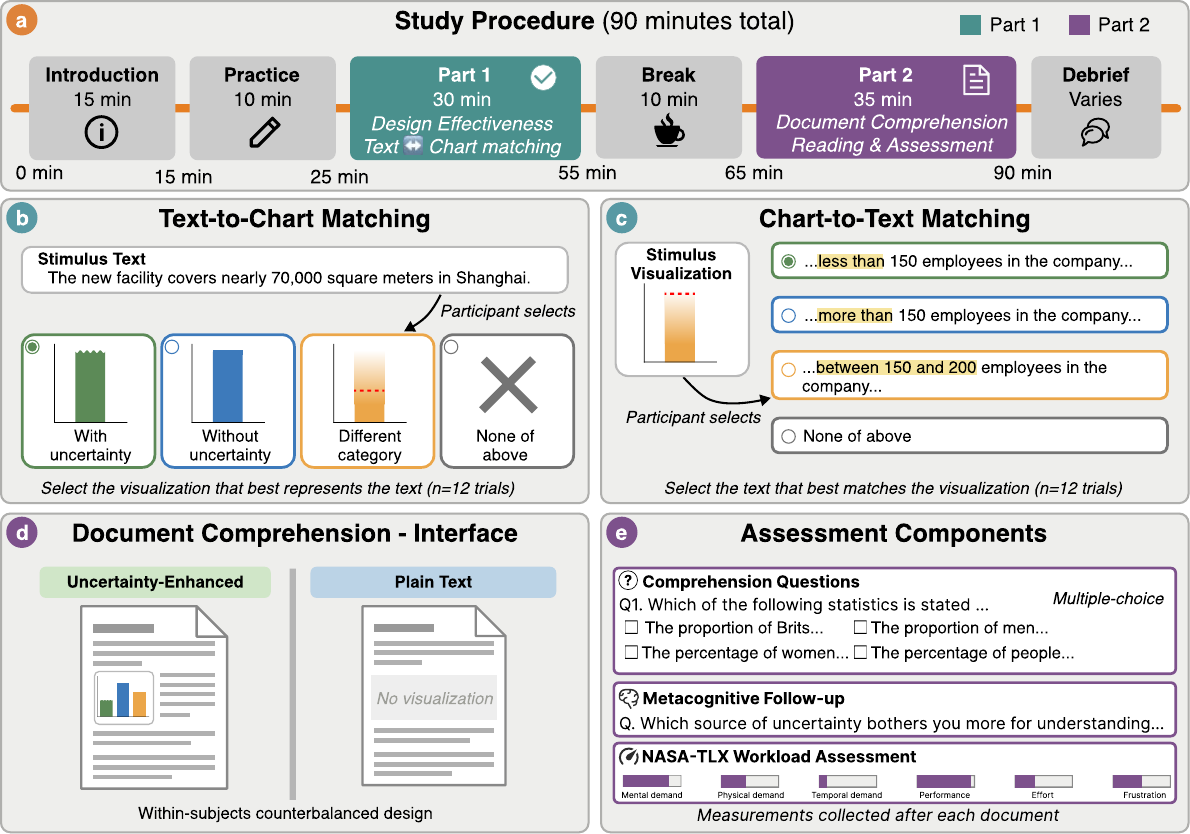}
    \caption{Two-part evaluation methodology assessing semantic preservation and comprehension impact. a, Study procedure timeline (90 minutes total): 15-minute introduction, 10-minute practice, 30-minute Part 1 (bidirectional matching tasks), 10-minute break, 35-minute Part 2 (document reading and assessment), followed by variable-length debrief. b, Text-to-Chart Matching: participants select which visualization best represents stimulus text containing uncertainty expressions from four options (with uncertainty encoding, without uncertainty, different category, none). c, Chart-to-Text Matching: participants identify which text description best matches the stimulus uncertainty visualization from four options. Both tasks included 12 trials covering all uncertainty categories. d, Document Comprehension interface showing within-subjects counterbalanced design: Uncertainty-Enhanced condition embeds interactive uncertainty-aware visualizations within documents; Plain Text condition presents identical content without visualizations. e, Assessment components include multiple-choice comprehension questions, metacognitive follow-up probing uncertainty awareness, and NASA-TLX workload measurement across six dimensions. }
    \label{fig:study-design}
\end{figure*}

\subsection{Study Design and Goals}
To evaluate the effectiveness of uncertainty visualization encoding, this study addresses two research questions. 
First, how can we encode linguistic uncertainty expressions in visual forms that people can accurately decode? 
Second, how do uncertainty visualizations affect participants' comprehension and interpretation of uncertainty information in data-rich documents? 
These questions represent sequential validation stages: visualization effectiveness must precede utility assessment.

We designed a two-part evaluation study corresponding to each question (Figure~\ref{fig:study-design}). 
Part 1 evaluates semantic preservation through bidirectional matching tasks where participants match uncertainty expressions with appropriate visualizations and vice versa. 
Part 2 examines authentic document reading impact, measuring comprehension accuracy, cognitive load, and confidence when participants read documents with versus without uncertainty visualizations. 
This design validates both fundamental encoding effectiveness and practical utility in realistic reading contexts.

\subsection{Participants}
We recruited 12 participants (6 male, 6 female, ages 23-30 years). 
All participants had at least a bachelor's degree, with 12 holding graduate degrees. 
Participants were recruited through university mailing lists and social media, then screened for English proficiency and basic visualization literacy. 
Eligibility criteria required ability to understand data-rich documents containing linguistic uncertainty expressions and proficiency in reading bar charts, line charts, and pie charts. 
A brief qualification task assessed these skills before participation. All participants provided IRB-approved informed consent and received \$15 compensation.

\subsection{Materials and Apparatus}
We developed a proof-of-concept uncertainty visualization tool that automatically generates uncertainty-encoded visualizations from uncertainty expressions based on the visual encoding strategies defined in the taxonomy. 
The tool implements all four uncertainty categories with their specific visual treatments. 
The system uses large language models and the taxonomy to identify and classify uncertainty expressions from text, then generates visualizations across five chart types—vertical bar, grouped bar, stacked bar, pie, and line charts—maintaining semantic fidelity while preserving quantitative accuracy through dual-objective rendering.

For the two-part evaluation, we prepared two distinct sets of materials. 
Part 1 materials included 12 text-to-chart pairs and 12 different chart-to-text pairs covering all four uncertainty categories. 
Each pair included multiple options: uncertainty-aware visualization with proper encoding, plain chart without uncertainty indicators, alternative uncertainty category encodings, or "none suitable." 
Part 2 materials consisted of authentic data-rich documents from academic journals, research reports, and data journalism containing diverse linguistic uncertainty expressions spanning the four categories, presented in two versions: uncertainty-enhanced with \toolname{} visual encodings embedded within text, and plain text containing identical content without visual enhancements.
These documents were drawn directly from the 12-document corpus analyzed in the formative study.

For data collection and experiment administration, Part 1 was conducted using Qualtrics survey platform, which provided structured questionnaire delivery for bidirectional matching tasks, randomized stimulus presentation, and systematic response collection. 
The Qualtrics interface was configured to randomize order, balance conditions, and control for order effects. 
Part 2 utilized a custom-developed web-based reading interface that integrated the \toolname{} system directly into the reading experience. 
This web-based platform enabled uncertainty visualization rendering with all visual encoding strategies, supported document-chart integration and interactive exploration through question mark tooltips and hover-to-reveal original forms, and recorded detailed behavioral data including reading time, scroll patterns, visualization interaction events such as clicking question marks or hovering over darker and thicker borders, and response latencies.

\subsection{Tasks and Metrics}
The within-subjects design employed distinct methodological approaches for each part (Figure~\ref{fig:study-design}). 
Part 1 isolated encoding-decoding accuracy through controlled matching tasks, measuring correspondence between linguistic uncertainty and visual encodings without confounding document context effects. 
Part 2 embedded visualizations within authentic reading scenarios to assess comprehension impact, confidence, and cognitive workload.

Part 1 implemented bidirectional matching tasks (Figure~\ref{fig:study-design}b-c).
The text-to-chart task presented linguistic uncertainty expressions with four visualization options testing whether participants could identify appropriate visual treatments. 
The chart-to-text task presented visualizations with specific encodings, requiring participants to select matching linguistic expressions. 
This bidirectional design measured both encoding effectiveness (can people translate uncertainty expressions to correct visual forms?) and decoding accuracy (can people interpret visual treatments back to appropriate linguistic expressions?).

Part 1 employed four experimental conditions: (1) correctly-encoded visualizations with appropriate visual treatments, (2) standard charts without uncertainty indicators, (3) alternative encodings with wrong category treatments or mismatched intensity levels, and (4) "none suitable" option. 
Performance metrics included matching accuracy (percentage of correct matches) and semantic suitability ratings (7-point Likert scale assessing how well visualizations preserved semantic meaning).

Part 2 presented documents in two conditions: uncertainty-enhanced with visual encodings, and plain text without augmentation (Figure~\ref{fig:study-design}d). 
Participants read each document, interacted with encodings (clicking question marks for explanations, hovering over borders for original forms), then answered multiple-choice questions testing uncertainty-related comprehension and completed open-ended summaries synthesizing key findings. 
Metrics included NASA-TLX workload assessment across six dimensions (mental demand, physical demand, temporal demand, performance, effort, frustration), reading time, answer accuracy, and interpretation confidence ratings (Figure~\ref{fig:study-design}e).

\subsection{Procedure}
The 90-minute procedure followed a counterbalanced within-subjects design conducted remotely via video conferencing (Figure~\ref{fig:study-design}a). 
Participants completed both parts with a 10-minute break between to prevent fatigue. Sessions were video recorded and behaviorally logged with consent.

Part 1 began with orientation explaining the taxonomy categories and their visual encodings to establish shared understanding. 
Text-to-chart and chart-to-text tasks were counterbalanced, with each participant completing 12 pairs of each type in randomized order. 
For each item, participants selected the most appropriate option from four choices, then immediately rated semantic suitability on a 7-point scale. 
Qualtrics automatically recorded selections, ratings, and response times.

Part 2 presented documents in randomized sequence, counterbalanced between uncertainty-enhanced and plain text conditions. 
Participants read at their own pace with full access to interactive features. The system recorded reading time, scroll patterns, and interaction events. 
After each document, participants answered comprehension questions, completed summary tasks, and provided NASA-TLX workload and confidence ratings.

Participants concluded with a post-session questionnaire about their experience, the clarity and usefulness of visual encodings, and technical difficulties. 
All data were anonymized and securely stored with access restricted to authorized researchers. Random identifiers ensured anonymity while enabling cross-component data linking.

\section{Study Result Analysis} 
We present Part 1 findings on encoding accuracy first, followed by Part 2 document comprehension outcomes, and conclude with qualitative insights that explain the mechanisms underlying observed performance patterns.

\subsection{Part 1: Semantic Preservation Through Bidirectional Matching}
\subsubsection{Overall Matching Accuracy}

\begin{figure}
    \centering
    \includegraphics[width=\linewidth]{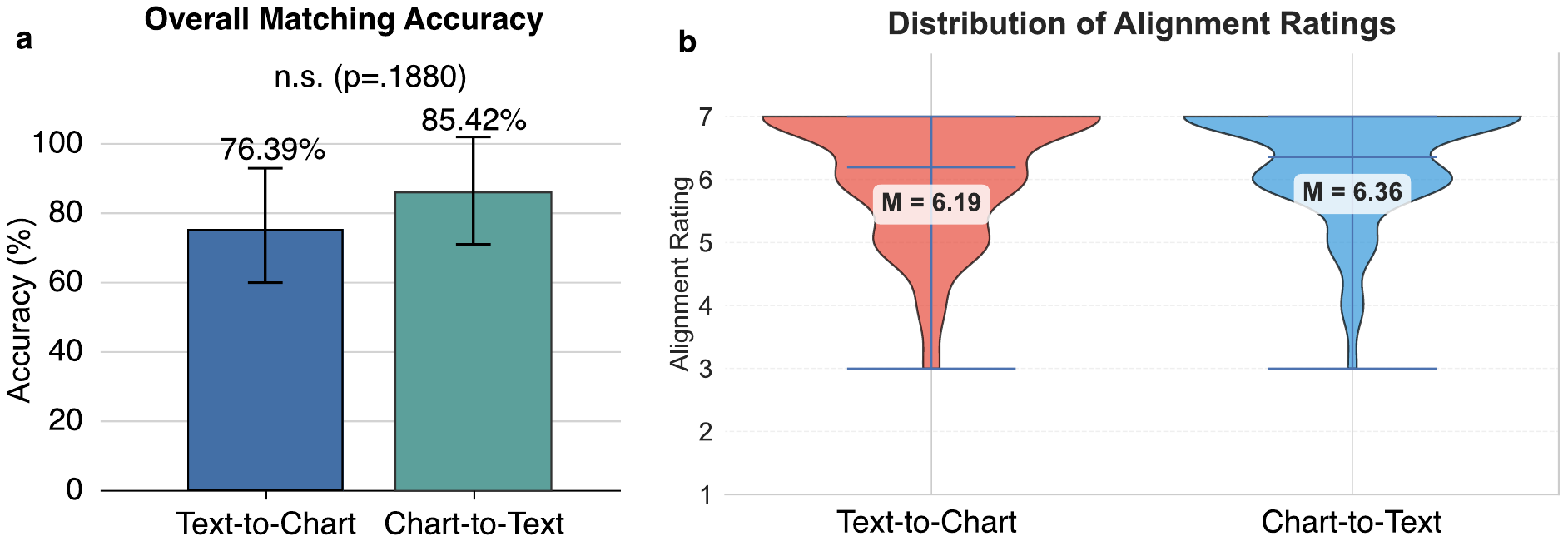}
    \caption{a, Single bar chart compares Text-to-Chart vs Chart-to-Text with error bars (SD). b, Two distribution plots (violin plots) showing 1-7 Likert scale responses for Text-to-Chart  and Chart-to-Text.}
    \label{fig:eval-part1-results}
\end{figure}

\textbf{RQ1: Can we encode linguistic uncertainty expressions in visual forms that people can accurately decode?
}
Part 1 addresses whether our uncertainty-aware visualizations successfully preserve the semantic meaning that authors express through linguistic uncertainty markers. 
We evaluated this through bidirectional matching tasks that test both encoding (translating text to appropriate visual forms) and decoding (interpreting visual treatments back to linguistic expressions). 
Success requires that participants can reliably map between textual uncertainty expressions such as "approximately 15\%" or "more than 8\%" and their corresponding visual encodings including squiggly borders, diagonal hatching patterns, darker and thicker borders, and explicit labels.

Chart-to-text matching outperformed text-to-chart matching, revealing an asymmetry in how participants encode and decode visual uncertainty. 
Participants achieved 85\% accuracy when interpreting visual encodings back to textual expressions (chart-to-text) compared to 76\% when selecting appropriate visualizations from text (text-to-chart) (Figure~\ref{fig:eval-part1-results}a).

This 9 percentage point gap reflects a fundamental asymmetry between two cognitively distinct processes. 
Text-to-chart matching required participants to mentally construct an expected visual form from a linguistic description, a generative process that demands prior familiarity with encoding conventions and is further complicated by individual differences in visual encoding preferences. 
Chart-to-text matching required participants to recognize and decode a directly perceived visual encoding, a recognition process that is cognitively less demanding and less dependent on prior convention knowledge. 
The higher chart-to-text accuracy therefore indicates that our visual encoding strategies successfully communicate uncertainty semantics once presented to readers, while the lower accuracy in text-to-chart matching suggests that explicit onboarding or guided familiarization with encoding conventions would help users more reliably anticipate visual encodings from linguistic uncertainty expressions.

\subsubsection{Text-to-Chart Performance Analysis}

\begin{table*}[t]
\centering
\caption{Text-to-Chart Question-Level Performance Breakdown (N=12 participants)}
\begin{center}
\begin{tabular}{lllrr}
\toprule
\textbf{Question} & \textbf{Chart Type} & \textbf{Uncertainty Category} & \textbf{Accuracy} & \textbf{Avg Suitability} \\
\midrule
Q1 & Bar & Inferential Derivation & 83\%& 7\\
Q2 & Bar & Inferential Derivation, Precision Boundaries & 75\%& 6\\
Q3 & Pie & Precision Boundaries & 92\%& 6\\
Q4 & Bar & Precision Boundaries & 83\%& 6\\
Q5 & Bar & Inferential Derivation, Precision Boundaries & 92\%& 6\\
Q6 & Pie & Inferential Derivation, Surface Form Normalization & 75\%& 6\\
\rowcolor{gray!20}
\textbf{Q7} & \textbf{Line} & \textbf{Inferential Derivation} & \textbf{50\%}& \textbf{6}\\
\rowcolor{gray!20}
\textbf{Q8} & \textbf{Line} & \textbf{Precision boundaries} & \textbf{42\%}& \textbf{5}\\
Q9 & Bar & Precision Boundaries & 67\%& 6\\
Q10 & Bar & Precision Boundaries & 83\%& 6\\
Q11 & Bar & Surface Form Normalization, Non-Inferable Gap & 83\%& 6\\
Q12 & Bar & Inferential Derivation & 92\%& 6\\
\midrule
\multicolumn{3}{l}{\textbf{Overall}} & \textbf{76\%}& \textbf{6}\\
\bottomrule
\end{tabular}
\end{center}
\begin{tablenotes}
\small
\item Suitability ratings measured on 7-point Likert scale (1=Strongly Disagree, 7=Strongly Agree). Gray rows indicate questions with accuracy below 60\%.
\end{tablenotes}
\label{tab:text-to-chart-performance}
\end{table*}

Text-to-chart matching revealed strong overall performance (76\%) but with notable variability across uncertainty categories and chart types (Table~\ref{tab:text-to-chart-performance}). 
Accuracy ranged from 92\% (Q3, Q5, Q12) to 42\% (Q8), with mean semantic suitability rating of 6 out of 7, indicating that while participants could generally identify appropriate encodings, certain combinations of uncertainty type and chart structure proved challenging. 
This variability suggests that encoding effectiveness depends not only on the uncertainty category itself but also on how visual marks represent that category within specific chart structures. 
The wide performance range highlights the need to examine which design elements succeeded and which require refinement.

Performance analysis revealed a critical chart type dependency, with bar and pie charts outperforming line charts in text-to-chart matching tasks. 
Questions using bar or pie charts (Q1, Q3, Q5, Q10, Q11, Q12) averaged 83\% accuracy, while line chart questions (Q7, Q8, Q9) averaged only 53\% accuracy, suggesting our uncertainty encodings were optimized for discrete categorical data rather than continuous trends. 
Questions Q7 (line chart, 50\% accuracy) and Q8 (line chart, 42\% accuracy) represent the most severe encoding failures, with Q8 performing barely above chance level. 
These results demonstrate that visual encoding strategies effective for bar and pie charts do not transfer directly to line charts without substantial modification.

Interview analysis identified three specific design flaws in line chart encodings that caused the Q7 and Q8 performance drops: gradient ambiguity, static trend representation, and processing overhead.
First, participants found gradients representing ranges (such as "less than X") visually ambiguous.
P8 stated "The gradient doesn't really mean anything for me," while P9 interpreted fading as visual noise rather than a bounded range. 
Second, when text implied temporal behavior ("kept below 6.5\% for 10 years"), participants rejected charts showing flat reference lines. 
P2 noted "Chart 1 doesn't look right, it implies no value," expecting visual fluctuation to represent change over time. 
The disconnect between textual descriptions of dynamic processes and static visual thresholds violated participants' expectations about how time series data should appear. 
Third, P8 described Q7 as "confusing due to the need to process a lot of explanations," with unclear phrasing like "over hundred production processes" adding cognitive burden beyond the uncertainty encoding itself. 
These three factors compounded to create a comprehension bottleneck specific to line chart representations.

Despite accuracy challenges, semantic suitability ratings remained high across all text-to-chart items, with 77\% of responses rated "Agree" (6) or "Strongly Agree" (7) on the 7-point scale (Figure~\ref{fig:eval-part1-results}b). 
Distribution analysis showed 59 "Strongly Agree" and 27 "Agree" responses out of 110 total ratings collected from correct matches, indicating that when participants selected the correct chart, they felt confident in its semantic alignment with the text. 
The high suitability ratings despite moderate accuracy suggest that errors stemmed from encoding confusion rather than fundamental disagreement with our semantic mappings. 
Participants who successfully decoded the visual encoding logic strongly endorsed the approach, while those who failed typically selected alternative options rather than rating the correct option poorly. 
This pattern indicates that the core visual encoding strategy resonates with participants once understood, but the initial learning barrier remains substantial for certain encoding types, particularly those involving gradients or continuous trends.

\subsubsection{Chart-to-Text Performance Analysis}

\begin{table*}[t]
\centering
\caption{Chart-to-Text Question-Level Performance Breakdown (N=12 participants)}
\begin{center}
\begin{tabular}{lllrr}
\toprule
\textbf{Question} & \textbf{Chart Type} & \textbf{Uncertainty Category} & \textbf{Accuracy} & \textbf{Avg Suitability} \\
\midrule
Q1 & Bar & Precision Boundaries & 100\%& 6\\
Q2 & Bar & Inferential Derivation, Precision Boundaries & 67\%& 6\\
Q3 & Pie & Inferential Derivation & 83\%& 7\\
Q4 & Bar & Inferential Derivation, Precision Boundaries & 92\%& 6\\
Q5 & Bar & Precision Boundaries & 100\%& 7\\
\rowcolor{gray!20}
Q6 & Pie & Inferential Derivation & 58\%& 7\\
Q7 & Line & Inferential Derivation, Precision Boundaries & 67\%& 5\\
Q8 & Line & Precision Boundaries & 100\%& 7\\
Q9 & Bar & Inferential Derivation, Surface Form Normalization & 67\%& 7\\
Q10 & Bar & Inferential Derivation, Surface Form Normalization & 92\%& 6\\
Q11 & Bar & Inferential Derivation & 100\%& 6\\
Q12 & Bar & Surface Form Normalization, Non-Inferable Gaps & 100\%& 7\\
\midrule
\multicolumn{3}{l}{\textbf{Overall}} & \textbf{85\%}& \textbf{6}\\
\bottomrule
\end{tabular}
\end{center}
\begin{tablenotes}
\small
\item Suitability ratings measured on 7-point Likert scale. Gray rows indicate questions with accuracy below 60\%.
\end{tablenotes}
\label{tab:chart-to-text-performance}
\end{table*}

Chart to text matching demonstrated higher and more consistent performance (85\% accuracy, mean suitability 6/7) compared to the reverse direction, with five questions achieving perfect 100\% accuracy (Table~\ref{tab:chart-to-text-performance}). 
Questions Q1, Q5, Q8, Q11, and Q12 showed unanimous correct responses across all 12 participants, while the lowest performer (Q6, 58\%) still exceeded the worst text to chart item by 16 percentage points. 
This robust performance indicates that participants could reliably interpret our visual encodings and translate them back to appropriate linguistic expressions. 
The consistency across questions suggests that when presented with uncertainty visualizations, participants possessed a stable mental model for decoding visual treatments into semantic meaning, even when they struggled to construct those same visualizations from textual descriptions in the reverse task.

Interview analysis revealed that chart to text tasks benefited from direct visual pattern recognition, eliminating the need for mental construction of visual features that burdened Text-to-Chart tasks. 
Participants employed an "elimination strategy" where they quickly identified obvious numerical mismatches before evaluating uncertainty semantics. 
P8 explained: "Usually I'll take a look at the bullet text first, then see if the number matches, and then try to infer the words around it." 
This sequential filtering allowed participants to narrow options rapidly by checking quantitative alignment before assessing qualitative uncertainty descriptions. The perceptual advantage of direct visual parsing proved decisive. 
P7 described immediate comprehension: "Look at numbers in chart, then match to text," avoiding the linguistic to visual translation bottleneck entirely. P
3 articulated how visualizations served as verification anchors: "I use the visualization because everything is much clearer. Without it, I don't have anything to double check my interpretation." 
The ability to perceive uncertainty markers directly through squiggly borders, diagonal hatching patterns, or gradient shading provided concrete visual evidence that participants could match to linguistic uncertainty, whereas constructing those markers from text required imagining abstract visual properties before comparison.

Specific visual encodings exhibited differential levels of intuitive interpretability, with squiggly borders and diagonal hatching patterns succeeding while gradients and "NA" labels systematically failed. 
Three encoding strategies demonstrated immediate recognition. Squiggly borders for approximation resonated strongly: P11 stated "I prefer where it's a squiggly line, at first glance I can tell it's roughly 80\%," while P6 noted that current national education system familiarized students with squiggle symbols for approximation. 
Diagonal hathcing ppatern or shaded bars for inferred data proved effective for transparency. 
P2 found this encoding clear: "I like the shaded part, it implies it's not objective truth and was inferred," while P5 appreciated "dashed lines were helpful to understand this data is not present directly." 
Interactive tooltips explaining calculations received universal approval from 11 of 12 participants. 
P3 noted "The pop up does help a lot, it actually helps reading the text easier," while P9 valued "transparency in how numbers were calculated."
Conversely, two strategies failed consistently. 
Gradients representing ranges confused 8 of 12 participants. 
P8 bluntly stated "The gradient doesn't mean anything for me," P1 questioned "Can I interpret it as probability density?" revealing fundamental misunderstanding, and P11 found "the flow line unclear where the range ended." 
The lack of explicit boundaries left participants uncertain about upper and lower limits. 
"NA" labels for missing data provoked unanimous rejection across all 12 participants. 
P1 called it "wrong, like trying to make up something," P9 insisted it "should just be blank," and P2 noted participants "conflate NA with zero or Not a Number," interpreting the label as system error rather than intentional data absence. 
These systematic failures highlight that visual metaphors cannot rely on implicit meaning but require either cultural familiarity (squiggles for approximation) or explicit definition through interaction (tooltips for calculations).

\subsection{Part 2: Document Comprehension Impact}
\subsubsection{Comprehension Accuracy and Reading Efficiency}

\begin{figure}
    \centering
    \includegraphics[width=\linewidth]{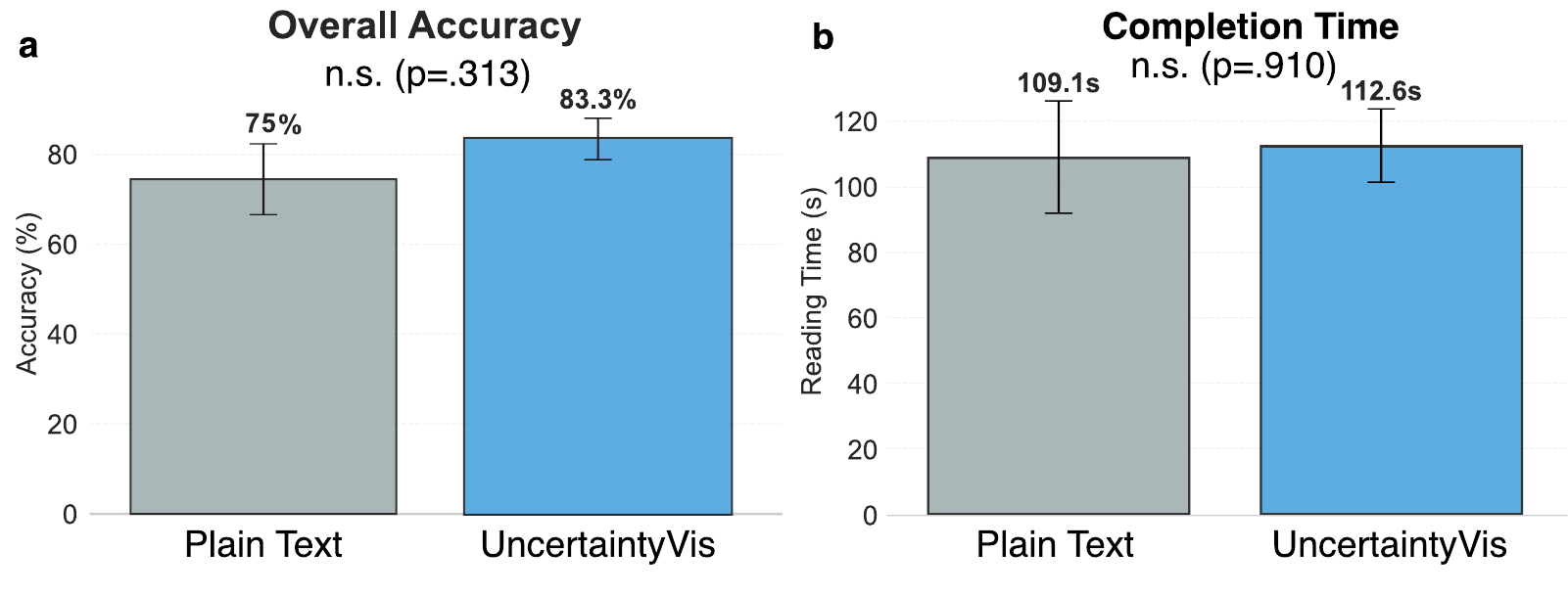}
    \caption{a, Single bar chart compares answer accuracy between Plain Text vs \toolname{} with error bars (SD). Single bar chart compares answer completion time between Plain Text vs \toolname{} with error bars (SD).}
    \label{fig:eval-part2-results}
\end{figure}

\textbf{RQ2: How do uncertainty visualizations affect readers' comprehension and interpretation of uncertainty information in data-rich documents?}

Part 2 examines whether uncertainty-aware visualizations provide practical benefits when readers engage with authentic data-rich documents. 
We measured three complementary dimensions of reading experience: comprehension accuracy (whether readers correctly interpret data and uncertainty), cognitive workload (the mental effort required to process documents), and user confidence (readers' subjective assessment of clarity and usefulness). 
Participants read documents in two conditions: uncertainty-enhanced visualizations that preserve linguistic uncertainty through visual encodings, and plain text alone without visual augmentation.

Uncertainty-enhanced documents yielded higher comprehension accuracy (83\%) compared to plain text (75\%), though this 8 percentage point improvement did not reach statistical significance (Wilcoxon signed-rank test, p=0.313, r=-0.088) (Figure~\ref{fig:eval-part2-results}a). 
Both conditions provided access to the source text, which contained sufficient information to answer questions. 
Uncertainty visualizations functioned as complementary verification tools rather than replacing textual information. 
Participants P3, P8, and P12 explicitly described using charts to "double-check" their text-based interpretations, explaining why accuracy gains remained modest while cognitive workload showed stronger patterns. 
P3 stated: "I use visualization to confirm my interpretation rather than relying on it alone," while P8 noted: "If I'm uncertain in the previous one, I could just look into the graph, then it could help me build up my confidence more compared to only text."

Reading times remained statistically equivalent between conditions (Plain: M=109.1s, Chart: M=112.6s, Wilcoxon signed-rank test, p=0.910), indicating that comprehension benefits occurred without imposing processing overhead (Figure~\ref{fig:eval-part2-results}b). 
The 3.5 second numerical difference favoring plain text falls well within expected variance and does not represent meaningful processing delay. 
This time-neutral result demonstrates that participants extracted insights at comparable speeds while viewing additional visual information. 
Visualizations did not impose the cognitive cost that critics might expect when adding visual elements to text-based documents. 
Following the interpretation framework established in prior text-to-visualization research, non-significant reading time represents a positive finding. 
Uncertainty visualization provides cognitive benefits without time penalties, making it a cost-free enhancement to document comprehension.

Interview evidence revealed that uncertainty visualizations eliminated mental calculation burden through modality substitution rather than adding visual processing overhead, explaining the time-neutral benefit. 
Multiple participants (P3, P8, P10, P12) explicitly described how charts removed the need to perform mental arithmetic. 
P10 stated: "Without charts, it's very difficult to find correct answers from scattered numbers. You need calculations across paragraphs." 
P3 explained that unified visual representation reduced cognitive load: "All data is in the same denomination. 
Easier comparison reduces mental load for calculations." The conversion of natural language frequencies like "one in five" to percentages in visualizations saved cognitive effort that would otherwise be spent on mental arithmetic. 
P3 used charts as a verification mechanism, stating visualization enabled "comparison group" confidence without re-reading text multiple times to verify interpretations. 
Participants traded text re-reading time for visual scanning time, achieving cognitive efficiency through modality shift rather than simple addition of processing demands. 
The result is comparable task completion speed with altered cognitive experience favoring reduced mental calculation and increased verification confidence.

\subsubsection{Cognitive Workload Assessment}

\begin{figure}
    \centering
    \includegraphics[width=\linewidth]{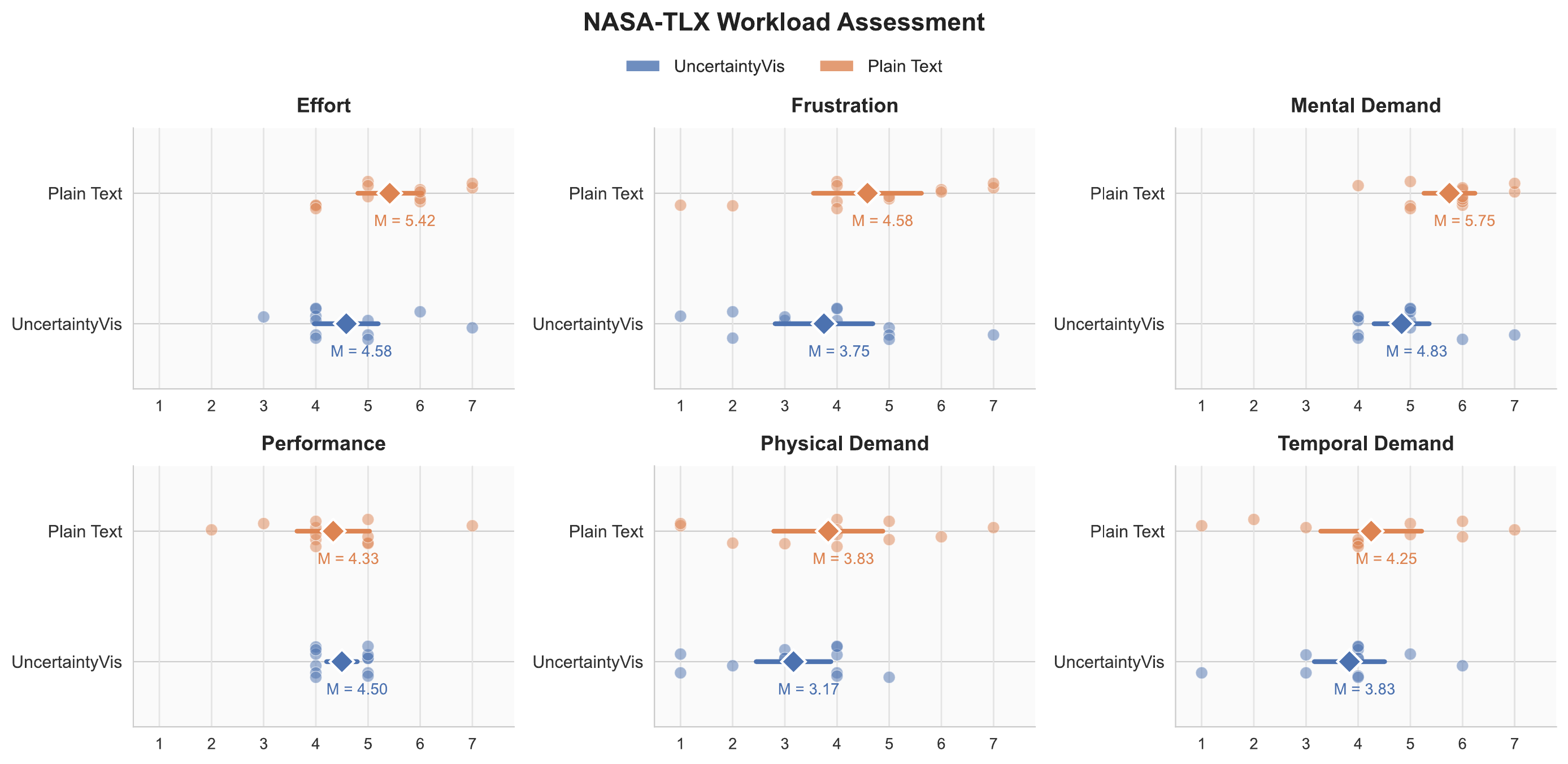}
    \caption{NASA-TLX Workload Assessment shows all six dimensions.}
    \label{fig:eval-NASA}
\end{figure}

Uncertainty-enhanced documents showed a trend toward reduced mental demand (Plain: M=5.75, Chart: M=4.83, Wilcoxon signed-rank test, p=0.054, effect size=0.460), approaching conventional significance thresholds (Figure~\ref{fig:eval-NASA}). 
While the p value narrowly exceeded the $\alpha$=0.05 threshold, the medium to large effect size indicates practical cognitive benefits that participants experienced during document processing. 
Effect size quantifies the magnitude of difference independent of sample size, providing complementary evidence to statistical significance testing. 
The 16\% reduction in perceived mental demand, combined with consistent patterns across other workload dimensions, indicates that uncertainty visualizations likely reduce cognitive burden.

Perceived effort showed a similar pattern trending toward reduction (Plain: M=5.42, Chart: M=4.58, paired t-test, p=0.073, effect size=0.769), with a large effect size indicating substantial practical differences in participants' subjective experience. 
The convergence of trending effects in both mental demand and effort, the two dimensions most directly related to cognitive processing, suggests a consistent pattern of cognitive efficiency gains rather than isolated statistical fluctuations. 
These near-significant findings on cognitive demand dimensions provide suggestive quantitative evidence that aligns with strong qualitative consensus regarding reduced calculation burden and increased processing confidence. 
The parallel trends across multiple cognitive load indicators strengthen the interpretation that uncertainty visualizations genuinely reduce processing demands, even though individual tests did not reach conventional significance levels. 
Large effect sizes in both dimensions indicate that when fully powered studies replicate this work, these trends will likely achieve statistical significance.

Remaining NASA-TLX dimensions showed no significant differences between conditions: frustration (p=0.255), physical demand (p=0.316), temporal demand (p=0.621), and performance self-assessment (p=0.750). 
The specificity of effects, with trending reductions in mental demand and effort but not frustration, time pressure, or performance evaluation, indicates that uncertainty visualizations address interpretive confidence and calculation burden rather than interface usability or overall task complexity. 
This targeted benefit profile suggests the intervention operates on cognitive interpretation processes rather than affecting global perceptions of task difficulty or interface quality. 
The absence of effects on temporal demand confirms that time-neutral reading speeds reflect genuine efficiency rather than participants rushing through chart-enhanced materials.

Interview analysis identified three specific mechanisms participants described as contributing to perceived cognitive efficiency, aligning with the trending mental demand and effort reductions observed quantitatively. 
First, unit conversion elimination reduced cognitive friction when text mixed natural language frequencies with percentages. P8, P11, and P12 described cognitive burden when encountering mixed representations requiring mental translation. 
P8 stated: "It requires more brain power. I would rather you say 20\% than one in ten."
Visualizations normalized all representations to percentages, eliminating this conversion burden that participants found particularly taxing during text-only reading. 
Second, pattern recognition substitution allowed visual processing to replace linguistic parsing. 
P2 described being a "visual person" who found text blocks "hard to process," while charts "take it out and put on screen so you don't have to process in your head." 
This substitution of visual pattern recognition for linguistic parsing potentially underlies the trending mental demand reduction, as participants leveraged perceptual systems rather than relying solely on verbal working memory. 
Third, ambiguity anxiety reduction emerged as participants felt more confident when visualizations provided verification mechanisms. 
P3 felt "uncertain without visualization because I don't have anything to double-check interpretation," while P6 valued transparency: "If there's uncertainty, showing it is appreciated because it gives a more objective view." 
This confidence boost may contribute to reduced perceived effort even when accuracy remains comparable, as participants experience less interpretive anxiety during the comprehension process.
These qualitative mechanisms provide complementary support for the marginal quantitative trends, suggesting real cognitive efficiency benefits that warrant further investigation with larger samples to achieve definitive statistical validation.

\subsubsection{Participant Preference}

\begin{figure}
    \centering
    \includegraphics[width=\linewidth]{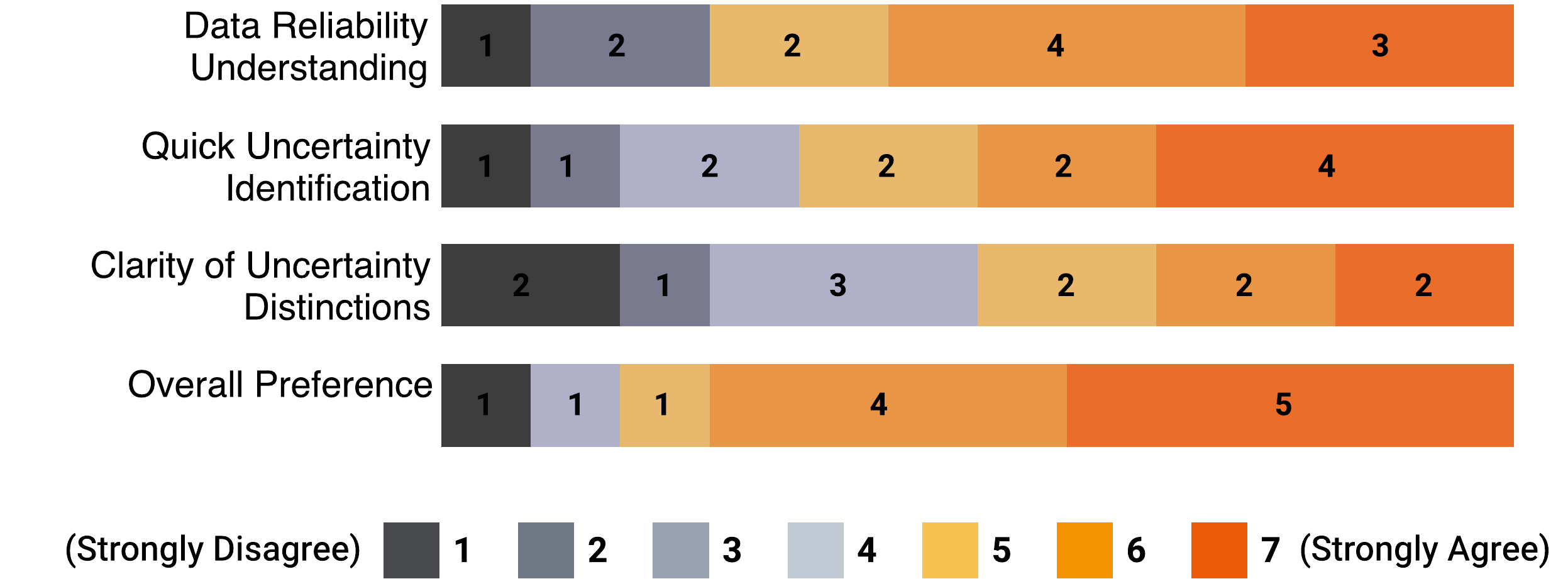}
    \caption{Participant Preference for \toolname{} shows four questions with stacked distribution bars (1-7 scale)}
    \label{fig:preference-distribution}
\end{figure}

Post-task preference surveys revealed strong endorsement for uncertainty visualization despite identified design flaws, with 75\% of participants rating overall preference at 6 or 7 on the 7-point scale (M=5.75, Figure~\ref{fig:preference-distribution} Overall Preference). 
Four dimensions assessed participant preference with consistently positive results. Data reliability understanding received moderate positive ratings (M=5.17), with 58.3\% of participants rating 6 or 7, indicating that the majority found uncertainty markers helpful for assessing source trustworthiness. 
Quick uncertainty identification showed similar patterns (M=5.17), with 50.0\% rating 6 or 7, though more distributed responses across the scale (16.7\% each at ratings 4 and 5) suggested moderate rather than universal consensus on identification speed benefits. 
Clarity of uncertainty distinctions received the lowest ratings (M=4.42), with only 33.4\% rating 6 or 7 and 25.0\% rating 4, reflecting the gradient and encoding ambiguities identified in interview analysis. 
Overall preference for uncertainty visualizations achieved the highest rating (M=5.75), with 75.0\% rating 6 or 7, indicating strong preference despite acknowledged imperfections. 
The gap between high overall preference and lower clarity ratings suggests participants valued the transparency intent and computational benefits even when specific visual encodings remained difficult to interpret. 

Qualitative evidence revealed that visualizations increased interpretive confidence independent of accuracy gains. 
P3 used charts as verification tools, stating: "Without the visualization, I don't have anything that I can double check my interpretation." 
P8 described confidence building effects: "If I'm uncertain in the previous one, I could just look into the graph, then it could help me build up my confidence more compared to only text." 
This psychological reassurance through transparency emerged as a distinct benefit beyond computational accuracy.

Interview analysis revealed that preference for uncertainty visualization followed a trimodal distribution, with nine adopters expressing positive or neutral views (P2, P3, P5, P6, P7, P8, P9, P10, P12), two non-responders showing indifference (P11, P4), and one active rejecter (P1).
Adopters valued charts for reducing mental calculation burden, with P12 stating: "I don't need to calculate by myself." P9 viewed visualizations as efficiency tools: "Make my life easier in terms of just trying to find information." P6 appreciated comprehensive representation: "If there's uncertainty, showing it is appreciated because it gives a more objective view of the full picture including uncertain parts." 
Non-responders relied on alternative strategies that diminished visualization value. 
P11 stated: "Text is enough for me. I'll just stick to text if it has all required information," preferring keyword scanning over visual interpretation. 
P4 found text easier "50\% of the time" because "numbers stand out in text" without requiring visual decoding. 
The single rejecter, P1, felt charts "destroyed trust," stating: "I spend more time understanding the chart than actually reading," finding visualizations "lossy" compared to exact figures in text. 
P1 preferred direct textual precision over visual abstraction, viewing the translation process as introducing interpretation errors. 
The 10 of 12 positive to neutral ratio (83\%) validates the core approach while highlighting the need for participant control to accommodate diverse processing styles. 
The presence of non-responders and a rejecter indicates that uncertainty visualization benefits are not universal, with individual differences in visual literacy, reading strategies, and trust in visual abstraction moderating effectiveness.

\subsection{Qualitative Findings: Visual Encoding Effectiveness and Design Insights}

\begin{table*}[t]
\centering
\caption{Qualitative Themes Synthesis from Semi-Structured Interviews (N=12 participants)}
\begin{center}
\begin{tabular}{p{3cm}p{2cm}p{6cm}p{4cm}}
\toprule
\textbf{Theme} & \textbf{Participants} & \textbf{Representative Quote} & \textbf{Design Implication} \\
\midrule

\textbf{Match-Verify Cognitive Strategy} & P7, P8, P10, P11 & \textit{"Usually I'll take a look at the bullet text first, then see if the number matches, and then try to infer the words around it." }(P8) & Visual encodings must mirror linguistic uncertainty expressions to maintain cognitive consonance between text and chart. \\
\midrule

\textbf{Successful Visual Encodings} & P2, P6, P11 (squiggly); P2, P5, P6 (hatching); P3, P9, P12 (tooltips) & \textit{"I prefer where it's a squiggly line. At first glance, I can tell it's roughly 80\%."} (P11) \textit{"I like the shaded part, it implies it's not objective truth and was inferred."} (P2) & Retain squiggly borders for approximation, diagonal hatching pattern for inference, and interactive tooltips for transparency. Increase visual salience. \\
\midrule

\textbf{Failed Visual Encodings} & P1, P8, P9, P11 (gradients); P1, P2, P8, P9 (NA labels) & \textit{"The gradient doesn't really mean anything for me."} (P8) \textit{"NA is kind of wrong, like trying to make up something."} (P1) & Replace gradients with explicit range markers. Eliminate "NA" labels; use blank space or hover tooltips. \\
\midrule

\textbf{Chart Type Dependency} & P2, P8 (line chart issues) & \textit{"Chart 1 (line chart) doesn't look right, it implies no value."} (P2) \textit{"You don't need a graph to say this at all."} (P8) & Bar and pie chart encodings succeed (83\% accuracy). Line chart encodings require redesign with confidence bands or interval markers (53\% accuracy). \\
\midrule

\textbf{Target Audience Personas} & P4, P7, P9 (time-poor); P6, P9 (accessibility); P3, P10 (unfamiliar) & \textit{"Imagine a CEO, you have tons of things to read, just show the charts."} (P4) \textit{"Dyslexic people can translate words into simple charts."} (P6) & Primary users: time-constrained professionals, accessibility seekers, domain novices. Provide user control for expert precision seekers. \\
\midrule

\textbf{Design Refinement Priorities} & P2, P8, P9, P11 (gradients); P1, P2, P8, P9 (NA); P7 (visibility) &\textit{ "Box plots for ranges better than bar charts."} (P9) \textit{"NA should just be blank."} (P9) & Critical improvements: replace gradients, eliminate NA labels, increase squiggly border salience, standardize color coding, develop line chart specific encodings. \\

\bottomrule
\end{tabular}
\end{center}
\begin{tablenotes}
\small
\item Themes emerged from thematic analysis of interview transcripts following both Part 1 (bidirectional matching) and Part 2 (document comprehension) tasks. Design implications translate participants' feedback into actionable system refinements.
\end{tablenotes}
\label{tab:qualitative-themes}
\end{table*}

Semi-structured interviews with all 12 participants revealed systematic patterns in visual encoding interpretation, cognitive strategies for uncertainty reasoning, and specific design elements requiring refinement. 
Thematic analysis identified five primary themes that explain the quantitative performance patterns and provide actionable design guidance (Table~\ref{tab:qualitative-themes}). 
First, participants employed a "Match-Verify" cognitive strategy when evaluating text-chart correspondence, prioritizing exact numerical alignment before assessing uncertainty marker appropriateness. 
Second, visual encodings demonstrated differential intuitiveness, with squiggly borders and diagonal hatching patterns succeeding while gradients and "NA" labels systematically failed. 
Third, chart type appropriateness emerged as a critical factor, with bar and pie chart encodings succeeding but line chart encodings requiring fundamental redesign. 
Fourth, based on participants' feedback and interviews, researchers analyzed and summarized four distinct target audience personas who would benefit most from uncertainty visualization, ranging from time-constrained executives to accessibility-seeking populations. 
Fifth, feedback converged on five high-priority design improvements addressing encoding clarity rather than aesthetic preference. 
These themes synthesize insights from both task performance observations and explicit participant reflections, connecting observed behaviors to underlying cognitive processes and revealing why certain design decisions succeeded or failed in practice.

\subsubsection{The Match-Verify Cognitive Strategy}
Participants consistently employed a two-stage "Match-Verify" strategy when evaluating text-chart correspondence, first seeking exact numerical alignment before assessing uncertainty marker appropriateness. 
In Stage 1 (Match), participants scanned for specific numbers as their initial filtering criterion. P7, P8, P10, and P11 described this numerical prioritization explicitly, with P8 stating: "Usually I'll look at the bullet text first, see if the number matches." 
If numerical values misaligned between text and chart, participants immediately rejected the chart regardless of uncertainty encoding quality, never advancing to evaluate whether visual uncertainty markers appropriately represented linguistic uncertainty expressions. 
In Stage 2 (Verify), only after confirming numerical correspondence did participants evaluate whether uncertainty markers such as squiggly borders or gradients aligned with linguistic hedging terms like "approximately" or "around." 
This sequential processing reveals that quantitative accuracy serves as a gatekeeper for qualitative uncertainty assessment. 
The critical insight emerges when text uses vague language but charts show precise visual elements. 
Participants hesitated even when the match was technically correct, with P10 noting: "Most difficult because I need to check again if these numbers are correct." 
The visual precision must mirror linguistic hedging for participants to experience cognitive consonance, suggesting that uncertainty is perceived as a holistic semantic property rather than an independent overlay on numerical data.

The Match-Verify strategy reveals that visual precision misalignment with linguistic uncertainty expressions creates cognitive dissonance, explaining why gradients and static reference lines failed even when semantically appropriate. 
P5 articulated this dissonance: "Strictly at point values, I could not fully agree because we're not sure if all these values are representative." 
Participants expect visual fuzziness through squiggly borders or shaded ranges when text employs hedge words, and they reject visually precise charts for linguistically approximate data. 
This behavior validates our core hypothesis that linguistic uncertainty requires distinct visual encoding rather than representing all data with uniform precision regardless of source text semantics. 
The strategy demonstrates that participants treat visual precision as a semantic signal carrying meaning about data certainty, not merely an aesthetic preference. 
When charts display sharp boundaries for approximate values, participants perceive a contradiction between what the text claims (uncertain) and what the visualization suggests (certain), leading to rejection of otherwise numerically accurate representations. 
This finding establishes that effective uncertainty visualization must maintain semantic alignment across modalities, with visual encoding choices directly communicating the epistemic status that authors express through linguistic hedging in original text.

\subsubsection{Differential Effectiveness of Visual Encodings}
Three visual encoding strategies demonstrated immediate intuitive recognition across the majority of participants: squiggly borders for approximation, diagonal hatching pattern for inference, and interactive tooltips for transparency. 
Squiggly borders achieved immediate understanding from 8 of 12 participants. 
P11 stated: "At first glance, I can tell it's roughly 80\%," while P2 noted that "the visual aspect of it implies the range of uncertainty." 
P6 attributed this intuitive recognition to educational context: "Singapore education system makes this intuitive for approximation." 
However, P7 identified a critical visibility constraint, noting that squiggly borders became difficult to perceive on small screens without zooming, revealing an accessibility challenge that must be addressed through increased stroke weight or amplitude. Diagonal hatching pattern or shaded bars for inferred data received positive reception from 7 of 12 participants. 
P2 found the transparency effective: "Shaded bars imply not objective truth and were inferred," while P6 appreciated that shading communicated "over or under estimation and the reference point." 
P5 valued the semantic distinction: "Dashed lines were helpful to understand this data is not present directly in the source." 
Interactive tooltips explaining calculation methodology achieved near universal approval, with 11 of 12 participants valuing this feature (only P11 remained indifferent). 
P3 stated: "The pop up helps reading the text a lot easier," while P9 praised the "transparency in how numbers were calculated." 
P10 found the "yellow question mark helpful for quick understanding," indicating that explicit explanatory elements succeed where implicit visual metaphors may fail.

Two visual encoding strategies systematically failed to communicate intended meaning and require immediate redesign: gradients for ranges and "NA" labels for missing data. 
Gradient encodings confused or were ignored by 8 of 12 participants. 
P8 bluntly stated: "The gradient doesn't mean anything for me," while P1 questioned fundamental interpretation: "Can I interpret it as probability density?" 
This misunderstanding reveals that participants lacked a shared mental model for gradient semantics. 
P11 found "the flow line unclear where the range ended," and P9 preferred alternative representations: "Fading throws me off. I prefer box plots for ranges." 
The root cause of gradient failure lies in the absence of clear boundaries. Participants require explicit upper and lower limits rather than continuous fades that leave endpoint ambiguity. 
"NA" labels for missing data provoked unanimous rejection across all 12 participants. 
P1 called the label "wrong, like trying to make up something," while P9 insisted it "should just be blank." 
P2 identified the core confusion: "Participants conflate NA with Zero or Not a Number," interpreting the label as a system error rather than intentional data absence. 
The semantic ambiguity of "NA" as both "Not Applicable" and "Not Available" compounded the problem, with participants unable to distinguish whether data was uncollected, uncollectable, or simply irrelevant. 
These systematic failures highlight that visual metaphors cannot rely on implicit meaning or ambiguous conventions but require either cultural familiarity, as with squiggly borders for approximation, or explicit definition through interactive elements, as with tooltips for calculation transparency.

Visual encoding effectiveness varied dramatically by chart type, with bar and pie charts succeeding (83\% average accuracy) while line charts failed (53\% average accuracy), revealing that our encodings were optimized for discrete categorical comparisons rather than continuous temporal trends. 
Bar and pie chart success factors included clear boundaries between categories, squiggly borders remaining visible on larger geometric shapes, and color differentiation aiding category distinction. 
These structural properties allowed uncertainty markers to function as intended, with participants easily perceiving modifications to familiar chart elements. 
Line chart failure factors demonstrated fundamental incompatibility between our encoding approach and continuous data representation. 
Static reference lines indicating thresholds like "kept below 6.5\%" were perceived as meaningless, with P2 and P8 rejecting charts that showed flat lines when text implied temporal variation. 
P2 stated: "Chart 1 (line chart) doesn't look right, it implies no value," expecting visual fluctuation to represent behavior over time. 
Gradients applied along continuous lines were misinterpreted as transparency effects rather than uncertainty bounds. 
Small squiggly dots on scatter plots became invisible without zooming, with P4, P6, and P12 missing these markers entirely until magnification revealed them. 
The design implication is clear: line charts require alternative uncertainty encodings specifically designed for continuous data, potentially including confidence bands, error ribbons, or discrete interval markers rather than adaptations of bar chart strategies. 
The chart type dependency demonstrates that uncertainty visualization cannot apply a universal encoding scheme across all visualization types but must develop type-specific strategies that respect the perceptual and semantic properties of each chart structure.

\subsubsection{Target Audience Identification}
Participants identified four distinct user personas who would benefit most from uncertainty-aware visualizations, with strongest consensus around time constrained professionals and accessibility seeking populations. 
The first persona, "Time Poor Skimmers," includes executives and busy professionals requiring rapid insight extraction without full text reading. 
P4, P7, P9, and P11 described this group, with P4 stating: "Imagine a CEO who has tons of things to read and just show the charts to make life easier." 
P7 self identified with this persona: "People who don't really read all the information because they're in a rush of time." 
This persona prioritizes efficiency over comprehensive understanding, using visualizations as shortcuts to extract key statistics from dense documents. 
The second persona, "Accessibility Seekers," encompasses users with reading disabilities or those finding large text blocks cognitively draining. 
P4, P6, P8, and P9 identified this group, with P6 specifically noting: "Dyslexic people can translate all those words into a simple chart," while P9 stated it is "good for those with reading issues because it's easier to see in pictures." 
This persona benefits from modality shift, leveraging visual processing to compensate for linguistic processing challenges. 
The third persona, "Context Unfamiliar Users," includes laypeople lacking domain background or non-native speakers struggling with technical terminology. 
P3 and P10 described this group, with P10 noting: "People not familiar with the background or who have trouble understanding some English words." 
This persona requires scaffolding to bridge knowledge gaps that domain experts navigate easily through textual descriptions alone. 
The fourth persona, "Data Storytellers," encompasses journalists and communicators presenting narratives to broad audiences. 
P2 positioned the tool for "journalists when doing storytelling with data," contrasting this with professional documents requiring precise legal language. 
This persona values visual communication for public engagement rather than technical documentation.

P1 and P11 represented the "expert precision seeker" nonbeneficiary profile, prioritizing raw textual data over visual abstraction and rejecting uncertainty visualization as introducing interpretive noise. 
P1 articulated strong rejection: "I wouldn't want any uncertainty visualization because I spend more time understanding the chart than actually reading. 
Text gives more direct information." P1 viewed visual encoding as a "lossy" translation that obscured rather than clarified information, preferring unmediated access to exact numerical values and linguistic expressions. 
P11 demonstrated nonresponse rather than active rejection: "Text is enough for me. I'll stick to text if it has the required information." 
P11's efficiency at keyword scanning rendered visualizations redundant, finding them neither helpful nor harmful but simply unnecessary for task completion. 
These responses reveal that visualization benefits are not universal across all cognitive styles or expertise levels. 
Expert users who have developed efficient text processing strategies and who value precision over interpretation speed may find uncertainty visualization adds processing overhead without corresponding benefit. 
The 10 of 12 positive to neutral ratio (83\%) validates the core approach for the majority of participants while highlighting the need for opt in and opt out functionality to accommodate those preferring direct textual access.
The presence of both active rejecters and passive nonresponders indicates that effective deployment requires participant control, allowing individuals to select their preferred information modality rather than imposing a single visualization enhanced format on all readers regardless of their processing preferences or domain expertise.

\subsubsection{Design Refinement Priorities}
Interview synthesis identified five high priority design improvements with strong participant consensus, all addressing fundamental encoding clarity rather than aesthetic preference. 
First, gradients must be replaced with explicit range markers. P2, P8, P9, and P11 agreed that current gradient fading lacks clear boundaries, preventing participants from determining precise upper and lower limits. 
P9 recommended: "Box plots for ranges are better than bar charts," suggesting error bars or explicit "Lower: X, Upper: Y" annotations as alternatives that provide definitive range endpoints. 
Second, "NA" labels must be eliminated and replaced with blank space. 
P1, P2, P8, and P9 unanimously rejected the current approach, with P9 stating it "should just be blank" and recommending a dash character or explicit "Data Not Available" tooltip on hover if explanation is required. 
The systematic misinterpretation of "NA" as zero or system error rather than intentional data absence renders this label counterproductive to transparency goals. 
Third, squiggly border salience must be increased to address visibility constraints. 
P7 identified that squiggly borders became imperceptible on small screens, requiring increased stroke weight or amplitude to ensure recognition without zooming. 
Testing minimum size thresholds across device types will establish appropriate visual parameters. 
Fourth, color coding must be standardized across all visual encodings to maintain consistency between legends and chart elements. 
P11 and P12 noted confusion when legend colors showed solid fills while charts displayed diagonal hatching patterns, or when hatched regions in charts did not match legend representations exactly. 
Ensuring legend colors precisely match chart fills, including texture patterns, and using distinct colors for different uncertainty categories (approximation versus inference) will eliminate this source of ambiguity. 
Fifth, line chart specific encodings must be developed to address the systematic failures observed in Q7 and Q8. 
Current bar chart adaptations fail for temporal trends, achieving only 53\% accuracy compared to 83\% for bar and pie charts. 
Investigating confidence ribbons, interval markers, or other continuous uncertainty representations designed specifically for line chart structures represents a critical direction for future iteration. 
These five improvements emerged from converging evidence across multiple participants and task types, indicating they address fundamental rather than idiosyncratic issues. 
Implementing these refinements should substantially improve encoding clarity while maintaining the transparency and verification benefits that 83\% of participants valued in the current system.

The convergence of quantitative performance patterns and qualitative participants' insights validates our core hypothesis that linguistic uncertainty requires distinct visual encoding, while revealing specific refinements needed for robust deployment. 

The medium to large effect sizes on cognitive workload dimensions suggest meaningful practical benefits that may achieve statistical significance in larger samples. 
However, the non-significant accuracy improvement (p=0.313) and the presence of two non-responders (P1, P11, representing 16.7\%) indicate that benefits are not universal. 
Individual processing style variations and current sample size (n=12) constrain generalizability. 
The identified design flaws, specifically gradients for ranges, "NA" labels for missing data, and line chart encodings for continuous trends, provide concrete targets for iterative refinement. 
The 83\% positive reception rate combined with convergent qualitative and quantitative evidence validates pursuing systematic improvement rather than abandoning the approach. 
The consistent pattern of cognitive efficiency benefits across multiple measures, supported by detailed mechanistic explanations from participant interviews, warrants replication with larger samples to achieve definitive statistical validation while implementing the five priority design improvements to address encoding clarity issues.
\section{Discussion}
\textbf{Statistical vs. Linguistic Uncertainty as Distinct Problems.}

\toolname{} reveals that linguistic uncertainty represents a fundamentally different problem from the statistical uncertainty traditionally addressed in visualization research. 
Statistical uncertainty is quantified through error bars, confidence intervals, and probability distributions, techniques designed to represent measurement error and sampling variability. 
These approaches address the question: given the data collection methods, what range of values might the true measurement occupy? 
Linguistic uncertainty poses a different question entirely: when an author writes 'approximately 15\% of students dropped out this year' rather than '15\%,' or 'more than 8\%' instead of a precise value, they signal that the exact figure is not critical to the narrative point being made. 
The approximate value carries sufficient meaning for the reader's purposes, and the linguistic hedge communicates this authorial judgment deliberately. 
Linguistic uncertainty is therefore a property of authorial expression: authors make deliberate word choices to signal imprecision, incompleteness, or derivation within their narratives, and these choices cannot be reduced to statistical distributions.

Our taxonomy of four categories (i.e., inferential derivation, precision boundaries, surface form normalization, and non-inferable gaps) is a distinct analytical framework to systematically categorize these authorial choices for a specific purpose: enabling the design of visual encodings that preserve rather than discard the semantic meaning authors encode through language. 
The taxonomy does not describe how authors think about uncertainty; it describes how visualization can organize and respond to the diversity of ways authors express it. 
Each category requires specialized visual encoding strategies that preserve the author's intended meaning rather than imposing mathematical frameworks designed for measurement error. 
This distinction establishes a new research domain at the intersection of natural language processing, human-computer interaction, and information visualization, where the challenge shifts from modeling uncertainty to preserving it as authors originally expressed it in data-rich documents.

\textbf{From Statistical Inference to Semantic Preservation}
We deliberately rejected computational inference approaches in favor of semantic preservation based on fundamental concerns about verifiability and author intent. 
Initial exploration attempted to leverage large language models to infer specific value distributions from uncertainty expressions—for instance, repeatedly sampling the model to estimate that "approximately 15\%" might represent a normal distribution centered at 15\% with some variance. 
This approach initially appeared promising because it would enable applying existing statistical uncertainty visualization methods to textual data. 
However, we abandoned this strategy when we recognized that the black-box nature of language models makes such inferred distributions fundamentally unverifiable. 
We cannot determine whether a generated distribution reflects the author's true intent or merely the model's training artifacts and hallucination tendencies. 
An author who writes "approximately 1\%" makes a conscious linguistic choice that differs semantically from claiming knowledge of an underlying statistical distribution—they signal imprecision without specifying its mathematical form. 
\toolname{} instead preserves these original linguistic markers through visual encodings that maintain direct traceability between text and visualization. 
Surface Form Normalization's darker and thicker borders reveal original textual forms on hover, while Inferential Derivation's yellow question mark badges display exact derivation logic when clicked. 
User feedback from our evaluation confirmed the value of this approach, with participants explicitly noting that the ability to verify transformation logic "helps validate and check the trustworthiness of the data." 
Semantic preservation prioritizes transparency and author fidelity over computational sophistication.

\textbf{The Explainability-Efficiency Balance in Uncertainty Visualization.}
User interactions with \toolname{} revealed a fundamental tension between providing detailed uncertainty provenance and supporting efficient at-a-glance comprehension. 
Participants valued the interactive explanations embedded in uncertainty-aware visualizations—hovering on converted values to reveal their original textual forms, clicking on inferred data points to display logical derivation steps through question mark badges. 
These interactions served an essential validation purpose, as users explicitly stated they wanted to "check the trustworthiness" of visualized data by examining how the \toolname{} interpreted uncertainty expressions. 
However, effective visualization requires immediate visual comprehension without mandatory interaction. 
A reader should grasp that a bar represents an approximation or an inferred value within seconds, not after clicking multiple elements to access explanatory tooltips. 
We addressed this tension through layered information architecture: primary visual encodings including diagonal hatching pattern for inferred values, squiggly borders for approximations, darker and thicker borders for normalized forms, and "N/A" labels for data gaps communicate uncertainty type at a glance, while secondary interactive layers provide detailed provenance on demand. 
This design choice raises broader questions about information revelation strategies in uncertainty-aware visualization. 
Which uncertainty characteristics merit immediate visual salience? When does additional detail enhance trust versus creating cognitive burden? 
The core principle remains clear: uncertainty visualization must balance transparency with usability, providing verification capabilities without requiring constant interaction that impedes natural reading flow.

\textbf{Uncertainty Visualization as Trust Infrastructure.}
The study revealed that uncertainty visualizations function as trust infrastructure rather than merely displaying information because they enable readers to judge whether to trust data claims and understand their limitations. 
Participants reported that seeing uncertainty in charts helps them validate and check data trustworthiness, challenging a common assumption about uncertainty communication. Conventional wisdom suggests that showing uncertainty makes readers less confident in the data, potentially undermining the authority of presented findings. 
Instead, we found the opposite: transparency increases trust. When readers can see that an author originally wrote "approximately 15\%" and verify how \toolname{} converted that phrase into a visual form, they trust the overall presentation more precisely because they understand its limitations.
This represents informed skepticism rather than blind acceptance—readers appreciate knowing what they don't know. \toolname{} suggests a different standard for data visualization in data-rich documents: charts should explicitly show their limitations, similar to how journalists cite sources to enable verification. 
One participant captured this potential transformation: "If the uncertainty visualization becomes a standard, it will really help to validate and check the trustworthiness of the data." 
This approach could help combat misinformation by making data claims verifiable rather than authoritative, shifting readers from passive consumers to active evaluators of evidence quality. 
The visual encoding strategies serve not merely as aesthetic choices but as mechanisms for establishing and maintaining reader trust through systematic transparency.

\textbf{Preliminary Patterns in Uncertainty Expression Across Domains}
Corpus analysis revealed preliminary patterns suggesting that different document genres may use linguistic uncertainty in different ways, though these observations must be interpreted with caution given the small number of documents per domain and the absence of length normalization across the corpus. 
Within our sample, medical documents tended to use precision boundaries more frequently, while economic documents showed a higher proportion of inferential derivation expressions.
These patterns may reflect genuine domain conventions, but they may equally reflect the specific stories being reported within each domain, individual author writing styles, or sampling variation inherent to a 12-document corpus. 
Despite these limitations, these preliminary patterns suggest a potentially productive direction for future visualization research. 
If domain-specific uncertainty expression tendencies can be confirmed through larger, systematically normalized corpora, future visualization systems could detect document type and adjust visual encoding strategies to align with the conventions that expert readers already expect within their fields. 
For instance, systems serving medical readers might foreground precision boundaries encodings, while systems serving economic analysts might prioritize inferential derivation transparency. 
We present these as hypotheses motivating future investigation rather than established findings.

\textbf{Methodology Choices and Positioning.}
The evaluation strategy made deliberate methodological choices to isolate visualization effectiveness and position \toolname{}'s contribution within the text-to-chart research landscape. 
We designed a two-part study where each part tests a different question with appropriate comparisons. 
Part 1 asked: do uncertainty-aware visualizations accurately preserve what the original text meant? 
We tested this through bidirectional matching tasks where participants chose between uncertainty-aware visualizations with proper visual encodings and standard visualizations without uncertainty indicators, or alternative encodings from different categories. 
This demonstrated whether visual encoding strategies successfully preserve semantic information that regular charts lose. 
Part 2 asked: does uncertainty visualization actually help people understand data-rich documents better? 
We tested this by having participants read documents with uncertainty-enhanced visualizations versus plain text alone, measuring comprehension accuracy, cognitive load, and confidence. 
This showed whether visualization adds value beyond just having textual information present. 
We made another important choice: pre-selecting which uncertainty expressions to visualize rather than having \toolname{} automatically detect them. 
This decision reflects that our main research question assesses visualization effectiveness for reading experience, not algorithm performance in detecting uncertainty-containing text. 
Automatic detection would confound results because evaluation failures could come from missed expressions, wrong classifications, or ineffective visual designs. 
By pre-selecting expressions, the study measured visualization quality, not detection algorithm performance. 
This approach distinguishes \toolname{} from existing text-to-chart systems like GistVis and ChartifyText. 
While those systems ask "can we generate charts from text?", we ask "can we preserve meaning during transformation?" GistVis checks whether charts accurately show extracted data; 
\toolname{} checks whether charts preserve semantic information that extraction typically destroys. 
When an author writes "approximately 15\%" instead of "15\%," they communicate more than a number—they signal epistemic positioning that our taxonomy captures through four distinct categories, each requiring different visual encoding strategies. 
This positions \toolname{}'s contribution not as adding features to existing systems, but as rethinking what text-to-chart translation should preserve.

\textbf{Future Directions.}
Future research can extend \toolname{}'s semantic preservation foundation across multiple dimensions. 
Automatic detection methods could identify uncertainty expressions in data-rich documents without manual pre-selection—now that visual encoding effectiveness has been validated, detection becomes a tractable engineering problem that could integrate with existing text-to-chart pipelines. 
Taxonomy expansion could address additional uncertainty types identified during corpus analysis. 
Epistemic confidence captures modal expressions indicating author certainty levels, such as "may," "likely," or "suggests," which communicate probabilistic reasoning without quantifying it. 
Rhetorical framing addresses how word choice influences numerical perception—"only 15\% of students dropped out" versus "as many as 15\% dropped out" frames identical data with different implications. 
These extensions would expand \toolname{} capability to handle more nuanced forms of linguistic uncertainty beyond the four categories currently implemented. 
Broader evaluation contexts could test effectiveness across different user groups including domain experts versus general readers, different decision contexts such as policy-making versus medical diagnosis versus business analytics, and different reading scenarios ranging from quick scanning to deep analysis. 
Cross-cultural variations in uncertainty expression and interpretation represent another important frontier, as linguistic uncertainty conventions may vary across languages and cultural contexts. 
Integration opportunities include combining uncertainty-aware visualization with existing document reading tools and platforms, and extending the approach to other data-rich formats including presentations, dashboards, and interactive reports. 
The foundation is now established: a semantic preservation approach prioritizing author intent, a validated taxonomy of linguistic uncertainty categories, effective visual encoding strategies for each category, and rigorous evaluation methodology demonstrating both technical accuracy and practical utility for comprehending data-rich documents.

\section{Conclusion}
Linguistic uncertainty in data rich documents requires fundamentally different treatment from the statistical uncertainty traditionally addressed in visualization research. 
While statistical uncertainty visualization focuses on measurement error and sampling variability through confidence intervals and error bars, linguistic uncertainty captures authors' deliberate epistemological choices expressed through natural language. 
Our four category taxonomy demonstrates that expressions like "approximately 15\%", "more than 8\%", and "significant improvement" carry distinct semantic meanings that automated chart generation systems currently discard. 
\toolname{} preserves these meanings through category specific visual encodings, maintaining semantic fidelity between textual expressions and visual representations. 
Our evaluation with 12 participants demonstrates that these encodings successfully communicate uncertainty semantics, with chart to text matching achieving 85.42\% accuracy and participants reporting reduced mental demand when using uncertainty-aware visualizations. 
While our study reveals specific design improvements needed for gradient encodings, missing data labels, and line chart representations, the core principle holds: preserving linguistic uncertainty as authors express it enables readers to make appropriately calibrated judgments about data reliability and completeness. 
Future work will focus on automatic detection of uncertainty expressions, taxonomy expansion to capture epistemic confidence and rhetorical framing, and systematic visual design refinements. 
These advances could establish new practices where automated visualization systems visually distinguish between precise and approximate values, stated and inferred data, and complete and incomplete information, rather than treating all numbers as equally certain. 
Just as journalists cite sources to enable verification, data visualizations in scientific publications, policy reports, and data journalism could systematically preserve the uncertainty semantics that authors encode through careful word choices, transforming readers from passive consumers into active evaluators of evidence quality.

\bibliographystyle{ACM-Reference-Format}
\bibliography{sample-base}

\appendix
\newpage
\appendix

\section{LLM Processing Prompts}
\label{appendix:prompts}

This appendix presents the two prompts used in the UncertaintyVis processing pipeline. 
Prompt A governs uncertainty expression extraction and structured data generation. 
Prompt B governs chart type recommendation based on extracted uncertainty categories.

\subsection{Prompt A: Uncertainty Expression Extraction}
\label{appendix:prompt-extraction}

Prompt A instructs the large language model to analyze a source sentence, 
identify uncertainty types according to the four-category taxonomy, 
and output a structured JSON dataset suitable for visualization generation.

\vspace{0.5em}
\noindent\textbf{System Role}

\begin{quote}
\textit{You are an expert in converting uncertain textual expressions into 
structured JSON datasets for uncertainty visualization. Your task is to analyze 
text, identify the four uncertainty categories defined below, and transform them 
into appropriate data structures.}
\end{quote}

\vspace{0.5em}
\noindent\textbf{Uncertainty Categories}

\begin{itemize}
    \item \textbf{Precision Boundaries}: Linguistic markers indicating approximation, 
    ranges, or bounded values. Sub-types include:
    \begin{itemize}
        \item \textit{Approximation}: about 15\%'', around 30\%'', approximately 50\%''
        \item \textit{Directional Up}: more than 20\%'', exceeded 40\%'', above 15\%''
        \item \textit{Directional Down}: less than 15\%'', below 30\%'', fewer than 40,000''
        \item \textit{Range}: 50--60 hectares'', between 4\% and 5\%''
    \end{itemize}
    
    \item \textbf{Surface Form Normalization}: Varied textual representations of 
    identical numerical values. Examples include text-to-number conversion 
    (half'' $\rightarrow$ 50\%, a quarter'' $\rightarrow$ 25\%), bracketed 
    explanations, and reference citations.
    
    \item \textbf{Inferential Derivation}: Quantitative information not explicitly 
    stated but calculable or logically deducible from stated values. 
    Sub-types include entity-inferred and value-inferred cases, both requiring 
    an \texttt{inference\_explanation} field in the output.
    
    \item \textbf{Non-Inferable Gaps}: Quantitative information that is fundamentally 
    absent and cannot be recovered through any analytical approach. 
    Encoded with \texttt{"value": null}.
\end{itemize}

\vspace{0.5em}
\noindent\textbf{Analysis Process}

The model follows five sequential steps:
\begin{enumerate}
    \item Identify explicit values and their associated uncertainty markers.
    \item Identify missing complementary data recoverable through calculation.
    \item Identify surface form expressions requiring normalization.
    \item Determine the appropriate chart type based on data structure.
    \item Calculate proximity levels for directional boundary encodings:
    \textit{Close} (3\% offset), \textit{Moderate} (5\% offset), 
    \textit{Large} (10\% offset).
\end{enumerate}

\vspace{0.5em}
\noindent\textbf{Output Format}

The model returns a structured JSON object conforming to the following schema:

\begin{verbatim}
{
  "chart_type": "[chart type from supported list]",
  "title": "Descriptive Title",
  "data": [
    {
      "category": "Category Name",
      "value": 50,
      "baseline_value": 40,
      "uncertainty_type": "precision_boundaries | 
                           surface_form_normalization | 
                           inferential_derivation | 
                           non_inferrable_gaps",
      "precision_marker": "approximation | directional up | 
                           directional down | directional between",
      "proximity_level": "close | moderate | large",
      "range": [50, 60],
      "original_form": "original text",
      "entity_inferred": true,
      "value_inferred": true,
      "inference_explanation": {
        "entity": "How entity was inferred",
        "value": "How value was calculated"
      }
    }
  ],
  "axes": {
    "x_axis": "X-axis Label",
    "y_axis": "Y-axis Label"
  }
}
\end{verbatim}

\vspace{0.5em}
\noindent\textbf{Key Transformation Rules}

\begin{itemize}
    \item Always calculate complementary data when possible 
    (e.g., if 27\% cancer and 15\% heart disease are stated, 
    infer 58\% as other causes'').
    \item Preserve original uncertainty; do not artificially increase precision.
    \item Add \texttt{inference\_explanation} for all derived values 
    to maintain transparency.
\end{itemize}

\subsection{Prompt B: Chart Type Recommendation}
\label{appendix:prompt-chart}

Prompt B instructs the large language model to recommend an appropriate chart type 
for a given sentence and its pre-identified uncertainty categories, 
based on data type compatibility and uncertainty visualization requirements.

\vspace{0.5em}
\noindent\textbf{System Role}

\begin{quote}
\textit{You are an expert in data visualization and uncertainty representation.}
\end{quote}

\vspace{0.5em}
\noindent\textbf{Supported Chart Types}

The following chart types are available for recommendation:
vertical bar chart, horizontal bar chart, grouped bar chart, stacked bar chart, 
line chart, scatter plot, stripe plots, pie chart, dot plots, 
and none (when no chart type is suitable).

\vspace{0.5em}
\noindent\textbf{Input}

\begin{itemize}
    \item \textbf{Sentence}: The source text excerpt containing the uncertainty expression.
    \item \textbf{Uncertainty Categories}: The pre-identified uncertainty categories 
    from Prompt A (Section~\ref{appendix:prompt-extraction}).
\end{itemize}

\vspace{0.5em}
\noindent\textbf{Analysis Criteria}

The model evaluates four factors when selecting a chart type:
\begin{enumerate}
    \item Data type compatibility: whether the data is temporal, categorical, 
    numerical, or proportional.
    \item Uncertainty visualization compatibility: whether the chart type 
    supports the required uncertainty encoding methods such as error bars, 
    confidence intervals, ranges, or missing data representations.
    \item Chart effectiveness for the specific uncertainty pattern present.
    \item Co-occurrence handling: how the chart type accommodates multiple 
    simultaneous uncertainty categories within a single data element.
\end{enumerate}

\vspace{0.5em}
\noindent\textbf{Output Format}

The model returns a structured JSON object with three fields:

\begin{verbatim}
{
  "chart_type": "exact match from supported chart types",
  "reasoning": "one sentence explaining the choice based on 
                data type and uncertainty patterns",
  "confidence": "high | medium | low"
}
\end{verbatim}

\newpage
\onecolumn


\section{Part 1 Study Materials: Bidirectional Matching Tasks}

\subsection{Bidirectional Matching Tasks Overview}

\begin{table}[h]
\centering
\caption{Text-to-Chart Task Materials (12 Questions)}
\label{tab:text-to-chart}
\begin{tabular}{p{0.4cm} p{6.5cm} p{3.5cm} p{2.5cm}}
\toprule
\textbf{Q} &
\textbf{Source Text} &
\textbf{Uncertainty Categories} &
\textbf{Chart Type} \\
\midrule

1 &
Since 1981, the leading cause of death in Japan has been cancer, which accounted for 27\% of total deaths in 2018, followed by heart disease at 15\% &
Inferential Derivation &
Vertical Bar \\
\addlinespace

2 &
China's auto output exceeded 30.16 million units in 2023, up 11.6 percent year-on-year &
Inferential Derivation; Precision Boundaries &
Vertical Bar \\
\addlinespace

3 &
As we explained in a story last month, less than 15\% of those seeking asylum were ultimately granted it in fiscal years 2022 and 2023 &
Precision Boundaries &
Vertical Bar \\
\addlinespace

4 &
And we get 50--60 hectares [120--150 acres] of green space that keeps the city cool, prevents flooding &
Precision Boundaries &
Vertical Bar \\
\addlinespace

5 &
The most ambitious of which involved removing half of the open-air parking spaces in the city, about 11,000 &
Inferential Derivation; Precision Boundaries &
Vertical Bar \\
\addlinespace

6 &
Half disapprove of the major changes he has proposed for government spending &
Surface Form Normalization; Inferential Derivation &
Pie Chart \\
\addlinespace

7 &
If this trend continues, Korea's population is estimated to halve by the year 2100 &
Inferential Derivation &
Line Chart \\
\addlinespace

8 &
While relative poverty is unlikely to be eradicated, in Europe, over the last decade, it has been kept below 6.5 percent (using the 40 per cent of median disposable income standard) &
Precision Boundaries &
Line Chart \\
\addlinespace

9 &
Smoking rates in men of around 80\% in 1970 now compare with around 30\% in 2013. In contrast, rates in women have remained constantly low, at around 15\% and 8\%, respectively &
Precision Boundaries &
Line Chart \\
\addlinespace

10 &
There are currently fewer than 40,000, according to the DOE, with a quarter of those in California &
Precision Boundaries &
Vertical Bar \\
\addlinespace

11 &
The 29-year-old Serbian centre, had 32 points, 16 rebounds and 16 assists plus four steals while Jamal Murray added 27 points for the Nuggets &
Surface Form Normalization; Non-Inferable Gaps &
Grouped Bar \\
\addlinespace

12 &
In 50 years time, the number of working age people will have halved, the pool eligible to take part in the country's mandatory military service will have shrunk by 58\% &
Inferential Derivation &
Grouped Bar \\

\bottomrule
\end{tabular}
\end{table}

\begin{table*}[t]
\centering
\caption{Chart-to-Text Task Materials (12 Questions).}
\label{tab:chart-to-text}
\small
\begin{tabular}{p{0.4cm} p{6.0cm} p{4.5cm} p{2.0cm}}
\toprule
\textbf{Q} &
\textbf{Source Text} &
\textbf{Uncertainty Categories} &
\textbf{Chart Type} \\
\midrule

1 &
The DHS data show 6.5 million encounters at the U.S.-Mexico border in that time frame, a figure that includes both the 5.8 million apprehensions between legal ports of entry and a little more than 700,000 migrants who arrived at ports of entry without authorization.'' &
Precision Boundaries, Inferential Derivation &
Stacked Bar \\
\addlinespace

2 &
DHS statistics show 2.8 million of the encounters at the southern border resulted in a removal or expulsion directly from CBP custody. Most of those removals — nearly 2.5 million — were immediate expulsions under Title 42.'' &
Precision Boundaries, Inferential Derivation &
Stacked Bar \\
\addlinespace

3 &
Total DHS repatriations through October amounted to 3.7 million, a figure that includes the 2.8 million removals directly from CBP, as well as removals by ICE.'' &
Inferential Derivation &
Stacked Bar \\
\addlinespace

4 &
There were 6.5 million encounters at the southern border from February 2021 through October, including a little more than 700,000 migrants who arrived without legal documentation at ports of entry.'' &
Precision Boundaries, Inferential Derivation &
Stacked Bar \\
\addlinespace

5 &
About 2.5 million people through October have been released into the U.S. That figure includes 2 million released by Border Patrol. The 2.5 million number also includes nearly 534,000 paroles processed at legal ports of entry.'' &
Precision Boundaries &
Stacked Bar \\
\addlinespace

6 &
China exported 5.22 million vehicles in 2023, becoming a leading car exporter in the world. Of all vehicle exports, about one-third were new-energy vehicles (NEVs), totaling 1.773 million units.'' &
Inferential Derivation &
Stacked Bar \\
\addlinespace

7 &
The facility will utilize many smart industrial robots allowing for the automation of over 100 production processes with an automation rate of 90 percent.'' &
Precision Boundaries, Inferential Derivation &
Stacked Bar \\
\addlinespace

8 &
Hesai has recently completed the construction of a new R\&D center of nearly 70,000 square meters in Shanghai for manufacturing named Maxwell.'' &
Precision Boundaries &
Vertical Bar \\
\addlinespace

9 &
Of Labour's 202 MPs (excluding Speaker Lindsay Hoyle), 104 are women — and of the Liberal Democrats' 11 MPs, seven are women. &
Surface Form Normalization, Inferential Derivation &
Grouped Bar \\
\addlinespace

10 &
Of those who say Trump has not accomplished much, 47 per cent pin the blame on him while a quarter blame congressional Republicans. Seven per cent say Democrats are to blame.'' &
Surface Form Normalization, Inferential Derivation &
Pie Chart \\
\addlinespace

11 &
In 2022, 98\% of births in South Korea were within marriage, while only 2\% were outside marriage.'' &
Inferential Derivation &
Pie Chart \\
\addlinespace

12 &
Damian Lillard added 24 points, nine assists and seven rebounds, while Malik Beasley added 20 points.'' &
Surface Form Normalization, Non-Inferable Gaps &
Grouped Bar \\

\bottomrule
\end{tabular}
\end{table*}

\subsection{Text-to-Chart: Full Question Set}
\label{appendix:text-to-chart-full}

\begin{figure}[H]
\centering
\begin{minipage}{\textwidth}
    \textbf{Question 1}\\[0.5em]
    \textit{Source Text:} Since 1981, the leading cause of death in Japan has been cancer,
    which accounted for 27\% of total deaths in 2018, followed by heart disease at 15\%.''\\[0.5em]
    \textbf{Uncertainty Categories:} Inferential Derivation \quad
    \textbf{Chart Type:} Vertical Bar Chart
\end{minipage}
\vspace{1em}

\textit{Select the chart that best represents the uncertainty in the source text:}\\[0.8em]

\begin{minipage}{0.5\textwidth}
    \centering
    \includegraphics[width=\textwidth]{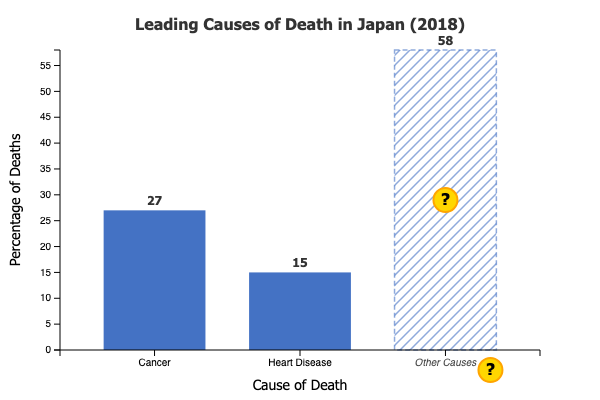}\\[0.3em]
    Uncertainty Visualization
\end{minipage}
\hfill
\begin{minipage}{0.5\textwidth}
    \centering
    \includegraphics[width=\textwidth]{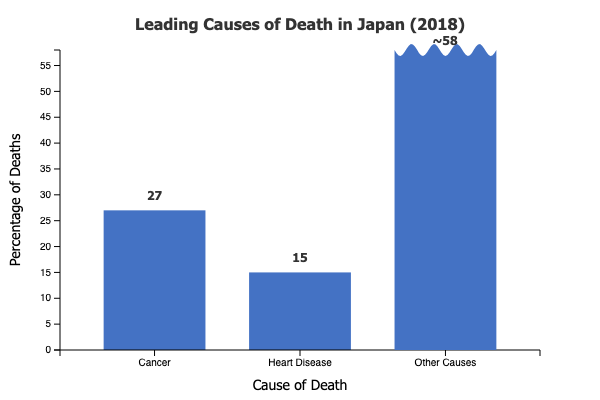}\\[0.3em]
    Alternative Visualization
\end{minipage}

\vspace{0.8em}

\begin{minipage}{0.5\textwidth}
    \centering
    \includegraphics[width=\textwidth]{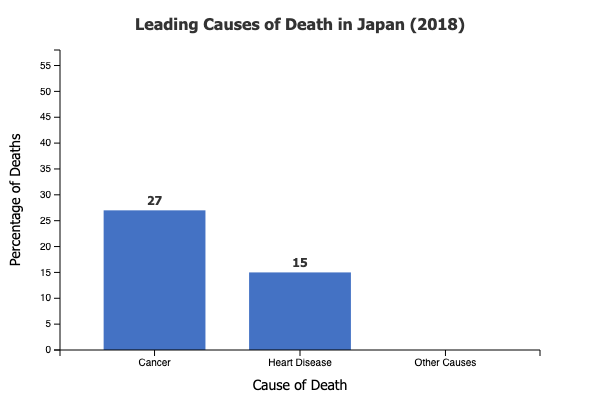}\\[0.3em]
    Certainty Visualization
\end{minipage}

\vspace{0.5em}
\hrule
\end{figure}

\begin{figure}[H]
\centering
\begin{minipage}{\textwidth}
    \textbf{Question 2}\\[0.5em]
    \textit{Source Text:} China's auto output exceeded 30.16 million units in 2023,
    up 11.6 percent year-on-year.''\\[0.5em]
    \textbf{Uncertainty Categories:} Inferential Derivation; Precision Boundaries \quad
    \textbf{Chart Type:} Vertical Bar Chart
\end{minipage}
\vspace{1em}

\textit{Select the chart that best represents the uncertainty in the source text:}\\[0.8em]

\begin{minipage}{0.5\textwidth}
    \centering
    \includegraphics[width=\textwidth]{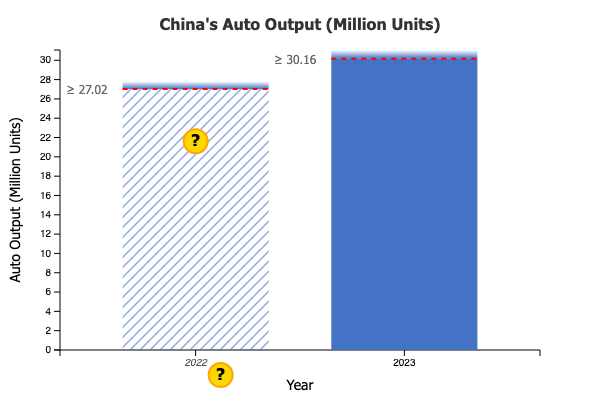}\\[0.3em]
    Uncertainty Visualization
\end{minipage}
\hfill
\begin{minipage}{0.5\textwidth}
    \centering
    \includegraphics[width=\textwidth]{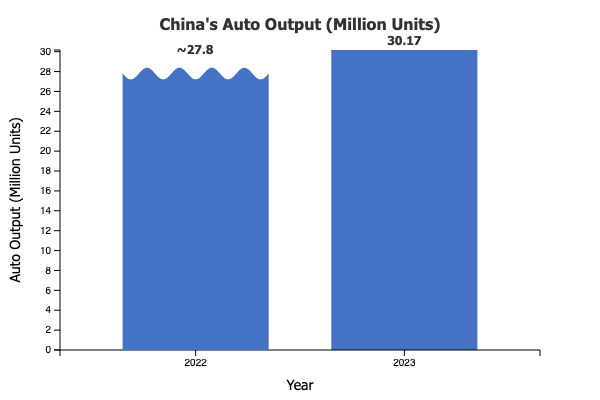}\\[0.3em]
    Alternative Visualization
\end{minipage}

\vspace{0.8em}

\begin{minipage}{0.5\textwidth}
    \centering
    \includegraphics[width=\textwidth]{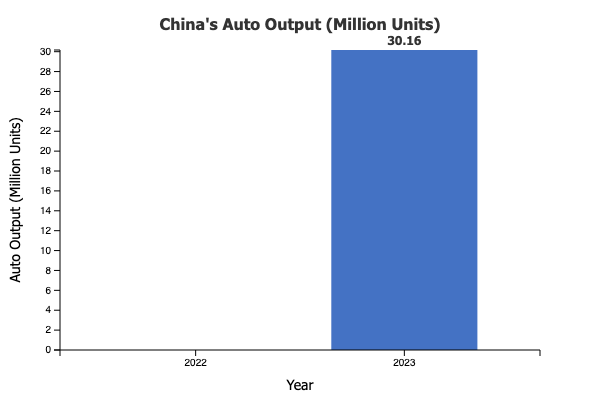}\\[0.3em]
    Certainty Visualization
\end{minipage}

\vspace{0.5em}
\hrule
\end{figure}

\begin{figure}[H]
\centering
\begin{minipage}{\textwidth}
    \textbf{Question 3}\\[0.5em]
    \textit{Source Text:} As we explained in a story last month, less than 15\% of those seeking
    asylum were ultimately granted it in fiscal years 2022 and 2023.''\\[0.5em]
    \textbf{Uncertainty Categories:} Precision Boundaries \quad
    \textbf{Chart Type:} Vertical Bar Chart
\end{minipage}
\vspace{1em}

\textit{Select the chart that best represents the uncertainty in the source text:}\\[0.8em]

\begin{minipage}{0.5\textwidth}
    \centering
    \includegraphics[width=\textwidth]{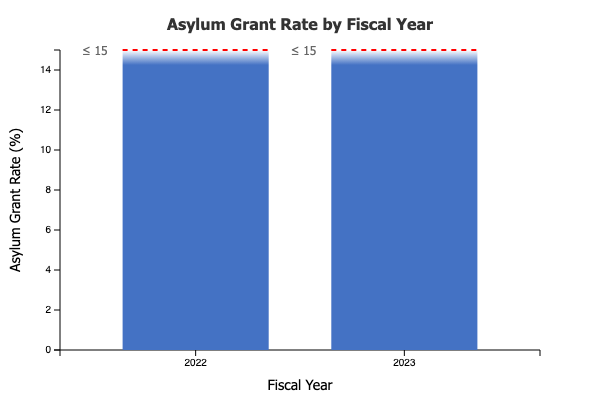}\\[0.3em]
    Uncertainty Visualization
\end{minipage}
\hfill
\begin{minipage}{0.5\textwidth}
    \centering
    \includegraphics[width=\textwidth]{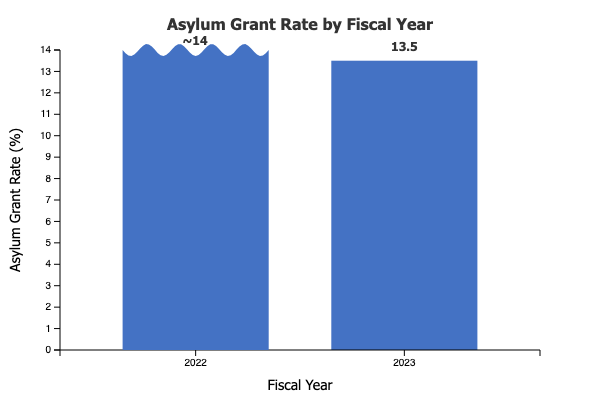}\\[0.3em]
    Alternative Visualization
\end{minipage}

\vspace{0.8em}

\begin{minipage}{0.5\textwidth}
    \centering
    \includegraphics[width=\textwidth]{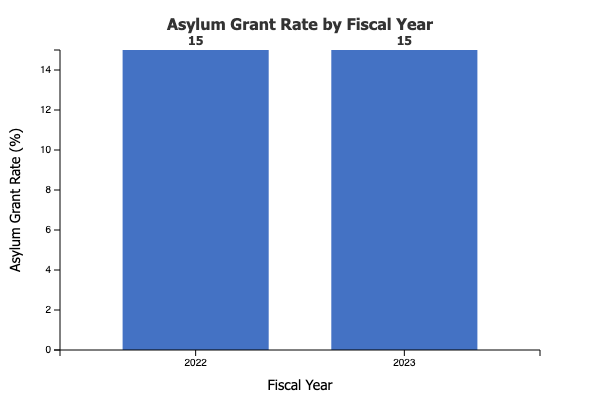}\\[0.3em]
    Certainty Visualization
\end{minipage}

\vspace{0.5em}
\hrule
\end{figure}

\begin{figure}[H]
\centering
\begin{minipage}{\textwidth}
    \textbf{Question 4}\\[0.5em]
    \textit{Source Text:} And we get 50--60 hectares [120--150 acres] of green space
    that keeps the city cool, prevents flooding.''\\[0.5em]
    \textbf{Uncertainty Categories:} Precision Boundaries \quad
    \textbf{Chart Type:} Vertical Bar Chart
\end{minipage}
\vspace{1em}

\textit{Select the chart that best represents the uncertainty in the source text:}\\[0.8em]

\begin{minipage}{0.5\textwidth}
    \centering
    \includegraphics[width=\textwidth]{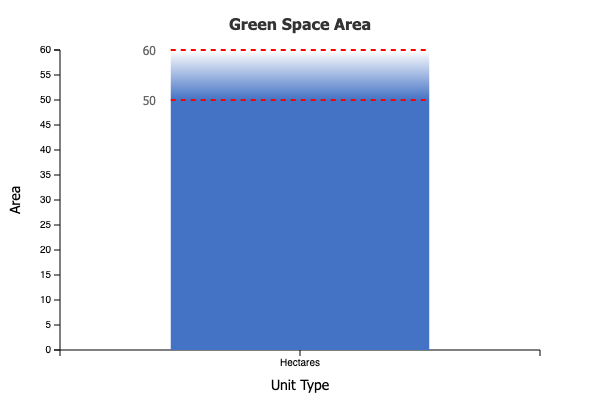}\\[0.3em]
    Uncertainty Visualization
\end{minipage}
\hfill
\begin{minipage}{0.5\textwidth}
    \centering
    \includegraphics[width=\textwidth]{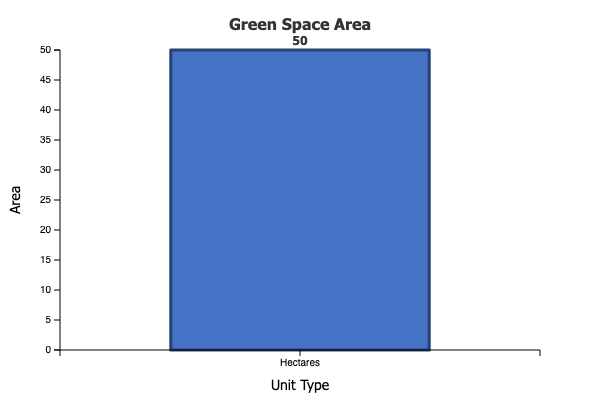}\\[0.3em]
    Alternative Visualization
\end{minipage}

\vspace{0.8em}

\begin{minipage}{0.5\textwidth}
    \centering
    \includegraphics[width=\textwidth]{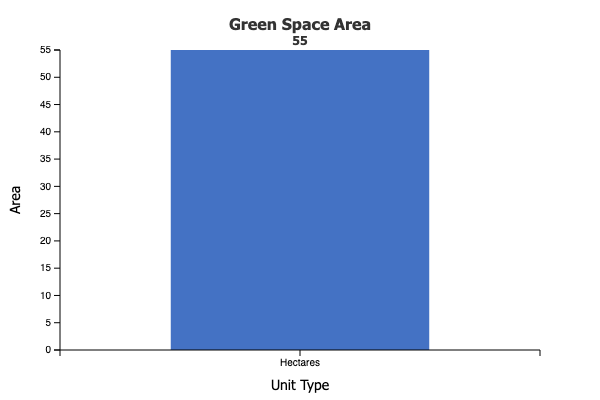}\\[0.3em]
    Certainty Visualization
\end{minipage}

\vspace{0.5em}
\hrule
\end{figure}

\begin{figure}[H]
\centering
\begin{minipage}{\textwidth}
    \textbf{Question 5}\\[0.5em]
    \textit{Source Text:} The most ambitious of which involved removing half of the
    open-air parking spaces in the city, about 11,000.''\\[0.5em]
    \textbf{Uncertainty Categories:} Inferential Derivation; Precision Boundaries \quad
    \textbf{Chart Type:} Vertical Bar Chart
\end{minipage}
\vspace{1em}

\textit{Select the chart that best represents the uncertainty in the source text:}\\[0.8em]

\begin{minipage}{0.5\textwidth}
    \centering
    \includegraphics[width=\textwidth]{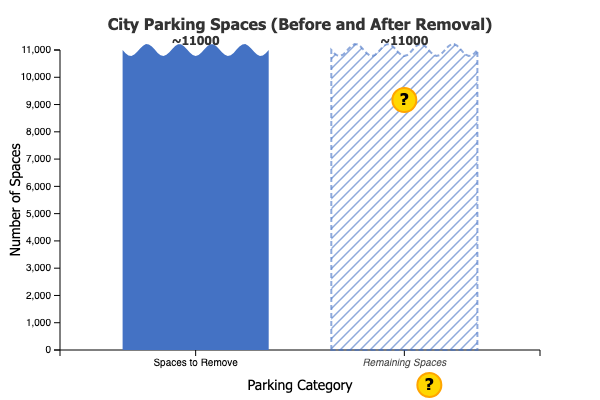}\\[0.3em]
    Uncertainty Visualization
\end{minipage}
\hfill
\begin{minipage}{0.5\textwidth}
    \centering
    \includegraphics[width=\textwidth]{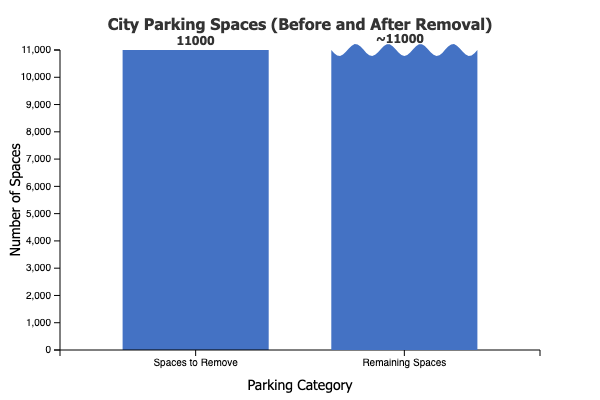}\\[0.3em]
    Alternative Visualization
\end{minipage}

\vspace{0.8em}

\begin{minipage}{0.5\textwidth}
    \centering
    \includegraphics[width=\textwidth]{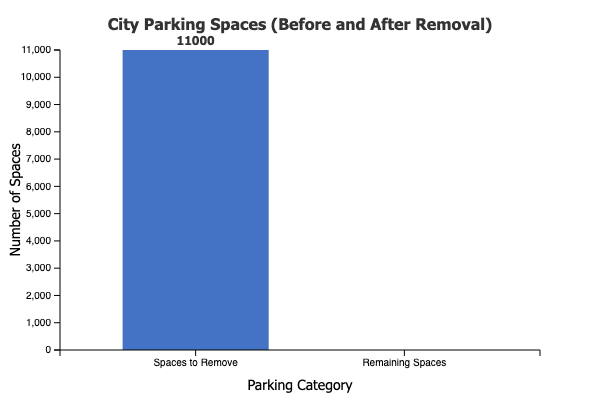}\\[0.3em]
    Certainty Visualization
\end{minipage}

\vspace{0.5em}
\hrule
\end{figure}

\begin{figure}[H]
\centering
\begin{minipage}{\textwidth}
    \textbf{Question 6}\\[0.5em]
    \textit{Source Text:} Half disapprove of the major changes he has proposed
    for government spending.''\\[0.5em]
    \textbf{Uncertainty Categories:} Surface Form Normalization; Inferential Derivation \quad
    \textbf{Chart Type:} Pie Chart
\end{minipage}
\vspace{1em}

\textit{Select the chart that best represents the uncertainty in the source text:}\\[0.8em]

\begin{minipage}{0.5\textwidth}
    \centering
    \includegraphics[width=\textwidth]{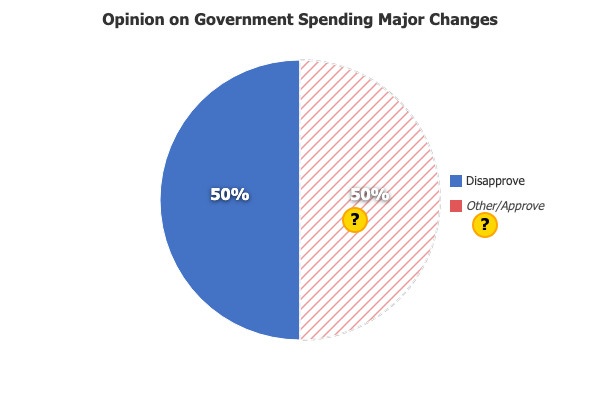}\\[0.3em]
    Uncertainty Visualization
\end{minipage}
\hfill
\begin{minipage}{0.5\textwidth}
    \centering
    \includegraphics[width=\textwidth]{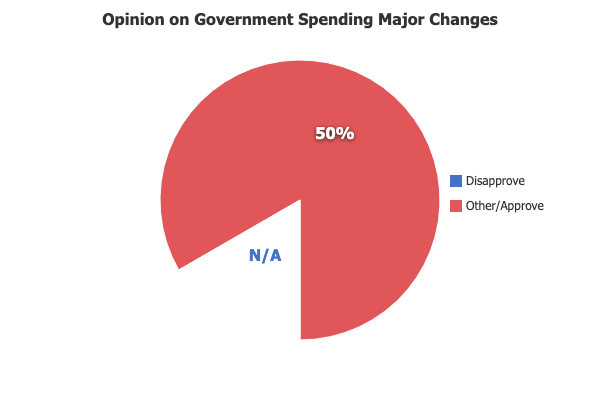}\\[0.3em]
    Alternative Visualization
\end{minipage}

\vspace{0.8em}

\begin{minipage}{0.5\textwidth}
    \centering
    \includegraphics[width=\textwidth]{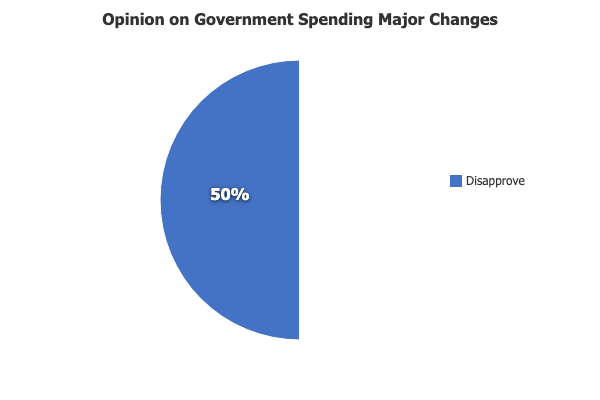}\\[0.3em]
    Certainty Visualization
\end{minipage}

\vspace{0.5em}
\hrule
\end{figure}

\begin{figure}[H]
\centering
\begin{minipage}{\textwidth}
    \textbf{Question 7}\\[0.5em]
    \textit{Source Text:} If this trend continues, Korea's population is estimated
    to halve by the year 2100.''\\[0.5em]
    \textbf{Uncertainty Categories:} Inferential Derivation \quad
    \textbf{Chart Type:} Line Chart
\end{minipage}
\vspace{1em}

\textit{Select the chart that best represents the uncertainty in the source text:}\\[0.8em]

\begin{minipage}{0.5\textwidth}
    \centering
    \includegraphics[width=\textwidth]{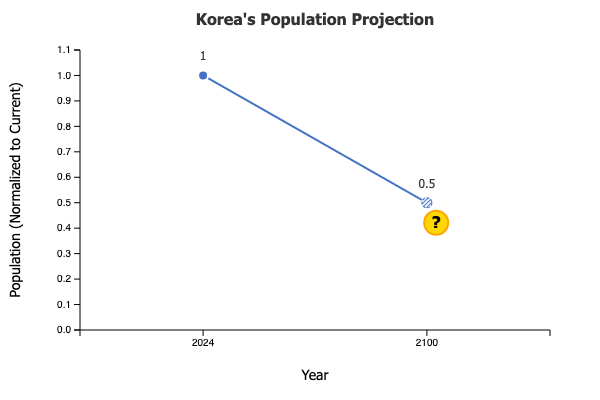}\\[0.3em]
    Uncertainty Visualization
\end{minipage}
\hfill
\begin{minipage}{0.5\textwidth}
    \centering
    \includegraphics[width=\textwidth]{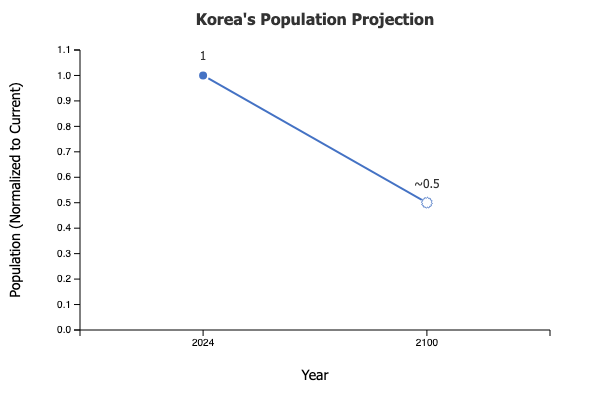}\\[0.3em]
    Alternative Visualization
\end{minipage}

\vspace{0.8em}

\begin{minipage}{0.5\textwidth}
    \centering
    \includegraphics[width=\textwidth]{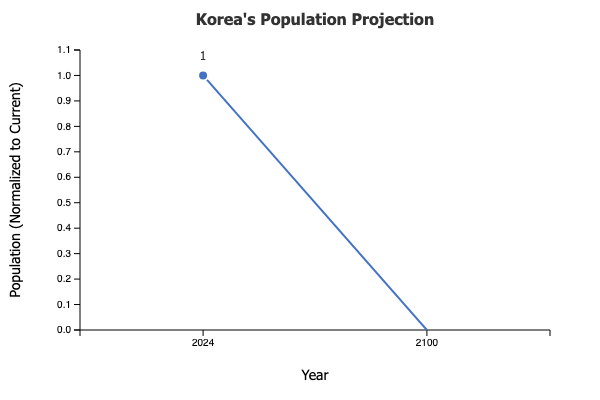}\\[0.3em]
    Certainty Visualization
\end{minipage}

\vspace{0.5em}
\hrule
\end{figure}

\begin{figure}[H]
\centering
\begin{minipage}{\textwidth}
    \textbf{Question 8}\\[0.5em]
    \textit{Source Text:} While relative poverty is unlikely to be eradicated, in Europe,
    over the last decade, it has been kept below 6.5 percent (using the 40 per cent of
    median disposable income standard).''\\[0.5em]
    \textbf{Uncertainty Categories:} Precision Boundaries \quad
    \textbf{Chart Type:} Line Chart
\end{minipage}
\vspace{1em}

\textit{Select the chart that best represents the uncertainty in the source text:}\\[0.8em]

\begin{minipage}{0.5\textwidth}
    \centering
    \includegraphics[width=\textwidth]{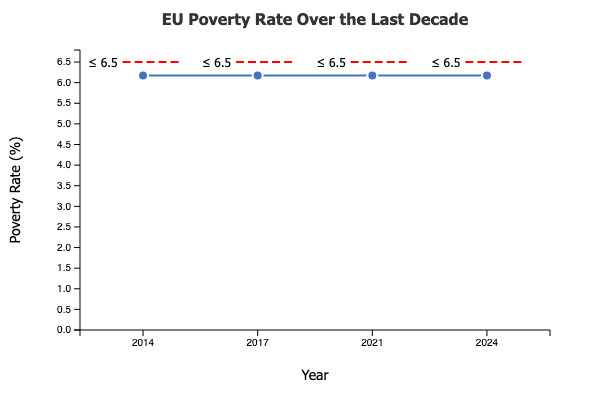}\\[0.3em]
    Uncertainty Visualization
\end{minipage}
\hfill
\begin{minipage}{0.5\textwidth}
    \centering
    \includegraphics[width=\textwidth]{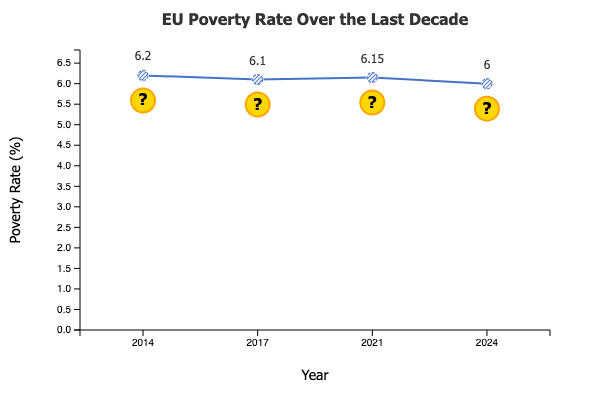}\\[0.3em]
    Alternative Visualization
\end{minipage}

\vspace{0.8em}

\begin{minipage}{0.5\textwidth}
    \centering
    \includegraphics[width=\textwidth]{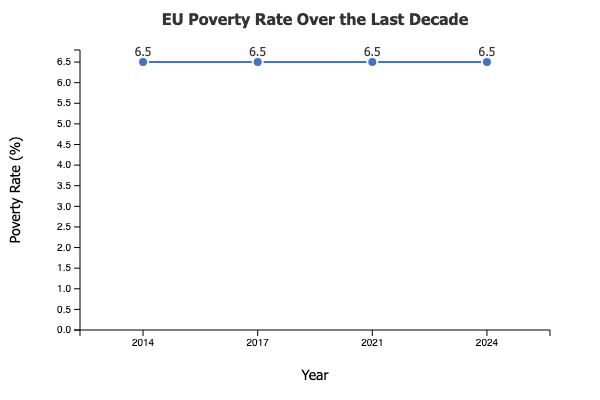}\\[0.3em]
    Certainty Visualization
\end{minipage}

\vspace{0.5em}
\hrule
\end{figure}

\begin{figure}[H]
\centering
\begin{minipage}{\textwidth}
    \textbf{Question 9}\\[0.5em]
    \textit{Source Text:} Smoking rates in men of around 80\% in 1970 now compare with
    around 30\% in 2013. In contrast, rates in women have remained constantly low,
    at around 15\% and 8\%, respectively.''\\[0.5em]
    \textbf{Uncertainty Categories:} Precision Boundaries \quad
    \textbf{Chart Type:} Line Chart
\end{minipage}
\vspace{1em}

\textit{Select the chart that best represents the uncertainty in the source text:}\\[0.8em]

\begin{minipage}{0.5\textwidth}
    \centering
    \includegraphics[width=\textwidth]{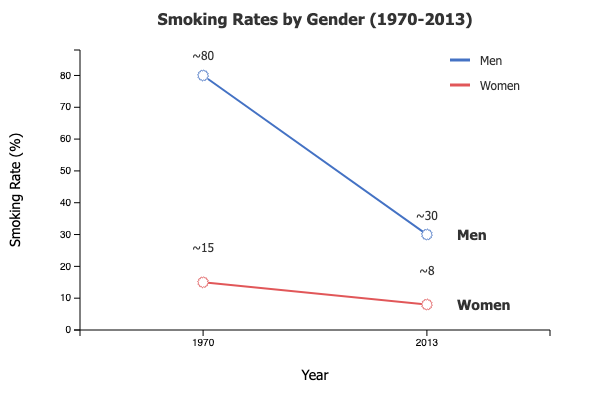}\\[0.3em]
    Uncertainty Visualization
\end{minipage}
\hfill
\begin{minipage}{0.5\textwidth}
    \centering
    \includegraphics[width=\textwidth]{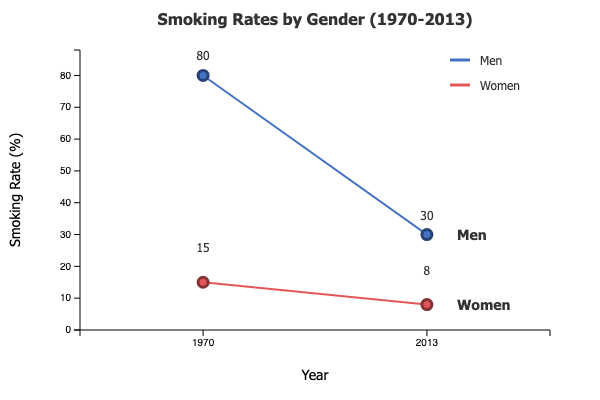}\\[0.3em]
    Alternative Visualization
\end{minipage}

\vspace{0.8em}

\begin{minipage}{0.5\textwidth}
    \centering
    \includegraphics[width=\textwidth]{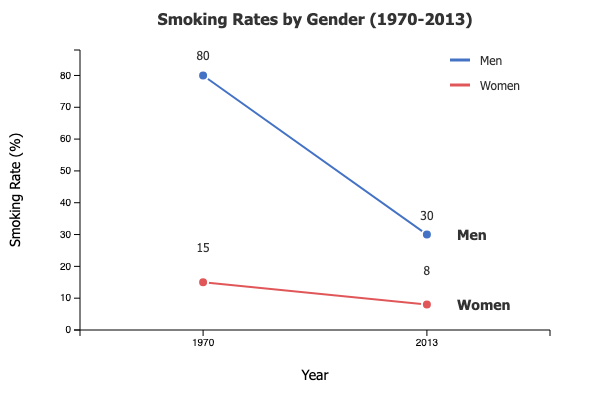}\\[0.3em]
    Certainty Visualization
\end{minipage}

\vspace{0.5em}
\hrule
\end{figure}

\begin{figure}[H]
\centering
\begin{minipage}{\textwidth}
    \textbf{Question 10}\\[0.5em]
    \textit{Source Text:} There are currently fewer than 40,000, according to the DOE,
    with a quarter of those in California.''\\[0.5em]
    \textbf{Uncertainty Categories:} Precision Boundaries \quad
    \textbf{Chart Type:} Vertical Bar Chart
\end{minipage}
\vspace{1em}

\textit{Select the chart that best represents the uncertainty in the source text:}\\[0.8em]

\begin{minipage}{0.5\textwidth}
    \centering
    \includegraphics[width=\textwidth]{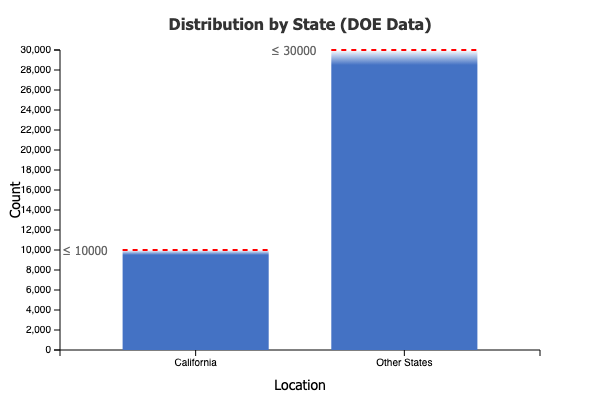}\\[0.3em]
    Uncertainty Visualization
\end{minipage}
\hfill
\begin{minipage}{0.5\textwidth}
    \centering
    \includegraphics[width=\textwidth]{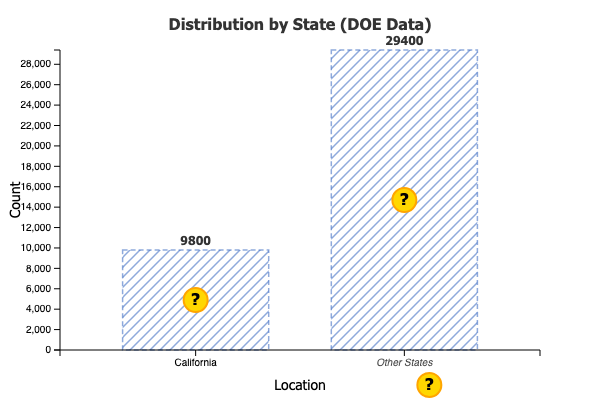}\\[0.3em]
    Alternative Visualization
\end{minipage}

\vspace{0.8em}

\begin{minipage}{0.5\textwidth}
    \centering
    \includegraphics[width=\textwidth]{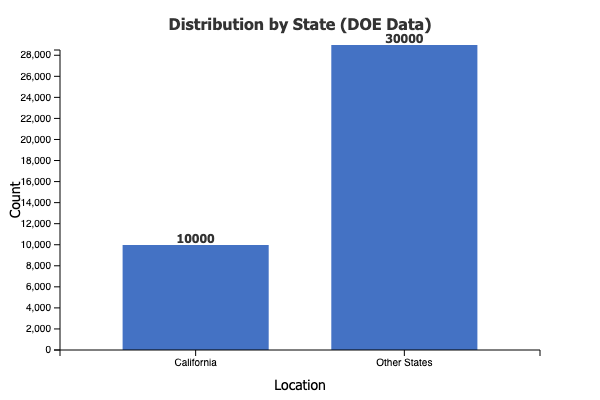}\\[0.3em]
    Certainty Visualization
\end{minipage}

\vspace{0.5em}
\hrule
\end{figure}

\begin{figure}[H]
\centering
\begin{minipage}{\textwidth}
    \textbf{Question 11}\\[0.5em]
    \textit{Source Text:} The 29-year-old Serbian centre, had 32 points, 16 rebounds
    and 16 assists plus four steals while Jamal Murray added 27 points for the Nuggets.''\\[0.5em]
    \textbf{Uncertainty Categories:} Surface Form Normalization; Non-Inferable Gaps \quad
    \textbf{Chart Type:} Grouped Bar Chart
\end{minipage}
\vspace{1em}

\textit{Select the chart that best represents the uncertainty in the source text:}\\[0.8em]

\begin{minipage}{0.5\textwidth}
    \centering
    \includegraphics[width=\textwidth]{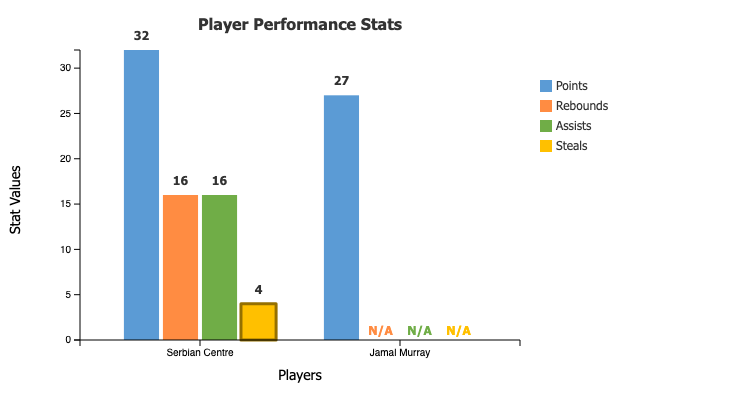}\\[0.3em]
    Uncertainty Visualization
\end{minipage}
\hfill
\begin{minipage}{0.5\textwidth}
    \centering
    \includegraphics[width=\textwidth]{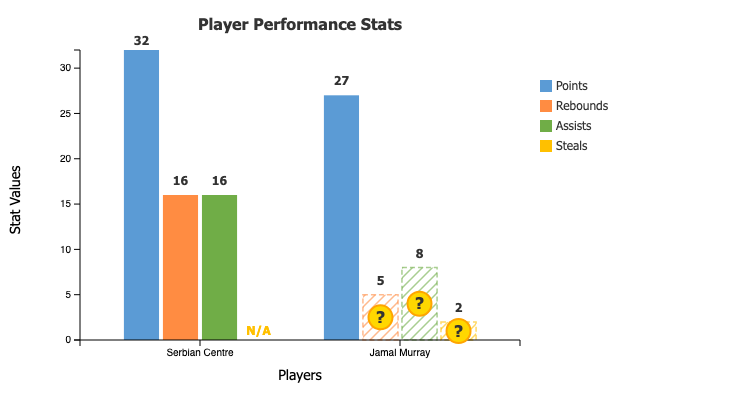}\\[0.3em]
    Alternative Visualization
\end{minipage}

\vspace{0.8em}

\begin{minipage}{0.5\textwidth}
    \centering
    \includegraphics[width=\textwidth]{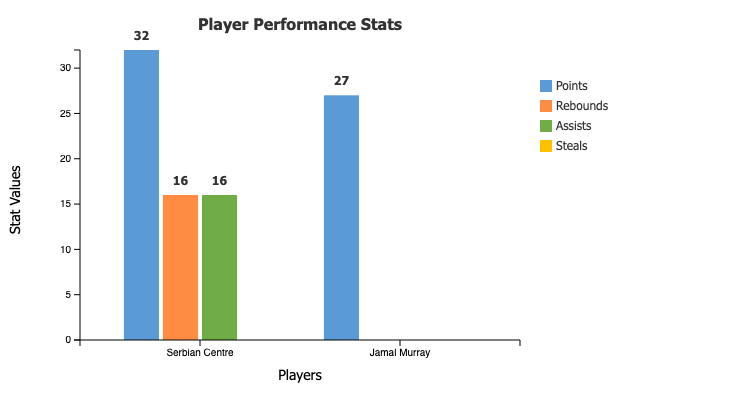}\\[0.3em]
    Certainty Visualization
\end{minipage}

\vspace{0.5em}
\hrule
\end{figure}

\begin{figure}[H]
\centering
\begin{minipage}{\textwidth}
    \textbf{Question 12}\\[0.5em]
    \textit{Source Text:} In 50 years time, the number of working age people will have
    halved, the pool eligible to take part in the country's mandatory military service
    will have shrunk by 58\%.''\\[0.5em]
    \textbf{Uncertainty Categories:} Inferential Derivation \quad
    \textbf{Chart Type:} Grouped Bar Chart
\end{minipage}
\vspace{1em}

\textit{Select the chart that best represents the uncertainty in the source text:}\\[0.8em]

\begin{minipage}{0.5\textwidth}
    \centering
    \includegraphics[width=\textwidth]{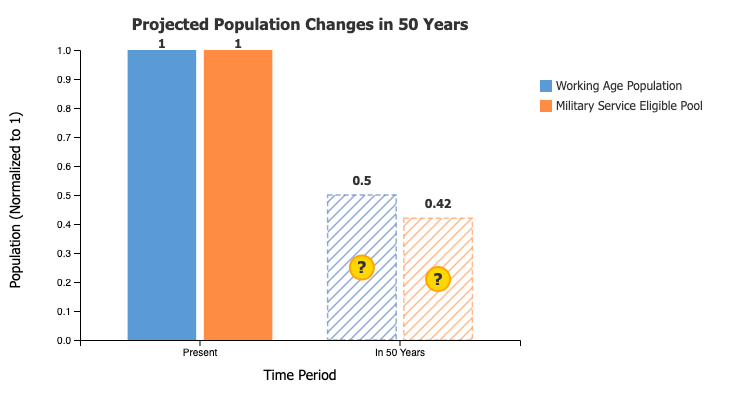}\\[0.3em]
    Uncertainty Visualization
\end{minipage}
\hfill
\begin{minipage}{0.5\textwidth}
    \centering
    \includegraphics[width=\textwidth]{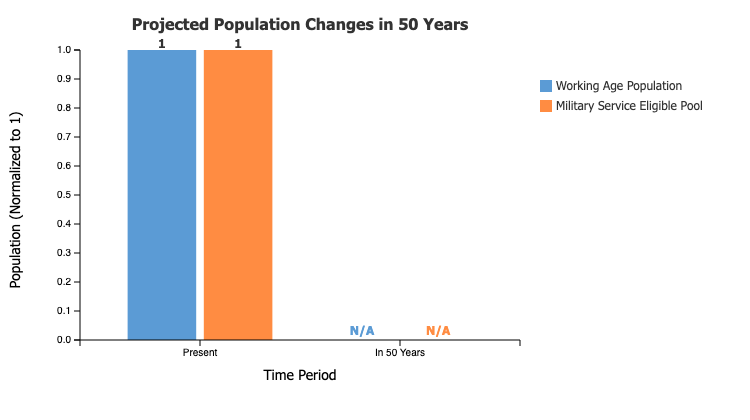}\\[0.3em]
    Alternative Visualization
\end{minipage}

\vspace{0.8em}

\begin{minipage}{0.5\textwidth}
    \centering
    \includegraphics[width=\textwidth]{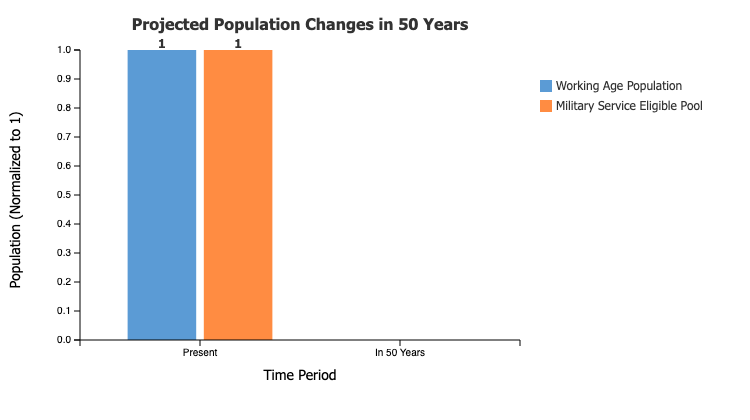}\\[0.3em]
    Certainty Visualization
\end{minipage}

\vspace{0.5em}
\hrule
\end{figure}

\subsection{Chart-to-Text: Full Question Set}
\label{appendix:chart-to-text-full}

For each question, participants were shown a chart and asked to select 
the sentence that best represents the uncertainty portrayed in the chart.

\vspace{1em}
\noindent\textbf{Question 1} \hfill

\noindent\textbf{Uncertainty Categories:} Precision Boundaries, Inferential Derivation \quad
\textbf{Chart Type:} Stacked Bar

\begin{center}
  \includegraphics[width=0.75\textwidth]{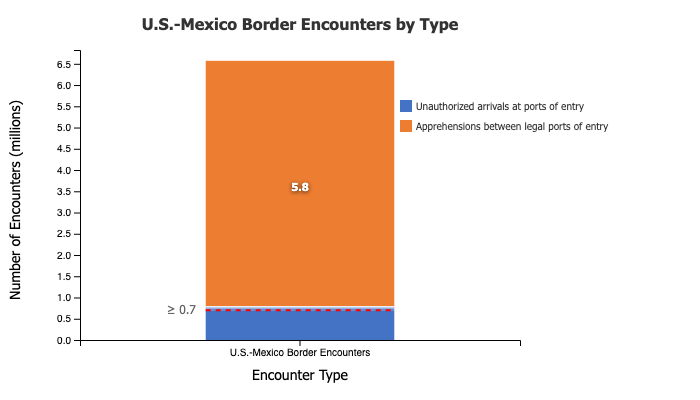}
\end{center}

\begin{itemize}
  \item (Source Text) The DHS data show 6.5 million encounters at the 
  U.S.-Mexico border in that time frame, a figure that includes both the 
  5.8 million apprehensions between legal ports of entry — the number 
  typically used for illegal immigration — and a little more than 700,000 
  migrants who arrived at ports of entry without authorization to enter 
  the U.S.

  \item (Alternative Text 1) The DHS data show 6.5 million encounters at the 
  U.S.-Mexico border in that time frame, a figure that includes both 
  approximately 5.8 million apprehensions between legal ports of entry — 
  the number typically used for illegal immigration — and migrants who 
  arrived at ports of entry without authorization totaling around 710,000.

  \item (Alternative Text 2) The DHS data show 6.5 million encounters at the 
  U.S.-Mexico border in that time frame, a figure that includes both the 
  5.8 million apprehensions between legal ports of entry — the number 
  typically used for illegal immigration — and 700,100 migrants who arrived 
  at ports of entry without authorization to enter the U.S.
\end{itemize}

\hrule
\vspace{1em}

\noindent\textbf{Question 2} \hfill

\noindent\textbf{Uncertainty Categories:} Precision Boundaries, Inferential Derivation \quad
\textbf{Chart Type:} Stacked Bar

\begin{center}
  \includegraphics[width=0.75\textwidth]{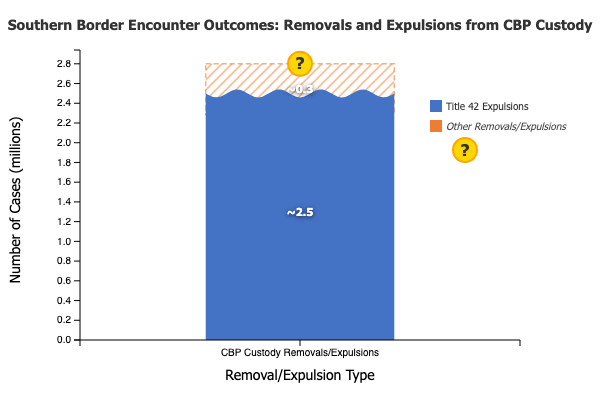}
\end{center}

\begin{itemize}
  \item (Source Text) But these claims ignore that DHS statistics show 
  2.8 million of the encounters at the southern border alone resulted in 
  a removal or expulsion directly from CBP custody, and all of the rest 
  of the migrants encountered are not simply released. Most of those 
  removals — approximately 2.5 million — were immediate expulsions 
  under Title 42.

  \item (Alternative Text 1) But these claims ignore that DHS statistics show 
  2.8 million of the encounters at the southern border alone resulted in 
  a removal or expulsion directly from CBP custody, and all of the rest 
  of the migrants encountered are not simply released. Most of those 
  removals — two point five million — were immediate expulsions under 
  Title 42, with 0.3 million other removals and expulsions.

  \item (Alternative Text 2) But these claims ignore that DHS statistics show 
  2.8 million of the encounters at the southern border alone resulted in 
  a removal or expulsion directly from CBP custody, and all of the rest 
  of the migrants encountered are not simply released. Most of those 
  removals — 2.5 million — were immediate expulsions under Title 42.
\end{itemize}

\hrule
\vspace{1em}

\noindent\textbf{Question 3} \hfill

\noindent\textbf{Uncertainty Categories:} Inferential Derivation \quad
\textbf{Chart Type:} Stacked Bar

\begin{center}
  \includegraphics[width=0.75\textwidth]{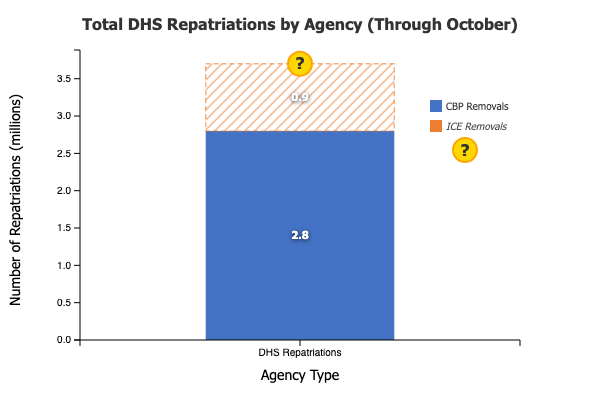}
\end{center}

\begin{itemize}
  \item (Source Text) Total DHS repatriations through October amounted to 
  3.7 million, a figure that includes the 2.8 million removals directly 
  from CBP, as well as removals by ICE.

  \item (Alternative Text 1) Total DHS repatriations through October amounted to 
  three point seven million, a figure that includes the two point eight 
  million removals directly from CBP, as well as approximately 0.9 million 
  removals by ICE.

  \item (Alternative Text 2) Total DHS repatriations through October amounted to 
  around 3.7 million, a figure that includes around 2.8 million removals 
  directly from CBP, as well as removals by ICE.
\end{itemize}

\hrule
\vspace{1em}

\noindent\textbf{Question 4} \hfill

\noindent\textbf{Uncertainty Categories:} Precision Boundaries, Inferential Derivation \quad
\textbf{Chart Type:} Stacked Bar

\begin{center}
  \includegraphics[width=0.75\textwidth]{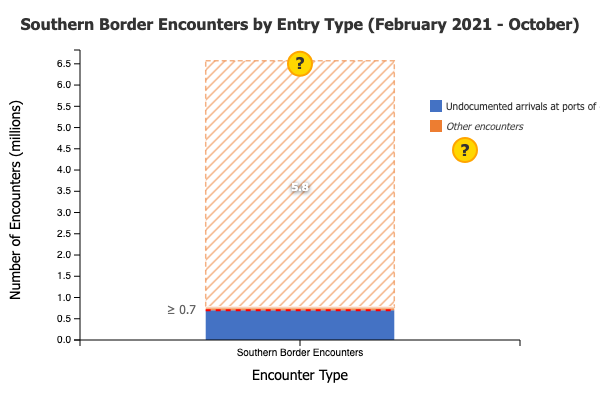}
\end{center}

\begin{itemize}
  \item (Source Text) As we said, there were 6.5 million encounters at 
  the southern border from February 2021 through October, including a 
  little more than 700,000 migrants who arrived without legal documentation 
  at ports of entry.

  \item (Alternative Text 1) As we said, there were 6.5 million encounters at 
  the southern border from February 2021 through October, where some 
  migrants arrived without legal documentation at ports of entry.

  \item (Alternative Text 2) As we said, there were 6.5 million encounters at 
  the southern border from February 2021 through October, including 
  700,000 migrants who arrived without legal documentation at ports 
  of entry.
\end{itemize}

\hrule
\vspace{1em}

\noindent\textbf{Question 5} \hfill

\noindent\textbf{Uncertainty Categories:} Precision Boundaries \quad
\textbf{Chart Type:} Stacked Bar

\begin{center}
  \includegraphics[width=0.75\textwidth]{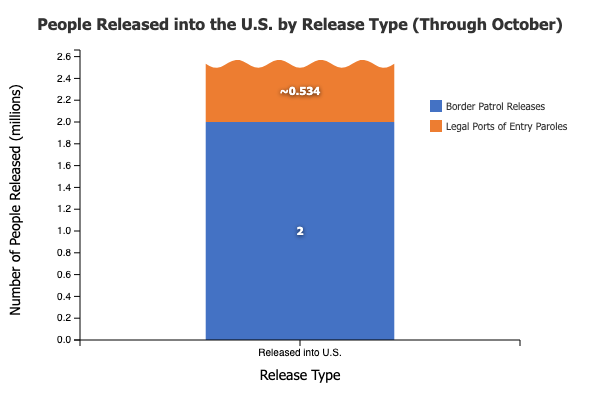}
\end{center}

\begin{itemize}
  \item (Source Text) About 2.5 million people through October have been 
  released into the U.S. That figure includes 2 million released by 
  Border Patrol, with a notice to appear in court or a notice to report 
  to ICE, or released through prosecutorial discretion or granted 
  humanitarian parole, which allows people into the country for a 
  temporary period. The 2.5 million number also includes approximately 
  534,000 paroles processed at legal ports of entry.

  \item (Alternative Text 1) About 2.5 million people through October have been 
  released into the U.S. That figure includes around 2 million released 
  by Border Patrol, with a notice to appear in court or a notice to report 
  to ICE, or released through prosecutorial discretion or granted 
  humanitarian parole, which allows people into the country for a 
  temporary period. The 2.5 million number also includes less than 534,000 
  paroles processed at legal ports of entry.

  \item (Alternative Text 2) 2.5 million people through October have been released 
  into the U.S. That figure includes 2 million released by Border Patrol, 
  with a notice to appear in court or a notice to report to ICE, or 
  released through prosecutorial discretion or granted humanitarian parole, 
  which allows people into the country for a temporary period. The 2.5 
  million number also includes 534,000 paroles processed at legal ports 
  of entry.
\end{itemize}

\hrule
\vspace{1em}

\noindent\textbf{Question 6} \hfill

\noindent\textbf{Uncertainty Categories:} Inferential Derivation \quad
\textbf{Chart Type:} Stacked Bar

\begin{center}
  \includegraphics[width=0.75\textwidth]{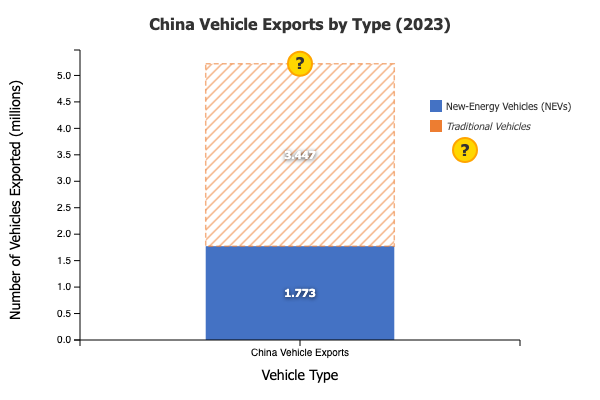}
\end{center}

\begin{itemize}
  \item (Source Text) China exported 5.22 million vehicles in 2023, 
  becoming a leading car exporter in the world. Of all vehicle exports, 
  one-third were new-energy vehicles (NEVs), totaling 1.773 million units.

  \item (Alternative Text 1) China exported 5.22 million vehicles in 2023, 
  becoming a leading car exporter in the world. Of all vehicle exports, 
  new-energy vehicles (NEVs) totaled 1.773 million units, with three 
  point four four seven million being traditional vehicles.

  \item (Alternative Text 2) China exported 5.22 million vehicles in 2023, 
  becoming a leading car exporter in the world. Of all vehicle exports, 
  more than 1.773 million units were new-energy vehicles (NEVs).
\end{itemize}

\hrule
\vspace{1em}

\noindent\textbf{Question 7} \hfill

\noindent\textbf{Uncertainty Categories:} Precision Boundaries, Inferential Derivation \quad
\textbf{Chart Type:} Stacked Bar

\begin{center}
  \includegraphics[width=0.75\textwidth]{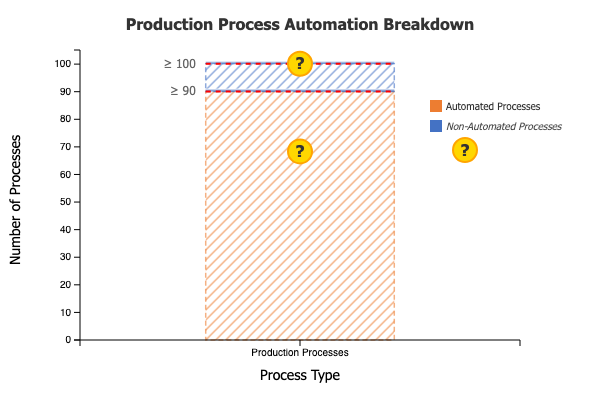}
\end{center}

\begin{itemize}
  \item (Source Text) The facility includes the world's most advanced 
  lidar testing lab and will utilize many smart industrial robots allowing 
  for the automation of over 100 production processes with an automation 
  rate of 90 percent.

  \item (Alternative Text 1) The facility includes the world's most advanced 
  lidar testing lab and will utilize many smart industrial robots allowing 
  for the automation of production processes with an automation rate of 
  ninety percent, though the number of non-automated processes remains 
  unclear.

  \item (Alternative Text 2) The facility includes the world's most advanced 
  lidar testing lab and will utilize many smart industrial robots allowing 
  for the automation of over 100 production processes with an automation 
  rate of less than 90 percent.
\end{itemize}

\hrule
\vspace{1em}

\noindent\textbf{Question 8} \hfill

\noindent\textbf{Uncertainty Categories:} Precision Boundaries \quad
\textbf{Chart Type:} Vertical Bar

\begin{center}
  \includegraphics[width=0.75\textwidth]{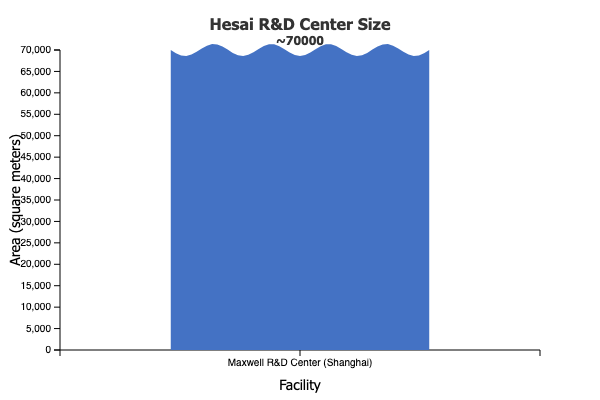}
\end{center}

\begin{itemize}
  \item (Source Text) Hesai, which also participated in the CES 2024, 
  has recently completed the construction of a new research and 
  development (R\&D) center of approximate 70,000 square meters in 
  Shanghai for manufacturing named Maxwell, the company told the 
  Global Times.

  \item (Alternative Text 1) Hesai, which also participated in the CES 2024, 
  has recently completed the construction of a new research and 
  development (R\&D) center of large size in Shanghai for manufacturing 
  named Maxwell, the company told the Global Times.

  \item (Alternative Text 2) Hesai, which also participated in the CES 2024, 
  has recently completed the construction of a new research and 
  development (R\&D) center of 70,000 square meters in Shanghai for 
  manufacturing named Maxwell, the company told the Global Times.
\end{itemize}

\hrule
\vspace{1em}

\noindent\textbf{Question 9} \hfill

\noindent\textbf{Uncertainty Categories:} Surface Form Normalization, Inferential Derivation \quad
\textbf{Chart Type:} Grouped Bar Chart

\begin{center}
  \includegraphics[width=0.75\textwidth]{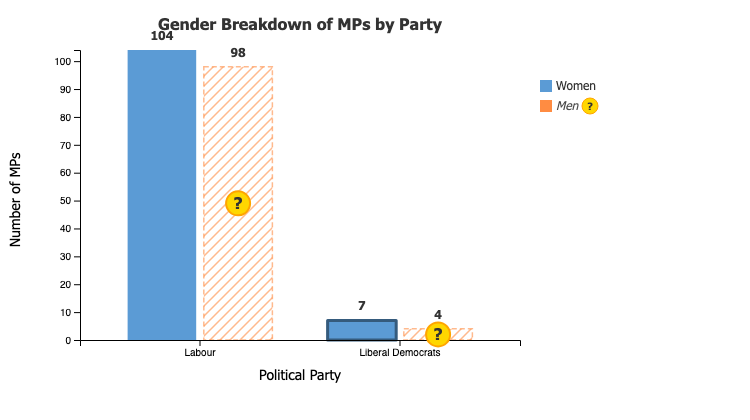}
\end{center}

\begin{itemize}
  \item (Source Text) Of Labour's 202 MPs (excluding Speaker Lindsay 
  Hoyle), 104 are women — and of the Liberal Democrats' 11 MPs, 
  seven are women.

  \item (Alternative Text 1) Of Labour's 202 MPs (excluding Speaker Lindsay 
  Hoyle), one hundred and five are women and ninety-eight are men — 
  and of the Liberal Democrats' 11 MPs, approximately 7 are women 
  with four being men.

  \item (Alternative Text 2) Of Labour's 202 MPs (excluding Speaker Lindsay 
  Hoyle), 104 are women — and of the Liberal Democrats' 11 MPs, 
  the majority are women.
\end{itemize}

\hrule
\vspace{1em}

\noindent\textbf{Question 10} \hfill

\noindent\textbf{Uncertainty Categories:} Surface Form Normalization, Inferential Derivation \quad
\textbf{Chart Type:} Pie

\begin{center}
  \includegraphics[width=0.75\textwidth]{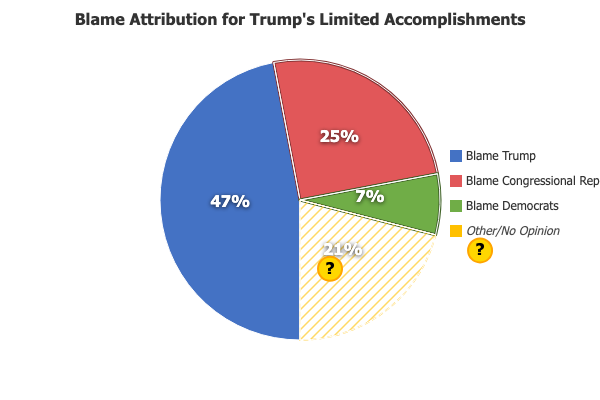}
\end{center}

\begin{itemize}
  \item (Source Text) Of those who say Trump has not accomplished much, 
  47 per cent pin the blame on him while a quarter blame congressional 
  Republicans. Seven per cent say Democrats are to blame.

  \item (Alternative Text 1) Of those who say Trump has not accomplished much, 
  47 per cent pin the blame on him while twenty-one per cent have other 
  opinions or no clear opinion, though some also mention congressional 
  Republicans and Democrats without specific attribution.

  \item (Alternative Text 2) Of those who say Trump has not accomplished much, 
  47 per cent pin the blame on him while 20 per cent blame congressional 
  Republicans. Seven per cent say Democrats are to blame.
\end{itemize}

\hrule
\vspace{1em}

\noindent\textbf{Question 11} \hfill

\noindent\textbf{Uncertainty Categories:} Inferential Derivation \quad
\textbf{Chart Type:} Pie

\begin{center}
  \includegraphics[width=0.75\textwidth]{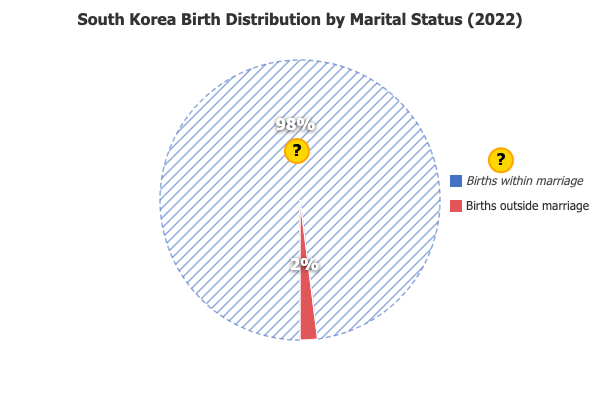}
\end{center}

\begin{itemize}
  \item (Source Text) In 2022, only 2\% were outside marriage.

  \item (Alternative Text 1) In 2022, a really small portion of births occurred 
  outside marriage.

  \item (Alternative Text 2) In 2022, the majority of births in South Korea were 
  within marriage, while only 2\% were outside marriage.
\end{itemize}

\hrule
\vspace{1em}

\noindent\textbf{Question 12} \hfill

\noindent\textbf{Uncertainty Categories:} Surface Form Normalization, Non-Inferable Gaps \quad
\textbf{Chart Type:} Grouped Bar Chart

\begin{center}
  \includegraphics[width=0.75\textwidth]{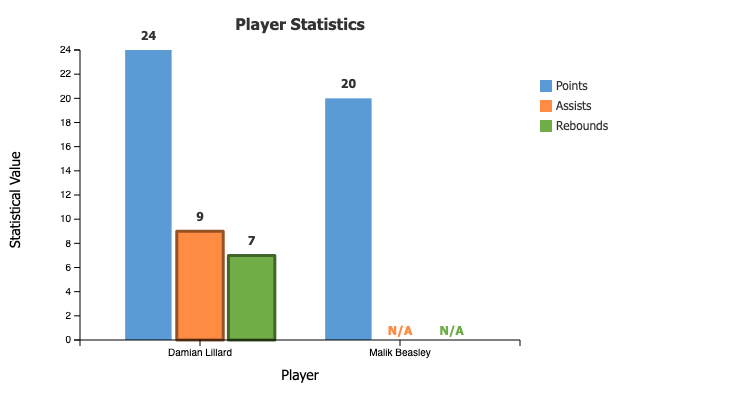}
\end{center}

\begin{itemize}
  \item (Source Text) Damian Lillard added 24 points, nine assists and 
  seven rebounds, while Malik Beasley added 20 points.

  \item (Alternative Text 1) Damian Lillard added 24 points, while Malik Beasley 
  added 20 points.

  \item (Alternative Text 2) Damian Lillard added 24 points, nine assists and 
  seven rebounds.
\end{itemize}

\end{document}